%% file: ms.tex
\pdfoutput=1

\documentclass[twocolumn,times,numberedappendix,appendixfloats]{aastex61}

\usepackage{epstopdf}	
\usepackage{graphicx}	
\usepackage{latexsym}	
\usepackage{bm}		
\usepackage[english]{babel}
\usepackage{booktabs}
\usepackage{amsmath}	
\usepackage[Symbol]{upgreek}	
\usepackage{natbib}	

\newlength{\txw}
\newlength{\txh}

\newcommand{\fesc}   {\ensuremath{f_{\rm esc}}}
\newcommand{\fescabs}{\ensuremath{f_{\rm esc}^{\rm abs}}}
\newcommand{\fescrel}{\ensuremath{f_{\rm esc}^{\rm rel}}}
\newcommand{\eg}     {e.g.,}
\newcommand{\ie}     {i.e.,}

\newcommand{\HST}    {\emph{HST}}
\newcommand{\JWST}   {\emph{JWST}}
\newcommand{\AV}     {\ensuremath{A_{V}}}

\newcommand{\emin}   {\ensuremath{e^{-}}}
\newcommand{\Lya}    {\ensuremath{\mbox{\textrm{Ly}}\alpha}}
\newcommand{\magarc} {mag\,arcsec$^{-2}$}

\newcommand{\mAB}    {\ensuremath{m_{\text{\tiny AB}}}}
\newcommand{\MAB}    {\ensuremath{M_{\rm AB}}}
\newcommand{\zmean}  {\ensuremath{\langle z\rangle}}
\newcommand{\Msol}   {\ensuremath{{\rm M}_{\odot}}}
\newcommand{\igm}    {\textnormal{\tiny \textsc{IGM}}}
\newcommand{\lyc}    {\textnormal{\fontsize{6}{6} \textsc{L}\text{y}\textsc{C}}}
\newcommand{\uvc}    {\textnormal{\tiny \text{UVC}}}
\newcommand{\SExtractor}{\textsc{SExtractor}}

\defcitealias{Windhorst2011}{W11}

\begin{document}

\title{Hubble Space Telescope Wide Field Camera 3 Observations of Escaping
Lyman Continuum Radiation from Galaxies and Weak AGN at Redshifts
\lowercase{$z$}\,$\simeq\,$2.3--4.1}

\correspondingauthor{Brent M. Smith}\author{Brent M. Smith}\affiliation{School of Earth \& Space Exploration, Arizona State University, Tempe, AZ 85287-1404, USA}\email{ Brent.Smith.1@asu.edu}
\author{Rogier A. Windhorst}
\affiliation{School of Earth \& Space Exploration, Arizona State University, Tempe, AZ 85287-1404, USA}
\author{Rolf A. Jansen}
\affiliation{School of Earth \& Space Exploration, Arizona State University, Tempe, AZ 85287-1404, USA}
\author{Seth H. Cohen}
\affiliation{School of Earth \& Space Exploration, Arizona State University, Tempe, AZ 85287-1404, USA}
\author{Linhua Jiang}
\affiliation{School of Earth \& Space Exploration, Arizona State University, Tempe, AZ 85287-1404, USA}
\affiliation{Kavli Institute for Astronomy and Astrophysics, Peking University, Beijing 100871, P. R. China}
\author{Mark Dijkstra}
\affiliation{Institute of Theoretical Astrophysics, University of Oslo, Postboks 1029, 0858 Oslo, Norway}
\author{Anton M. Koekemoer}
\affiliation{Space Telescope Science Institute, Baltimore, MD 21218, USA} 
\author{Richard Bielby}
\affiliation{University of Durham, South Road, Durham DH1 3LE, UK} 
\author{Akio K. Inoue}
\affiliation{Osaka Sangyo University, Osaka, Japan}
\author{John W. MacKenty}
\affiliation{Space Telescope Science Institute, Baltimore, MD 21218, USA} 
\author{Robert W. O'Connell}
\affiliation{Department of Astronomy, University of Virginia, Charlottesville, VA 22904-4325, USA}
\author{Joseph I. Silk}
\affiliation{The Johns Hopkins University, Baltimore, MD 21218, USA}

\shortauthors{Smith, B., et al.} 
\shorttitle{Escaping Lyman Continuum at $z$$\simeq$2.3--4.1}

\begin{abstract}
We present observations of escaping Lyman Continuum (LyC)
radiation from 34 massive star-forming galaxies and 12 weak AGN with reliably
measured spectroscopic redshifts at $z$$\simeq$2.3--4.1.  We analyzed Hubble
Space Telescope (\HST) Wide Field Camera 3 (WFC3) mosaics of the Early Release
Science (ERS) field in three UVIS filters to sample the
rest-frame LyC over this redshift range.  With our best current assessment of
the WFC3 systematics, we provide 1$\sigma$ upper limits for the average LyC
emission of galaxies at \zmean=2.35, 2.75, and 3.60 to $\sim$28.5, 28.1, and 30.7
mag in image stacks of 11--15 galaxies in the WFC3/UVIS F225W, F275W, and F336W,
respectively. The LyC flux of weak AGN at \zmean=2.62 and 3.32 are detected at
28.3 and 27.4 mag with SNRs of $\sim$2.7 and 2.5 in F275W and F336W for stacks of 7 and
3 AGN, respectively, while AGN at \zmean=2.37 are constrained to $\gtrsim$27.9\,mag
at 1$\sigma$ in a stack of 2 AGN.  The stacked AGN LyC light profiles are
flatter than their corresponding non-ionizing UV continuum
profiles out to radii of r$\lesssim$0\farcs9, which may indicate a radial dependence
of porosity in the ISM. With synthetic stellar SEDs fit to UV continuum measurements
longwards of \Lya\ and IGM transmission models, we constrain the absolute LyC
escape fractions to \fescabs$\simeq22^{+44}_{-22}$\% at \zmean=2.35 and $\lesssim$55\% at
\zmean=2.75  and 3.60, respectively. All available data for galaxies, including published
work, suggests a more sudden increase of \fesc\ with redshift at $z$$\simeq$2. Dust
accumulating in (massive) galaxies over cosmic time correlates with increased \ion{H}{1} column density, which may lead to reducing \fesc\ more suddenly at $z$$\lesssim$2. This may suggest that
star-forming galaxies collectively contributed to maintaining cosmic reionization at
redshifts $z$$\gtrsim$2--4, while AGN likely dominated reionization at $z$$\lesssim$2. 
\end{abstract}

\journalinfo{\textrm{(submitted to)} The Astrophysical Journal}

\section{Introduction} 
\label{sec:intro}

At the end of the cosmic dark ages, radiation emitted by the first luminous 
objects in the universe began to reionize the intergalactic medium (IGM).
The far-ultraviolet (FUV) ionizing radiation, specifically the Lyman continuum
(``LyC''; $\lambda$\,$\leq$\,912\AA), emitted by massive stars in the first
star-forming galaxies (SFGs), or accretion disks around supermassive black
holes in early Active Galactic Nuclei (AGN), may have initiated the epoch of
cosmic reionization \citep[e.g.,][]{Madau2004}.  Additional sources of LyC
radiation and high energy particles within galaxies, such as high mass X-ray
binaries, galactic outflows/inflows and superwinds, accretion onto dark matter
halos, massive pre-galactic Pop\,III stars, and young globular clusters may
have contributed to the reionization of the IGM as well
\citep[e.g.,][]{Ricotti2002, Sternberg2002, Mirabel2011, Kulkarni2014}.  This
LyC radiation would have formed bubbles of ionized hydrogen around these UV
bright galaxies, which then expanded and merged until the IGM became
completely ionized \citep[e.g.,][]{Gnedin2000, MiraldaEscude2000, Loeb2001,
Fan2002}.  This phase transition of the neutral IGM began somewhere in the
epoch $z$\,$\simeq$\,10--20 \citep{Hinshaw2013, Ade2014,
PlanckCollaboration2015}, and completed when the IGM was fully ionized by
$z$\,$\simeq$\,6 (\citealt{Mesinger2004};
\citealt{Fan2006},~\citeyear{Fan2006a}; \citealt{Schroeder2012};
\citealt{McGreer2014}; \citealt{Becker2015}). Observations of \Lya\
emitting galaxies also favor (volume averaged) neutral fractions in excess of
$\langle x_{\rm HI}\rangle>0.3$ at $z$$\sim$7 \citep[e.g.,][]{Dijkstra2011,
Jensen2012, Mesinger2014, Choudhury2015}.

Because neutral hydrogen and dust are opaque to FUV radiation, LyC photons can
only escape from galaxies in regions where the surrounding \ion{H}{1} column
density, $\mathrm{N_H}$, and dust extinction are low.  Thus, in order for a
fraction of the produced LyC photons to escape (\fesc), the interstellar
medium (ISM) in the galaxy and its surrounding circumgalactic medium must be
cleared.  This can be accomplished by supernova winds \citep{Fujita2003},
which can also suppress the formation of low mass stars and increase the
formation of LyC producing high mass stars, and can be further enhanced by AGN
outflows \citep{Silk2009}.  High star-formation rates can also increase the
porosity of the ISM \citep{Clarke2002}.  Semi-analytical models of
\citet{Dove2000} show that LyC emitted by OB associations can become trapped
in super-bubbles until they expand outside of the disk.  Once the surrounding
medium is either cleared or fully ionized, it becomes transparent to LyC
radiation, which can then escape through these regions of the galaxy, or be
Thomson scattered by free electrons and/or dust.  The escaping LyC can then be
observed along some lines-of-sight, which can be distributed randomly in a
galaxy, and is in some cases offset from the galactic center
\citep[e.g.,][]{Iwata2009, Vanzella2010, Vanzella2012}.

Stacks of ground-based spectra have shown that AGN produce more LyC than
star-forming Lyman Break Galaxies \citep[LBG;][]{VandenBerk2001,Shapley2003},
though LBGs selected via drop-out techniques may have fainter LyC emission due
to their selection compared to other UV bright SFGs \citep{Vanzella2015}.
Rest-frame UV spectra of AGN taken with \HST\ and the Far-Ultraviolet
Spectroscopic Explorer (\emph{FUSE}; \citealt{Moos2000}) have shown
significant detections of escaping LyC flux at
0.5\,$\lesssim$\,$z$\,$\lesssim$\,2.5
\citep[e.g.,][]{Telfer2002,Scott2004,Shull2012,Lusso2015}, but only upper
limits of \fesc\,$\lesssim$\,1--2\% from galaxies at the same redshifts
\citep[e.g.,][]{Bridge2010,Siana2010,Rutkowski2015,Sandberg2015}.  AGN
contributed the majority of LyC photons to the ionizing background from their
peak epoch at $z$\,$\simeq$\,2 until today, and maintain the ionized state of
the IGM \citep{Cowie2009}.  However, because AGN are much more rare than
galaxies, and their space density decreases at $z$\,$\gtrsim$\,2
\citep{Silverman2008,Ebrero2009,Aird2015}, AGN likely did not reionize the IGM
at $z$\,$\gtrsim$\,3 \citep{Willott2010,Gilkman2011,Masters2012}, though they are
believed to be the only sources responsible for \ion{He}{2} reionization at
$z$\,$\simeq$\,3 \citep{Haardt2012,Worseck2014}.  Therefore, SFGs are regarded
as the most likely candidates that started the reionization of the IGM at
$z$\,$\gtrsim$\,6 \citep[but see, e.g.,][]{Madau2015}.

Since higher IGM opacity at $z$\,$\gtrsim$\,6 prevents a direct study of LyC
emission from SFGs at this epoch, we must study lower redshift analogs in
order to understand the sources of reionization of the IGM.  Despite many
attempts, rest-frame FUV observations of SFGs at
0.5\,$\lesssim$\,$z$\,$\lesssim$\,2.5 have so far not yielded significant
detections of escaping LyC flux \citep[e.g.,][]{Ferguson2001, Giallongo2002,
Fernandez-Soto2003, Malkan2003, Inoue2005, Siana2007, Siana2010,Cowie2009,
Bridge2010,Grazian2016, Rutkowski2015, Sandberg2015, Guaita2016}.  Ground-based
spectra \citep{Steidel2001,Shapley2006,Cooke2014,deBarros2015} and
optical narrow-band and broadband imaging of SFGs at
$3$\,$\lesssim$\,$z$\,$\lesssim$\,4 \citep{Iwata2009,
Vanzella2010,Vanzella2012,Boutsia2011, Nestor2011, Nestor2013,Mostardi2013}
have revealed evidence for escaping LyC photons along several sight-lines,
with \fesc\,$\simeq$\,1--40\% despite higher IGM opacities at these higher
redshifts \citep{Haardt1996,Haardt2012}.  Furthermore, \citet{Vanzella2012}
estimate \fesc\ for one LBG (GDS\,J033216.64$-$274253.3 at $z$=3.795) to be
$>$25\%, although ground-based measurements of escaping LyC may be
contaminated with non-ionizing flux from blended lower redshift foreground
interlopers due to the lower resolution of ground-based seeing
\citep{Vanzella2010a, Nestor2013, Mostardi2015, Siana2015}.  Spectroscopy of
gamma-ray burst afterglows from 2\,$<$\,$z$\,$<$\,8 have also been used to
constrain \fesc\ to $<$6\% at these redshifts \citep{Chen2007, Fynbo2009,
Wyithe2010}.

Observations of some local starburst galaxies have shown significant, yet
varying \fesc\ values \citep{Leitherer1995, Hurwitz1997, Tumlinson1999,
Deharveng2001, Heckman2001, Borthakur2014, Izotov2016}, although
\citet{Hanish2010} find that local starburst galaxies do not exhibit higher
escape fractions compared to ordinary local SFGs.  \ion{H}{2} regions in
nearby galaxies have been observed to release 40--75\% of the LyC photons
produced by massive stars into the local IGM \citep{Ferguson1996,
Leitherer1996, Oey1997, Zurita2002}.  In the local group, \citet{Bland-Hawthorn1999}
and \citet{Putman2003} find LyC escape fractions of only 1--2\%.

In the hierarchical formation scenario of galaxy assembly, the inflow of cold
gas and merging of high redshift compact galaxies plays a role in the
formation of massive young starburst galaxies.  The number density of those
Luminous Compact Blue Galaxies is also known to increase strongly with
redshift \citep{Lilly1998, Mallen-Ornelas1999}.  \emph{FUSE} observations of
analogous nearby, young, starbursting dwarf galaxies have shown that the
\fesc\ values of these galaxies reach $\lesssim$\,4\%, and can collectively
contribute a significant fraction to the ionizing background at high redshift
(\citealt{Bergvall2006, Grimes2007, Grimes2009, Leitet2011},
\citeyear{Leitet2013}).  Most theoretical models also predict that low mass
galaxies abundant at high redshifts are more likely to have higher \fesc\
values than the larger galaxies at low to moderate redshift
\citep{Razoumov2010, Yajima2011, Wise2014}.  Hence, it is likely that
different classes of objects dominated reionization at different cosmic
epochs, \ie\ the combined FUV output from (dwarf) SFGs may have started to
reionize the IGM at $z$\,$\gtrsim$\,6.5--7, then, along with more massive
galaxies, completed and maintained its ionized state at $z$\,$\lesssim$\,6
until AGN started to dominate at FUV wavelengths at $z$\,$\lesssim$\,2--2.5.

In this work, we describe our analysis of \HST\ rest-frame UV observations of
LyC escaping from massive SFGs and weak AGN at $z$\,$\simeq$\,2.3--4.1 in
three UVIS filters with the Wide Field Camera 3 (WFC3), taken shortly after
installation onto \HST. We compare the measured LyC fluxes of our sample to their
modeled intrinsic LyC fluxes using stellar population synthesis models and
Monte Carlo (MC) simulated line-of-sight IGM transmission models.

This paper is organized as follows.  In \S\ref{sec:data}, we describe the data
that we used for our analysis and how it was reduced.  In \S\ref{sec:sample}
we give our assessment of the available spectra for our galaxy samples and
their reliability and completeness.  In \S\ref{sec:stacking}, we outline the
method we implemented to create the stacked LyC images of our samples of
galaxies, how we perform photometry on the stacks, the observed LyC flux that
we measure, and the significance of these detections.  In \S\ref{sec:fesc}, we
introduce the stacked LyC escape fraction, how we calculated the \fesc\
values, their implications, and the observed and modeled radial profiles of
the escaping LyC from our samples.  In \S\ref{sec:discussion} and
\S\ref{sec:conclusion}, we discuss our results and present our conclusions.
We use Planck (\citeyear{PlanckCollaboration2015}) cosmology throughout:
$H_{0}$ = 67.8 km\,s$^{-1}$\,Mpc$^{-1}$, $\Omega_{\rm m}$=0.308 and
$\Omega_\Lambda$=0.692.  All flux densities (referred to as ``fluxes''
throughout) quoted are in the AB magnitude system \citep{Oke1983}, unless
stated otherwise.

\section{WFC3/UVIS and ACS/WFC Observations and Data Reduction}
\label{sec:data}

\subsection{ERS/GOODS-S WFC3/UVIS and ACS/WFC Data} \label{sec:observation}
\noindent\begin{figure*}[th!] \centerline{
\includegraphics[height=0.28\txh]{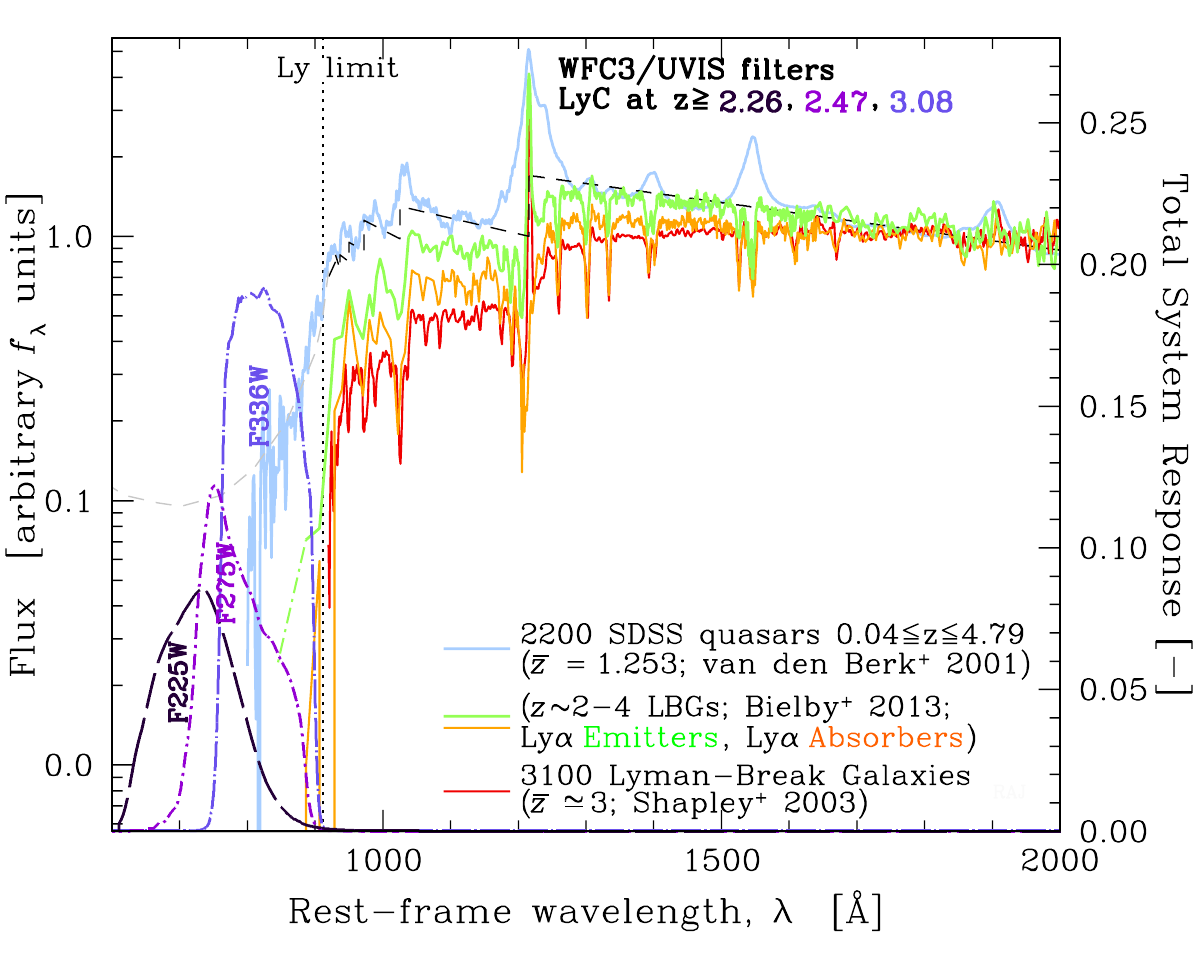}\ \
\includegraphics[height=0.28\txh]{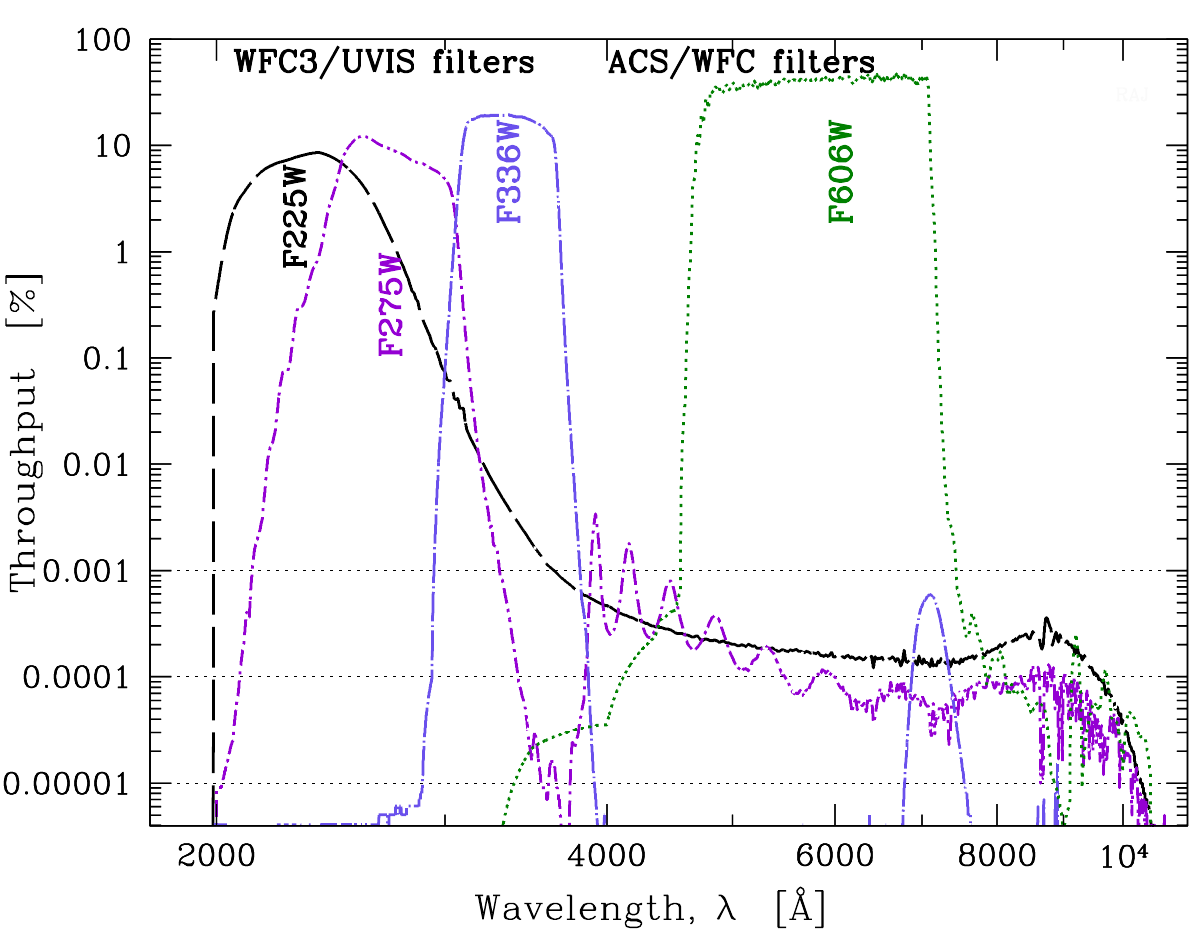} }
\caption{\noindent\small [\emph{a}] Example composite rest-frame FUV
spectra of SDSS QSOs at \zmean\,$\sim$\,1.3 (\citealt{VandenBerk2001}
[\emph{blue}]) and of LBGs at $z$\,$\simeq$\,2--4 (\citealt{Bielby2013}
[\emph{green} and \emph{orange}]; \citealt{Shapley2003} [\emph{red}]). The
WFC3/UVIS F225W, F275W, and F336W filter transmission curves
are ideally positioned to capture Lyman continuum emission
($\lambda$\,$<$\,912\AA) at $z$\,$\ge$\,2.26, $z$\,$\ge$\,2.47, and
$z$\,$\ge$\,3.08, respectively. The combined SEDs of
SDSS QSOs suggest a strong LyC signal, whereas the SEDs of LBGs suggest
fainter LyC flux. [\emph{b}] Total system throughput curves (observed
wavelengths) of the same WFC3/UVIS and ACS/WFC filters \citep{Dressel2015,
Avila2015} are shown on a logarithmic scale to emphasize their out of band
transmission (``red-leak''). These WFC3 UVIS filters were designed to
simultaneously maximize throughput and minimize such red-leaks. In
Appendix~\ref{sec:red-leak} we assess the percentage of non-ionizing UVC flux
with $\lambda$\,$>$\,912\AA\ leaking in the filter. We include the ACS/WFC
F606W filter as a UVC filter reference. \label{fig:figure1}} \end{figure*}
\indent Our UV data was taken with \HST's WFC3/UVIS camera in the Early Release
Science (ERS) field in September 2009 \citep[hereafter W11]{Windhorst2011}, less
than four months after Shuttle Servicing Mission SM4 that installed WFC3 onto \HST, at
a point when the WFC3/UVIS CCDs have not yet suffered from significant CTE degradation.
Complementary optical ACS/WFC data was taken in July 2002-May 2003 as part of the Great
Observatories Origins Deep Survey (GOODS; \citealt{Dickinson2003}). 
Fig.~\ref{fig:figure1}\emph{a} and \ref{fig:figure1}\emph{b} show the
WFC3/UVIS F225W, F275W, and F336W filter transmission curves, which are ideally
positioned to capture LyC emission with negligible red-leak at $z$\,$\gtrsim$\,2.26,
$z$\,$\gtrsim$\,2.47, and $z$\,$\gtrsim$\,3.08, respectively (see
Appendix~\ref{sec:red-leak} for a detailed discussion on red-leak). The corresponding 
rest-frame non-ionizing ultraviolet continuum (UVC) imaging of our galaxies were taken
with ACS/WFC in the F606W, F606W, and F775W, respectively.  These filters
sample rest-frame $\lambda_{\rm eff}$$\sim$1400-1800\AA\ for each of our
redshift intervals.  When we model the rest-frame UVC absolute magnitudes
(\MAB) from the spectral energy distribution (SED) fits, we integrate over the
interval 1500$\pm$100\AA\ (see \S~\ref{sec:fescintro}).  We also utilized
photometry from WFC3/IR F098M, F125W, and F160W imaging in the ERS field
\citepalias{Windhorst2011} and CANDELS WFC3/IR F105W, F125W, and F160W
\citep{Grogin2011,Koekemoer2011} photometric catalogs in GOODS-South
\citep{Guo2013} for object selection and SED fitting (see
\S\ref{sec:selection} and \S\ref{sec:fescintro}).

Table~\ref{table:table1} summarizes the data in the ERS and GOODS South fields
available for studying LyC emission, and the redshift range over which each of
these filters can sample LyC emission with negligible contamination from non-ionizing
flux.  Each lower redshift bound was carefully chosen such that
\emph{no} light with $\lambda$\,$>$\,912\AA\ is sampled below the filter's red
edge (defined at 0.5\% of the filter's peak transmission, as tabulated in
\citet{Dressel2015} and references therein).  The upper redshift bound of each
bin in Table~\ref{table:table1} occurs at the redshift where the next redder
filter can trace LyC emission more sensitively.  Fig.~\ref{fig:figure1}[a]
suggests that the observed escaping LyC emission strongly declines towards
shorter wavelengths.  For this reason, the broadband filters we use are most
sensitive to LyC emission at the low redshift end of each of the three
redshift ranges of Table~\ref{table:table1}.\vspace{-5pt}\vspace{1pt}
\noindent\begin{figure*}[t!] \centerline{
\includegraphics[width=0.9\txw]{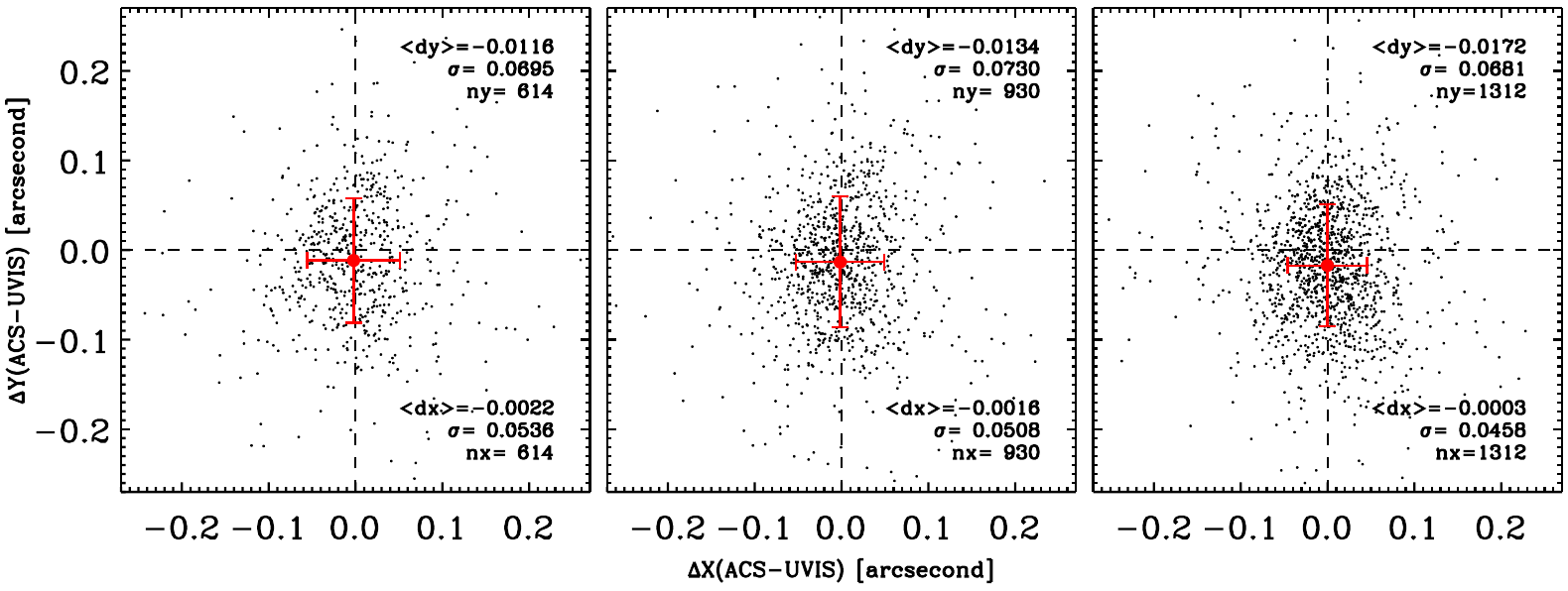} } \caption{\noindent\small
Residual astrometric errors in the improved WFC3/UVIS ERS mosaics in F225W
[\emph{left}], F275W [\emph{middle}], and F336W [\emph{right}], respectively,
as measured relative to the ACS/WFC F435W ERS mosaic. The new WFC3/UVIS
geometric distortion corrections provided a significantly improved
registration of all 8 WFC3/UVIS tiles to the ACS/WFC F435W mosaics compared to
the 2009 ERS mosaics of \citetalias{Windhorst2011}. The measured residual
systematic offsets are $\langle\Delta X\rangle$\,$\lesssim$\,0$\farcs$0022 and
$\langle\Delta Y\rangle$\,$\lesssim$\,0\farcs0172 compared to ACS (indicated
by the dashed lines), and random deviations less than 0$\farcs$054 in X and
0$\farcs$073 in Y. \label{fig:figure2} } \end{figure*}
\subsection{Image Calibration, Drizzling and Astrometric Accuracy}
\label{sec:datareduction}

The photometric and astrometric calibration and drizzling of the
ACS/WFC mosaics are described in \citet{Giavalisco2004}. The initial
astrometric calibration of the WFC3/UVIS ERS mosaics, described in
\citetalias{Windhorst2011}, left systematic offsets between individual WFC3
CCDs of up to $\sim$\,5 drizzled pixels ($\sim$\,0$\farcs$5) compared to the
GOODS v2.0\footnote{
\url{http://archive.stsci.edu/pub/hlsp/goods/v2/h\_goods\_v2.0\_rdm.html}}
F435W mosaics, especially at the edges of each UVIS mosaic tile (see
Appendix~A of \citetalias{Windhorst2011}). These offsets occurred in part due
to the way the ERS UVIS exposures were taken, but were primarily due to the
rather uncertain geometric instrument distortion correction (IDC) tables
available at the time \citep{Kozhurina-Platais2009}. The lack of UV bright
astrometric reference sources in the shallow ($\sim$\,900--1400\,s) individual
exposures further prevented accurate registration and drizzling of the F225W,
F275W, and F336W images. This issue was resolved with the improved IDC tables
of \citet{Kozhurina-Platais2013} and \citet{Kozhurina-Platais2014}. Using
these new IDC tables, we re-drizzled the UVIS ERS images into mosaics at a
plate scale of 0$\farcs$03\,pix$^{-1}$.
\placetable{1} 
\noindent\begin{table}[b!] 
\centering \caption{\small Summary of \HST\ WFC3/UVIS Images and Image Stacks
in the ERS Field \label{table:table1}} 
\setlength{\tabcolsep}{2pt}
\begin{tabular}{lcccccr}
\toprule\\[-21pt]  Filter & $\lambda\ /\ \Delta\lambda^{\rm a}$ & ${z_{\!\lyc}}^{\rm b}$
&     $\!\!$Obs.\,Date & ${t_{\rm exp}}^{\rm c}$ & PSF$^{\rm d}$ &
$\!\!$\emph{SB}(obs)$^{\rm e}$$\!\!$ \\[-1pt]
\midrule\\[-21pt] 
F225W&2359\,/\,467 & 2.26--2.47  & 2009 Sep 7--11 & 5,688 & 0$\farcs$087 & 29.80\\[-5pt] 
F275W&2704\,/\,398 & 2.47--3.08  & 2009 Sep 7--11 & 5,688 & 0$\farcs$087 & 29.82\\[-5pt] 
F336W&3355\,/\,511 & 3.08--4.35  & 2009 Sep 7--12 & 2,778 & 0$\farcs$088 & 29.76\\
\bottomrule 
\end{tabular}
\begin{minipage}{0.48\txw}{\linespread{1.3}\small $^{\rm a}$Central wavelength / bandwidth of
filter in \AA; $^{\rm b}$Redshift range over which rest-frame LyC emission can
in principle be sampled. The high end of each bin occurs at the redshift where
the next redder filter can better sample LyC emission at the same or higher
redshift.; $^{\rm c}$Average integration time of the mosaics in seconds;
$^{\rm d}$ Typical stellar PSF FWHM.; $^{\rm e}$ Measured
1$\sigma$ surface brightness sensitivity limit of our mosaics for a source of
uniform SB in a 2$\farcs$00 diameter aperture in AB \magarc\ (see
\citetalias{Windhorst2011} and Table~\ref{table:tableA1})}.
\end{minipage} 
\end{table}

Fig.~\ref{fig:figure2} shows the residual astrometric errors of the new ERS
mosaics (which we refer to as ERS ``v2.0'') for the F225W, F275W, and F336W
filters, measured relative to the ACS/WFC F435W mosaics. Residual systematic
offsets from the ACS/WFC F435W GOODS v2.0 mosaics for the 8 WFC3/UVIS ERS
tiles are now measured to be $\langle\Delta
X\rangle$\,$\lesssim$\,0$\farcs$0022 (0.024\,pix) and $\langle\Delta
Y\rangle$\,$\lesssim$\,0\farcs0172 (0.19\,pix), with 1$\sigma$ random
deviations less than 0$\farcs$054 (0.60\,pix) in X and 0$\farcs$073
(0.81\,pix) in Y. Any remaining systematic astrometric offsets are at the
sub-pixel level, and are sufficiently small that they no longer affect our SB
sensitivity to LyC flux, nor do they add contamination from neighboring
sources that can potentially blend in with the LyC signal due to astrometric
uncertainties.

\subsection{WFC3/UVIS Residual Sky-Background} 
\label{sec:sky}
\noindent\begin{figure*}[t!] \centerline{
\includegraphics[width=0.32\txw]{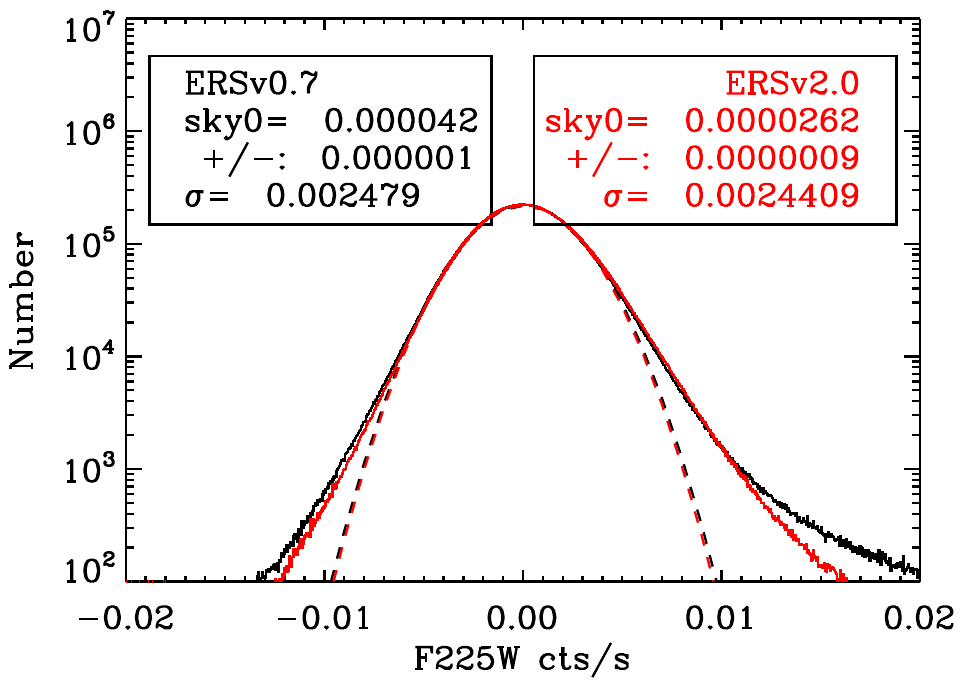}
\includegraphics[width=0.32\txw]{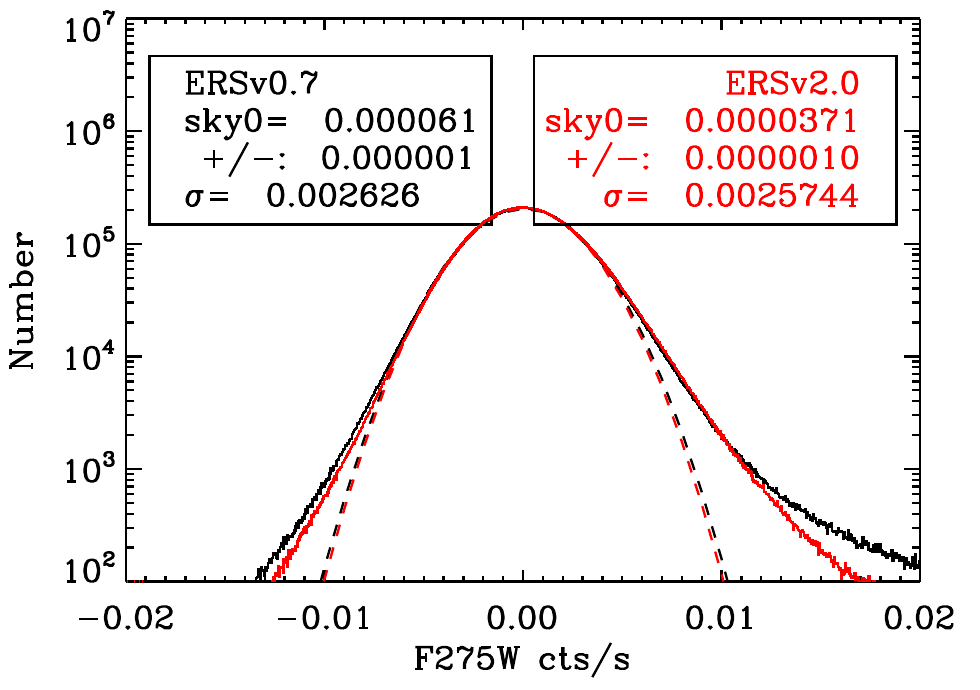}
\includegraphics[width=0.32\txw]{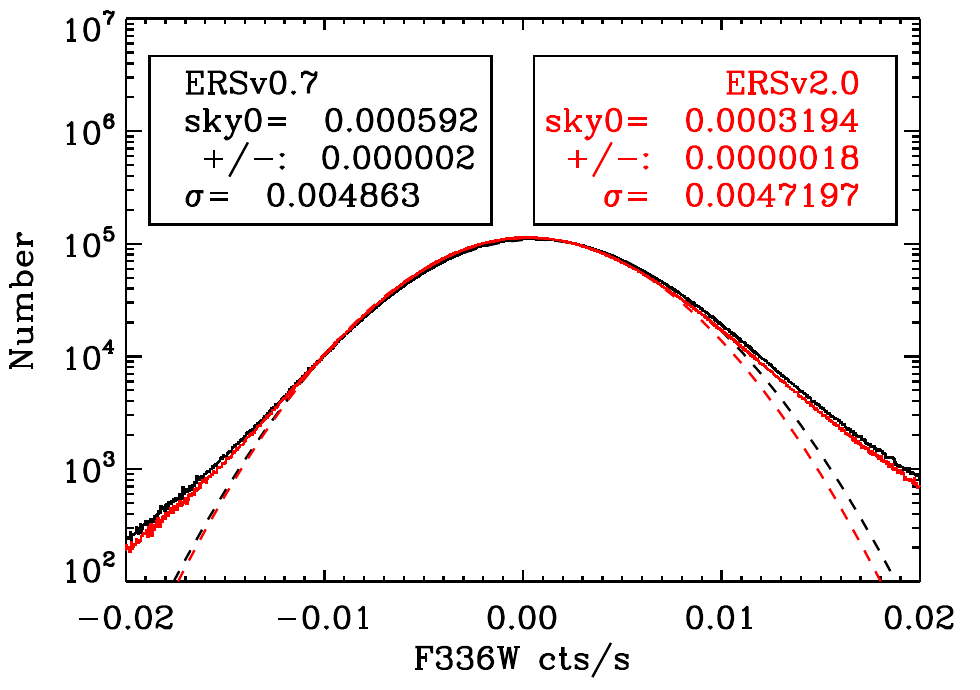} }
\caption{\noindent\small Residual sky-background levels in the drizzled
WFC3/UVIS ERS mosaics in the F225W [\emph{left}], F275W [\emph{middle}], and
F336W [\emph{right}] filters, determined as described in \S\ref{sec:sky}. The
best fit to the 2009 ERS mosaics of \citetalias{Windhorst2011} ("v0.7") is
shown in black, while the improved mosaics discussed here ("v2.0") are shown
in red. Dashed lines show Gaussian fits to the two histograms. Each panel
lists the best fit \emph{residual sky-background level (in ADU/sec)},
equivalent to 30.29, 29.99, and 28.15 \magarc, respectively, and the
uncertainties thereon across the entire WFC3 ERS mosaic.  \label{fig:figure3}}
\end{figure*}
The original ground-based WFC3/UVIS thermal vacuum flats left residual
gradients and patterns in the sky-background at the 5--10\% level
(\citealt{Sabbi2009}; \citetalias{Windhorst2011}). For the reprocessing of the
ERS data, we use the flat-fields from \citet{Mack2013}, which include on-orbit
``delta-flat'' corrections to the ground-based thermal vacuum flat-fields,
significantly reducing the large scale flat-field patterns seen in earlier ERS
mosaics.

\citetalias{Windhorst2011} measured zodiacal sky-background SB levels in the
ERS of 25.46, 25.64, and 24.82 \magarc\ in the WFC3/UVIS F225W, F275W,
and F336W respectively. When drizzling the data,
this sky-background is subtracted \citep[see][]{Koekemoer2013}. For the 5688,
5688, and 2788\,s total exposure times in these filters, this
corresponds to a subtraction of $\sim$\,0.00219, 0.00202, and
0.00704\,\emin/0$\farcs$09 pixel. To determine the best fit \emph{residual} sky
background level across the WFC3/UVIS ERS mosaics, we follow
\citet{Freedman1981}, who define the sampled bin width for optimal histogram
fitting as $2\times$IQR$\times$N$^{-1/3}$, where IQR is the inner quartile
range of the pixel distribution (\ie\ the range within 75\% and 25\%, or
$\pm$1.349$\sigma$/2 for a Gaussian distribution), and $N$ is the total number
of pixels used in the image to construct that histogram. We fit the logarithm
of the sky pixel histogram between $-$3$\sigma$ and $+$1$\sigma$ to a Gaussian
function by least squares to obtain the peak value of the fitted functions.

Fig.~\ref{fig:figure3} shows the sky pixel histograms and best fit
\emph{residual} sky-background levels in the WFC3/UVIS F225W, F275W, and F336W
ERS mosaics of \citetalias{Windhorst2011} in black, while the red curves and
measurements indicate the best fit to the data in the current, improved v2.0
mosaics. The slight narrowing of the negative tail of the Gaussian noise
distributions in the new mosaics reflects the better flat-fielding. Our best
fit \emph{residual} sky background values and uncertainties thereon are
(2.62$\pm$0.09)$\times$10$^{-5}$, (3.71$\pm$0.10)$\times$10$^{-5}$, and
(31.94$\pm$0.18)$\times$10$^{-5}$ \emin/s in the F225W, F275W, and F336W
filters, respectively, which corresponds to residual sky SB levels of 30.29,
29.99, and 28.15 \magarc\ left in the UVIS images \emph{after} drizzling,
which subtracted the sky-background to first order. Compared to the observed
ERS sky-backgrounds measured in \citetalias{Windhorst2011}, these residual sky
SB level values are 4.84, 4.35, and 3.33\,mag fainter than the UV sky (1.2\%,
1.8\%, 4.7\% of the UV sky), respectively. These residual sky-background
levels can be accurately determined locally and subsequently subtracted, which
we employed in our sub-image stacking technique to further increase our
sensitivity to extended, low SB LyC signal (see \S\ref{sec:method}).

\section{Spectroscopic Redshifts and Sample Selection}
\label{sec:sample}

\subsection{Spectroscopic Sample Selection}
\label{sec:selection}

In order to obtain accurate estimates of LyC escape fractions as low as
$\fesc$$\lesssim$1.0\%, we must require the interloper fraction to be
very small. Thus, any potentially contaminating, low redshift, interloping
galaxies that might create a false-positive LyC signal must be identified and
removed from our sample. We therefore require each galaxy that we include in
our analysis to have a highly reliable spectroscopic redshift.

Several wide field ground-based spectroscopic surveys have been performed in
the GOODS fields, including the ERS region, at low and high redshift with the
Very Large Telescope (VLT) \citep[e.g.,][]{Cristiani2000, LeFevre2004,
Szokoly2004, Wolf2004, Vanzella2008, Popesso2009, Wuyts2009, Balestra2010,
Silverman2010, Fiore2011, Kurk2012, LeFevre2015, Tasca2017} as well as \HST\
\citep{Momcheva2016}. We retrieved the reduced 1-dimensional FITS spectral data from
the ESO archives\footnote{\url{http://archive.eso.org/}, \url{http://www.eso.org/s
ci/activities/garching/projects/goods/MasterSpectroscopy.html}, and \url{http://cesam.lam.fr/vuds/DR1/}}
along with their corresponding 3D-HST grism spectra\footnote{\url{http://3dhst.research.yale.edu/Home.html}}
for galaxies at $z$\,$\geq$\,2.26 in order to verify the designated redshifts of these
galaxies. We plotted each spectrum at both the original spectral resolution
and smoothed with a $\sigma$=\,3\,\AA\ Gaussian convolution kernel to reduce
the noise in the spectrum for inspection. Typical spectral emission or
absorption features for SFGs and AGN were indicated, and portions of the
spectra around these features were magnified for closer examination.
Specifically, these emission or absorption features include the Lyman Break at
912\AA, \Lya\,1216\AA, \ion{Si}{2}\,1260\AA, \ion{O}{1}\,1304\AA,
\ion{C}{2}\,1335\AA, \ion{Si}{4}\,1398\AA, \ion{C}{4}\,1549\AA, and
\ion{C}{3}]\,1909\AA, and when present, \ion{C}{2}]\,2326\AA,
\ion{Fe}{2}\,2344\AA, and sometimes \ion{N}{5}\,1240\AA, \ion{Fe}{2}\,2600\AA,
\ion{Mg}{2}\,2798\AA, \ion{O}{2}\,3727\AA, [\ion{Ne}{3}]\,3869\AA, \ion{He}{2}\,4686\AA,
\ion{H}{0}$\beta$\,4861\AA, and [\ion{O}{3}]\,4959+5007\AA. In addition, we included
high contrast cutout images of the galaxies in the filter sampling the rest-frame LyC
emission, and all available longer wavelength filters for inspection and removal of
contaminating objects.

Five of us (BMS, RAW, SHC, RAJ, and LJ) visually inspected all spectra
and unanimously selected the highest quality spectra available from the
spectroscopic surveys and compose our spectroscopic sample of galaxies
and AGN. We found that including objects with spectra that had less
reliable redshifts improved our formal SNR, but likely added contaminating
flux rather than true escaping LyC flux. Hence, we \emph{only} included galaxies
with the highest quality spectra that coincided with their predicted
emission/absorption lines exactly.

Note that 12 of the 46 objects in our spectroscopic sample are
galaxies hosting a \emph{weak} AGN, as evidenced by the (broad) emission lines in their spectra, for example \Lya, \ion{N}{5}, \ion{Si}{4}, \ion{C}{4}, \ion{He}{2}, \ion{C}{3}], and \ion{Mg}{2}. We also cross-correlated the positions of our
galaxy sample with Chandra 4 Ms and Very Large Array 1.4 GHz source catalogs
to identify possible obscured/type II AGN using their radio/X-ray luminosities
and photon indices \citep[e.g.,][]{Xue2011,Fiore2012,Miller2013,Rangel2013,Xue2016}.
We identified five of the 12 galaxies hosting AGN from their X-ray
emission. In our analysis, we will consider the subsamples of 34 galaxies
without AGN signatures, and 12 galaxies with weak AGN, both separately and
combined.
\noindent\begin{figure*}[t!]\centerline{
\includegraphics[width=\txw]{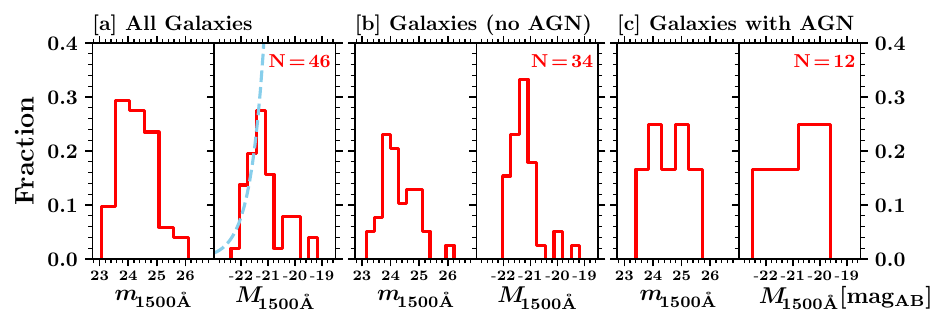} } \caption{\noindent\small
[\emph{a}] Absolute and apparent magnitude distributions at the rest-frame
1500$\pm$100\AA\ of the spectroscopic samples for
all galaxies. [\emph{b}] Same, for just the galaxies without AGN activity.
[\emph{c}] Same, for galaxies with indications of AGN activity. These
magnitudes were derived from the observed SED fits (see
\S\ref{sec:fescintro}), and therefore do not require k-corrections. The blue
dashed curve indicates the  slope of the luminosity function of \zmean=3.46
galaxies at \MAB=--20.8, equal to 0.84\,dex/mag. \label{fig:figure4}}
\end{figure*}

\subsection{Completeness and Representativeness of the Spectroscopic Samples}
\label{sec:completeness}

Initially, our sample of galaxies was limited to those with \emph{known}
spectroscopic redshifts. Our selection of galaxies with high quality spectra
in GOODS-S, which solely determined which objects were included in our analysis,
reduced our sample to galaxies that can be observed with ground-based spectroscopy
at a high signal-to-noise ratio (SNR). This, of course, can bias our results and
their subsequent interpretations, \eg\ if \fesc\ is a strong function of
luminosity (\MAB), dust extinction (\AV), metallicity, and/or age. We
therefore must consider how representative the characteristics of our selected
galaxy samples are in order to understand differences in the results of our
analyses of the populations.

In Fig.~\ref{fig:figure4} we plot the distribution of observed
apparent magnitudes (\mAB), and the corresponding absolute magnitudes (\MAB)
of the rest-frame non-ionizing UVC flux ($\lambda_{\rm eff}$=1500$\pm$100\AA)
of our samples. We consider all galaxies (Fig.~\ref{fig:figure4}\emph{a}),
galaxies hosting weak AGN (Fig.~\ref{fig:figure4}\emph{b}), and galaxies
without AGN (Fig.~\ref{fig:figure4}\emph{c}). These values were derived from
the apparent flux of the galaxies at the same rest-frame wavelengths, using
their best fit SED models (see \S\ref{sec:fescintro} and
Appendix~\ref{sec:extinction}), so no k-correction is necessary to directly
compare the \MAB\ values of the galaxies at various redshifts.

If the spectroscopic samples were complete, their apparent magnitude
distributions would resemble the galaxy count distributions of the full $V$
and $i$ band mosaics (\citealt{Giavalisco2004}; \citetalias{Windhorst2011}) to
a given AB magnitude limit, since these filters sample the UVC emission indicated
in Fig.~\ref{fig:figure4}, and because the spectroscopic samples were $r$ band or
$i$ band selected. Their \MAB\ distribution would also reflect the galaxy UV
luminosity function slope at their effective \MAB\ to the effective
completeness limits at these redshifts, which typically sample rest-frame
wavelengths $\lambda_{\rm eff}$\,$\simeq$\,1500\,--\,1700\AA\
\citep[e.g.,][]{Reddy2009, Finkelstein2015}.

It is clear from Fig.~\ref{fig:figure4} that our spectroscopically selected
samples are incomplete for \mAB\,$\gtrsim$\,24.0\,mag, both for galaxies with
and without weak AGN. For \mAB\,$\lesssim$\,24.0\,mag, the distributions are
consistent with the expected slope of the galaxy counts from
\citetalias{Windhorst2011}, so the selected samples may be representative for
LyC studies, but only for these brighter fluxes. We also note that our
selection of galaxies with high SNR spectra will have favored the broad
emission lines of (weak) AGN, and \Lya\ emission or strong absorption line
galaxies, while LBGs and other galaxies without prominent spectroscopic
features are less likely to have yielded the highly reliable redshifts
required to be included in our highest fidelity sample, even for
\mAB\,$\lesssim$\,24.0\,mag.

The UVC luminosities of the galaxies in our sample span
--22.2\,$\lesssim$\,\MAB\,$\lesssim$\,--19.0\,mag, with an average of
\MAB\,$\simeq$\,--21.1$^{+0.9}_{-0.5}$\,mag (1$\sigma$), indicative of
predominantly luminous galaxies about as bright as $M^{*}$ at
2.5\,$\lesssim$\,$z$\,$\lesssim$\,4 \citep[e.g.,][]{Hathi2010}, or of galaxies
hosting \emph{weak} AGN. Since this is the only sample for which reliable
redshifts currently exist, this is strictly the only luminosity range over
which the measurements and analyses of any escaping LyC emission that follows
will be valid. These galaxies may also be more luminous than galaxies
that contributed to reionization at $z$$>$7 \citep[e.g.,][]{Bouwens2012}.

\noindent\begin{figure*}[t!]
\centerline{ \includegraphics[width=\txw]{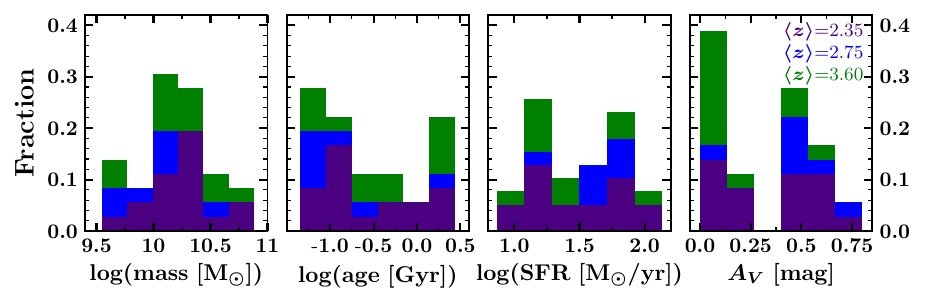} }
\caption{\noindent\small Stacked BC03 SED fit parameter distributions of the
spectroscopic samples for galaxies without AGN. The purple, blue, and green
bars represent the \zmean=2.35, \zmean=2.75, and \zmean=3.60 samples,
respectively. \label{fig:figure5}} \end{figure*}
Our sample also does not fully sample the parameter space of the age,
stellar mass, star-formation rate (SFR), and \AV\ values for galaxies at these
redshifts, indicated by their SED fits (see \S\ref{sec:fescintro}). These
biases are also more prominent in the individual redshift subsamples.
Fig.~\ref{fig:figure5} shows the distribution of these parameters for the
spectroscopic sample of galaxies without AGN. These galaxies more or
less evenly sample the mass and SFR distribution, which are are generally
quite massive and indicative of active star-formation, with masses ranging
from $\sim$\,10$^{9.5}$--$10^{10.9}$
($\langle$mass$\rangle$=$10^{10.2\pm0.3}$)\,\Msol\ and SFRs from
$\sim$\,10$^{0.8}$--$10^{3.1}$
($\langle$SFR$\rangle$=10$^{1.5\pm0.4}$)\,\Msol/yr, respectively. Their ages
and \AV\ distributions range from
$\sim$\,10$^{7.7}$--$10^{9.4}$\,yr
($\langle\mathrm{t_{age}}\rangle$\,$\simeq$\,10$^{8.2^{+0.9}_{-0.4}}$\,yr) and
\AV\,=\,0.0--0.8\,mag ($\langle\AV\rangle_{\rm med}\!\!\simeq\!\!0.3\pm$0.3\,mag),
respectively. We note that the
variation in these parameters from sample to sample is most likely due to the
selection of the spectroscopic sample, rather than any real correlation in
redshift, as the higher redshift galaxies were generally selected in the
redder ACS filters.

The incomplete sampling of these various parameters should be taken into
account when interpreting the \fesc\ values for these individual subsamples.
In order to obtain a more representative sample of galaxies in each redshift
bin, we must include more galaxies that sample the full range of these
parameters at their respective redshifts, with average parameters that reflect
the true averages for all galaxies at these redshifts, and probe fainter
luminosities. This should be a focus of future deeper spectroscopic surveys,
either from the ground or with JWST.
\section{LyC Emission Stacking and Photometry} 
\label{sec:stacking}
\subsection{Sub-Image Stacking for each LyC Filter} 
\label{sec:method}

Since absolute LyC escape fractions have been measured to be very low, and the
detected LyC emission to be very faint or not detected at all at
$z$\,$\lesssim$\,3 \citep[e.g.,][]{Steidel2001, Shapley2006, Iwata2009,
Siana2010, Mostardi2013}, we apply a technique of sub-image stacking of the
\textit{observed} LyC emission from multiple galaxies to increase the total
SNR and sensitivity to the faint, potentially low SB LyC flux from individual
galaxies. Stacking LyC emission from galaxies at similar redshifts can be used
to quantify the \textit{average} LyC emission from galaxies at their average
redshift. This method also reduces small scale residual systematic errors in
the stacked sub-images left from bias, dark current, sky-subtraction, flat-fielding,
and/or any gradients from variations in exposure time or photon
noise between exposures that might remain in the background of drizzled
mosaics, as effects from random systematics are averaged out in a stack
(see Appendices \ref{sec:stacktests}---\ref{sec:systematics}). We
create our stacks of the LyC emission for our samples as following.

For each galaxy, we extracted 151$\times$151 pixel
(4\farcs53$\times$4\farcs53) sub-images from the WFC3/UVIS mosaics in the 
respective filter that samples the LyC emission of each galaxy.
The size of these cutouts provided sufficient sampling of the photon
statistics in the sub-images for fitting the pixel count-rate distribution,
while minimizing the potential area of neighboring sources of non-ionizing flux.
Each sub-image was centered on the RA and Dec of the centroid of the
individual galaxy indicated in the {\it 3D-HST} photometric catalog
\citep{Skelton2014}.

We then created \SExtractor\ \citep{Bertin1996} segmentation maps from
$\chi^2$ images \citep{Szalay1999} generated from all available \HST\ data
for each LyC sub-image in order to identify \emph{all} neighboring objects
detected at a $\geq$\,1$\sigma$ threshold above the local sky. We then exclude
\emph{all} surrounding detections outside of a central circular aperture with
a 0$\farcs$5 radius (r\,$\simeq$\,17\,pix) found in the LyC segmentation maps.
We preserve all flux from any objects inside this central aperture when we
stack the sub-images, while also excluding those detected on the border of
the central aperture. This masking was applied to ensure that all potential
sources of non-ionizing flux from lower redshift neighbors along the line-of-sight
are removed before stacking. On average, $\sim$3$\pm$2 objects were removed
from each sub-image. We note that this procedure would not be possible when
stacking LyC emission of galaxies using ground-based observations alone, as
effects from seeing can blend neighboring non-ionizing contaminants with the
true LyC signal \citep[e.g.,][]{Nestor2013, Siana2015, Mostardi2015}. Each
individual masked sub-image was inspected visually to verify that no surrounding
objects indeed remained in the sub-images, including those seen only at longer
wavelengths in the 10 band ERS mosaics (see \S\ref{sec:data}). Thus, it is
possible that the rigorous removal of surrounding flux can sometimes result in
the removal of more extended (i.e., at r$\geq$0\farcs5) LyC flux from the stacked
images if this were detectable at $\geq$\,1$\sigma$ above the local sky-background
(see \S\ref{sec:stackofstacks} and \S\ref{sec:obsradprof}).

\noindent\begin{figure}[htbp] 
\centerline{\includegraphics[width=0.44\txw]{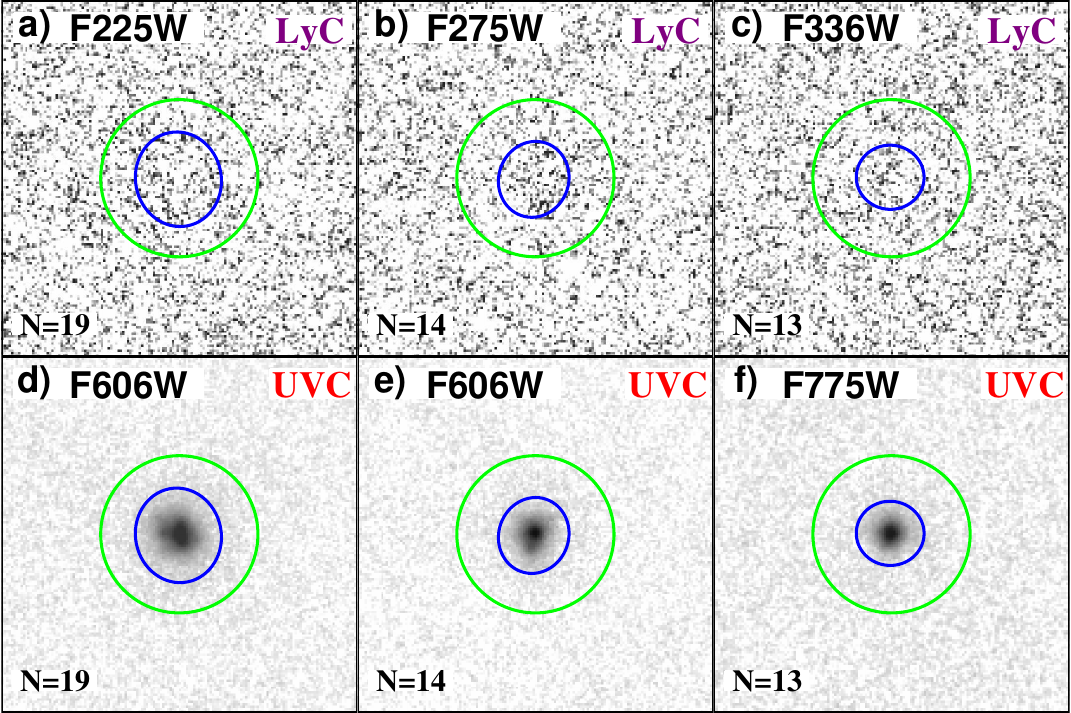}}
\caption{\noindent\small Sub-image stacks for the three different redshift bins
in our sample of \emph{all} galaxies, sampling LyC emission in: [\emph{a}]
F225W at 2.28\,$\le$\,$z$\,$\le$\,2.45, [\emph{b}] F275W at
2.47\,$\le$\,$z$\,$\le$\,3.08, and [\emph{c}] F336W at
3.13\,$\le$\,$z$\,$\le$\,4.15; and corresponding UVC
($\sim$1400\,$\lesssim$\,$\lambda_0$\,$\lesssim$\,1800\AA) emission in:
[\emph{d}] F606W, [\emph{e}] F606W, and [\emph{f}] F775W.
Note that the objects contributing to panels [\emph{d}] and [\emph{e}] differ,
since they correspond to different redshift bins. Blue ellipses indicate the
\SExtractor\ \texttt{MAG\_AUTO} UVC detected matched apertures, while green
apertures are 2\farcs0 diameter circles for comparison. All sub-images are
151$\times$151 pixels (4\farcs53$\times$4\farcs53) in size.
\label{fig:figure6}} \end{figure}
\noindent\begin{figure}[htbp] 
\centerline{\includegraphics[width=0.44\txw]{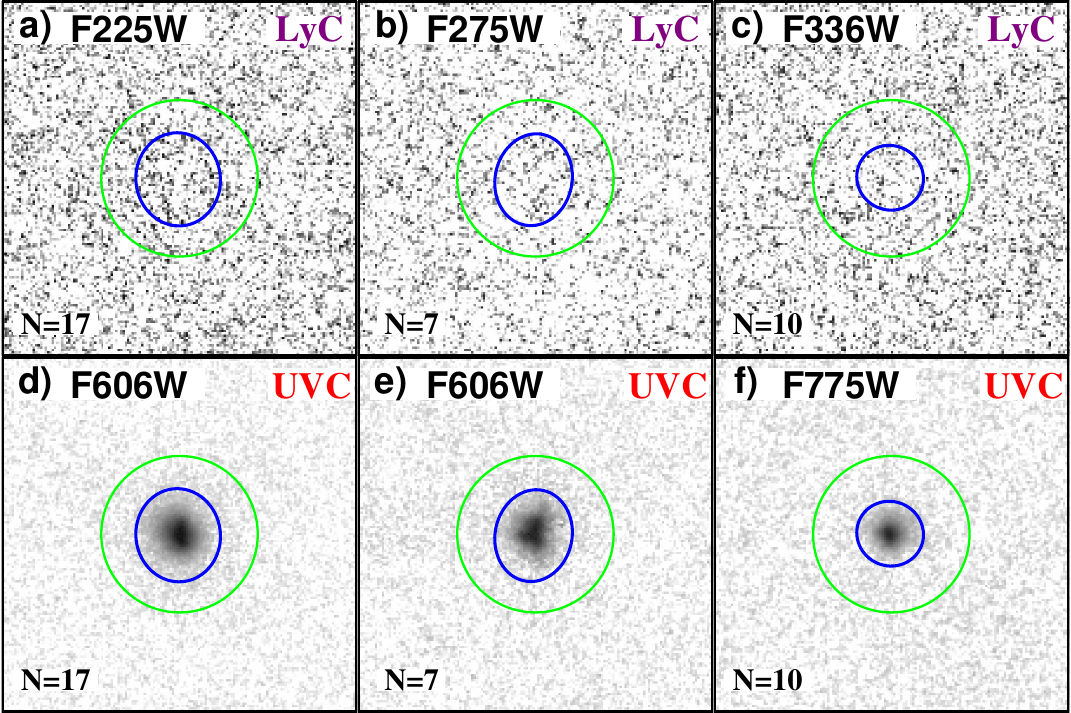}}
\caption{\noindent\small As Fig.~\ref{fig:figure6} for galaxies
\textit{without} AGN (\ie\ no obvious  signs of nuclear activity from their
spectra or X-ray/radio luminosities/photon indices. \label{fig:figure7}} 
\end{figure}
\noindent\begin{figure}[htbp] 
\centerline{\includegraphics[width=0.44\txw]{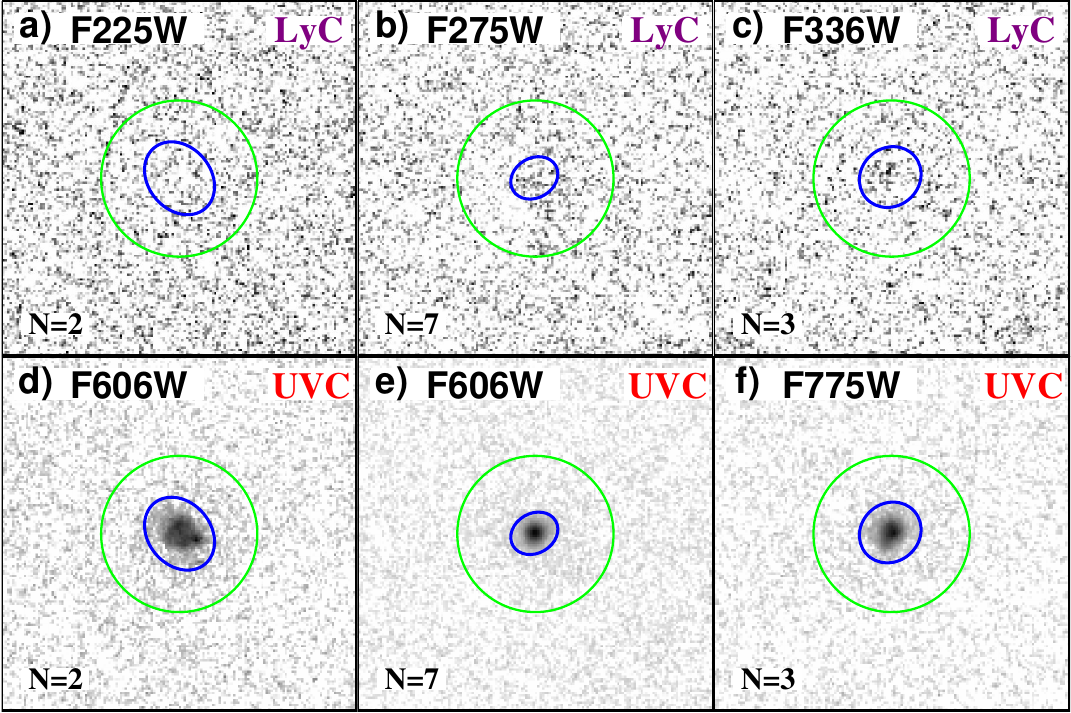}}
\caption{\noindent\small As Fig.~\ref{fig:figure6} for only the galaxies
hosting weak AGN.  \label{fig:figure8}} \end{figure}
We also subtracted a constant from each of the sub-images before object
removal in order to bring the mode of the sky-background of
the images as close to zero as possible. We calculated the mode of
background level from the count-rate histogram of the surrounding pixels outside
the central circular aperture, binning their values according to the Freedman-Diaconis
rule (see \S\ref{sec:sky}). We then fit each sky histogram to a Gaussian
function by least squares, taking the mode of the fitted Gaussian as the
background constant. This local sky-background removal was applied in order to
sum the actual LyC flux  \emph{above} the background from each sub-image,
rather than LyC+background, as variations in background levels between sub-images
can suppress the flux contribution from the faintest LyC emission in
the stack. The subtraction also removed any residual small scale gradients
between the sub-images left from bias/sky-subtraction, flat-fielding, and/or
exposure time/noise variations in the mosaics.

\noindent\begin{table*}[ht!] \centering \caption{LyC Stack
Photometry\label{table:table2}} \setlength{\tabcolsep}{6.8pt}
\begin{tabular}{lccrcccccr}     \toprule \\[-21pt] Filter        & $z$-range
& \zmean           & $N_{\rm obj}\!\!$ &  m$_{\!\lyc}$  & ABerr$_{\!\lyc}$
& SNR$_{\!\lyc}$   & A$_{\uvc}$     & m$_{\uvc}$        &  SNR$_{\uvc}$
\\[-5pt] \multicolumn{1}{c}{(1)}       & (2)          & (3)          &
(4)       & (5)          & (6)          &                      (7)       & (8)
& (9)          &  \multicolumn{1}{c}{(10)}     \\[-1pt] \midrule \\[-21pt]
\multicolumn{10}{l}{\sc All Galaxies:}\\[-5pt]
F225W & 2.276--2.450 & 2.352 & 19 & $>$28.26 & \nodata & (1.00)$^{\dagger}$ & 1.034 & 24.41 & 426.7 \\[-5pt]
F275W & 2.470--3.076 & 2.685 & 14 & 28.11 & 0.45 & 2.41 & 0.681 & 24.76 & 323.8 \\[-5pt]
F336W & 3.132--4.149 & 3.537 & 13 & $>$28.62 & \nodata & (1.00)$^{\dagger}$ & 0.553 & 24.63 & 247.5 \\[-5pt]
\multicolumn{10}{l}{\sc Galaxies without AGN:} \\[-5pt]
F225W & 2.276--2.449 & 2.350 & 17 & $>$27.91 & \nodata & (1.00)$^{\dagger}$ & 1.015 & 24.36 & 423.8 \\[-5pt]
F275W & 2.566--3.076 & 2.752 & 7 & $>$28.12 & \nodata & (1.00)$^{\dagger}$ & 0.932 & 24.46 & 268.3 \\[-5pt]
F336W & 3.132--4.149 & 3.603 & 10 & $>$30.73 & \nodata & (1.00)$^{\dagger}$ & 0.555 & 24.75 & 192.7 \\[-5pt]
\multicolumn{10}{l}{\sc Galaxies with AGN:} \\[-5pt]
F225W & 2.298--2.450 & 2.374 & 2 & $>$27.91 & \nodata & (1.00)$^{\dagger}$ & 0.637 & 25.21 & 85.0 \\[-5pt]
F275W & 2.470--2.726 & 2.618 & 7 & 28.26 & 0.41 & 2.66 & 0.253 & 25.12 & 232.7 \\[-5pt]
F336W & 3.217--3.474 & 3.316 & 3 & 27.42 & 0.44 & 2.47 & 0.486 & 24.38 & 158.7 \\[-2pt]
\bottomrule\vspace*{-12pt} \end{tabular}
\begin{minipage}{\txw}{\small Table columns: (1): WFC3 filter used;
(2): Redshift range of galaxies included in LyC/UVC stacks;  (3): Average
redshift of stack;  (4): Number of galaxies with high quality spectroscopic
redshifts used in the stacks; (5): Observed total AB magnitude of LyC emission from
stack (\SExtractor\ \texttt{MAG\_AUTO} aperture matched to UVC, indicated by the blue
ellipses in Figs.~\ref{fig:figure6}--\ref{fig:figure8}; (6): 1$\sigma$ error of average
LyC AB-mag (7): Measured SNR of the LyC stack flux within matched UVC aperture ($^{\dagger}$
indicates a 1$\sigma$ upper limit); (8): Area (in arcsec$^2$) of the UVC aperture;
(9): Observed total AB magnitude of the UVC stack; (10): Measured SNR of the UVC stack.} 
\end{minipage} \end{table*}
We then stacked the processed sub-images of all the galaxies in each UVIS/ACS
image with spectroscopic redshifts, where LyC can be observed in their
respective filter, using the average of the pixel count rates of the
sub-images, weighted by their corresponding \textsc{AstroDrizzle}
\citep{Fruchter2010, Gonzaga2012} weight maps. We did this by summing the weighted pixel
values of the processed sub-images, normalized by the sum of their weights
(\ie\ $\langle f_j\rangle\!\!=\!\!{ {\sum\limits_i W_i f_{i,j}} / {\sum\limits_i
W_i} }$, where $f_j$ represents the flux in counts per second measured in
pixel $j$ for sub-image $i$ and $W_i$ is the weight map for sub-image $i$).
We then created stacked weight maps for each LyC stack by summing the inverse
of the pixel values of the corresponding region in the weight maps, where the
galaxy sub-images were extracted, then inverting the sum to generate the
stacked weight maps (\ie\ $W_j\!=\!1/{\sum\limits_i{1/W_{i,j}}}$, where $W_j$
is the weight for pixel $j$ in sub-image $i$). These weight maps give the
relative weight of each pixel in the LyC stacks, and are used only for
quantifying all photometric errors in the observations. We created
stacks for the {\it total} sample of galaxies and separate stacks for the galaxy and
AGN samples, since each sample likely produces the majority of their LyC
photons by different mechanisms, which must be taken into account when
determining \fesc\ for these galaxies.

Since many \fesc\ values quoted in the literature are calculated relative to
the rest-frame non-ionizing UVC flux measured from
1500\,$\lesssim$\,$\lambda_0$\,$\lesssim$\,1700\,\AA\ (see
\S\ref{sec:fescintro}), we created corresponding UVC stacks for each LyC stack
from sub-images extracted from the ACS/WFC mosaics of the ERS/GOODS-S fields
that sample the UVC emission of our galaxies. For the redshift intervals that
sampled LyC emission in the F225W, F275W, and F336W filters that sample
the UVC emission correspond to F606W, F606W, and F775W, respectively.

The galaxies stacked in the WFC3/UVIS F225W filter contain co-added sub-images
frames of 19 galaxies over the redshift range 2.276\,$\le$\,$z$\,$\le$\,2.450
(\zmean\,=\,2.352), the F275W stack contains 14 galaxies at
2.470\,$\le$\,$z$\,$\le$\,3.076 (\zmean\,=\,2.685), and the F336W stack contains
13 co-added galaxies at 3.132\,$\le$\,$z$\,$\le$\,4.149 (\zmean\,=\,3.537).
These stacks, as well as the corresponding UVC stacks, are shown in Fig.~\ref{fig:figure6}.
Stacks for the subsamples of galaxies with weak AGN and galaxies without AGN are shown in
Fig.~\ref{fig:figure7} and \ref{fig:figure8}, with elliptical apertures
indicating regions where photometry was performed. 

The deepest galaxy counts in $J$ and $H$-band of \citetalias{Windhorst2011}
give us an estimate the total number of contaminating objects that could be
present in our r$\simeq$0$\farcs$5 radius LyC apertures
(Fig.~\ref{fig:figure6}--\ref{fig:figure8}). To the ERS limit of $J$,
$H$$\lesssim$27.55--27.25 mag, respectively, there are
$\lesssim$5.2$\times$10$^5$ galaxies deg$^{-2}$ (\citetalias{Windhorst2011}), yielding a $\lesssim$3\%
probability of finding one unrelated foreground object in, or overlapping
with, the LyC aperture. For our sample of 46 galaxies, this would amount to
$\lesssim$2 interlopers. Due to the possible interloper's proximity to the LyC
candidate, it is not always possible to obtain reliable spectroscopic or
photometric redshifts for these neighbors (see \S\ref{faintcontam} for a discussion on contamination from interlopers). Nonetheless, in these few cases,
light from the nearby neighbors was masked out with \SExtractor\ segmentation
maps. This was then repeated for \emph{all} other objects in the 151$\times$151 pixel image
sections outside the central r$\simeq$0\farcs5 aperture, to exclude
contaminating objects in the photometry in the central aperture, and to assure
that accurate measurements of the surrounding sky could always be obtained.
\subsection{Rest-frame Lyman Continuum Photometry}
\label{sec:photometry}

The results from our photometry measured in the apertures shown in
Fig.~\ref{fig:figure6}--\ref{fig:figure8} are summarized in
Table~\ref{table:table2}. Because the LyC flux escaping from galaxies in these
stacks is very faint, we perform all of our photometry on the LyC stacks with
\SExtractor\ using detection images in dual-image mode. As our reference
images, we use the corresponding non-ionizing UVC stacks to measure any
possible escaping LyC flux detectable within the aperture of brighter UVC
counterpart.

\noindent\begin{figure}[hb!]
\centerline{\includegraphics[width=.5\txw]{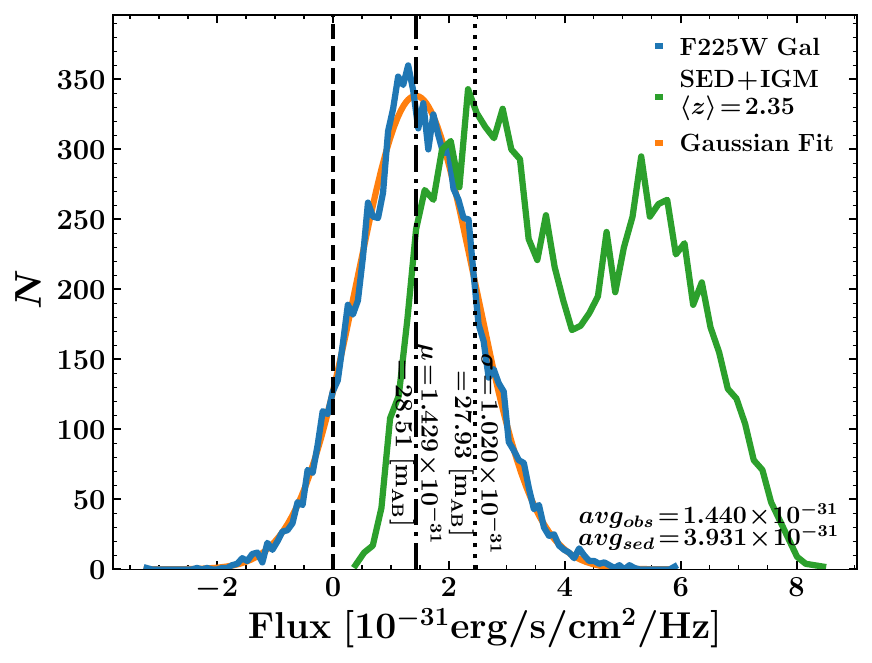} }
\caption{\noindent\small Example flux distribution for the F225W galaxies \emph{without} AGN
stack used for our photometric analysis listed in Table~\ref{table:table2}. Each pixel
in the stack was given a mean based on the pixel value in the stacked F225W image, 
and a variance from the sum of the sky-background variance and the square of the 
corresponding pixel value in the stacked RMS map. The blue distribution was generated 
by summing the pixel flux distributions inside the blue aperture from Fig.~\ref{fig:figure7} 
for each realization of the stack. The orange line is the Gaussian curve fit 
to the blue distribution. The mean and +1$\sigma$ values are shown as vertical dash-dot and dotted lines, respectively. The green distribution is the modeled intrinsic flux using 
the stacked best fit SED convolved with the IGM transmission models of \citet{Inoue2014}
and fitting error. The average value of the blue and green distributions is indicated as $avg_{obs}$ and $avg_{sed}$, respectively.
\label{fig:figure9}} \end{figure}
We used the individual RMS maps and the sky-background variance for each
sub-image in a given stack to create 10,000 random variations of each pixel
for each stack based on the combinations of these uncertainties, in order to
assess photometric errors and upper limits. This approach allows us to
generate flux distributions of the stack photometry based on systematic
uncertainties within the data itself. We measure the flux in the UVC matched
aperture for each realization of the stack and plotted them as shown in
Fig.~\ref{fig:figure9}. We quote the mean and 1$\sigma$ value of the flux
distributions in Table~\ref{table:table2}, or the 1$\sigma$ value as the upper-limit
for non-detections. We convert the flux measured by \SExtractor\ to AB magnitudes,
using the infinite aperture zeropoints listed on the STScI instrument
websites\footnote{\url{http://www.stsci.edu/hst/wfc3/phot\_zp\_lbn}}.

From these distributions, we measure an average LyC flux from galaxies
and AGN at \mAB\,$\simeq$\,28.11\,mag, with a SNR value at
$\sim$\,2.41 for the F275W stack. We measure 1$\sigma$ upper bounds of
\mAB\,$>$\,29.02 and 28.62\,mag for the F225W and F336W stacks, respectively.
For only the galaxies without AGN (Fig.~\ref{fig:figure7}), we place
1$\sigma$ upper bounds for the flux measured in the F225W, F275W and F336W
stacks at \mAB\,$>$\,27.91, 28.12 and 30.73\,mag, respectively.
The flux from galaxies with AGN was measured at \mAB\,$\simeq$\,28.26 and
27.42\,mag, with SNR\,$\sim$\,2.66 and 2.47 for the F275W and F336W stacks,
respectively, and we placed a 1$\sigma$ upper bound to the F225W stack flux at
\mAB\,$>$\,27.91\,mag (see Table~\ref{table:table2}). Our photometry indicates that
the AGN stacks are brighter than galaxies without AGN and have higher SNR,
despite having fewer contributing sub-images in the stacks.

We note that, although some LyC flux
might exist at the $\sim$1$\sigma$ level outside the measurement apertures, we
do not incorporate this flux into our measurement, as this would require us to
increase our aperture size and add extra noise in the aperture, which would
increase the uncertainty of our measurements, as well as the interloper
contribution.

We performed a series of critical tests on our data to ascertain the
robustness and validity of our stacking procedures and LyC detections, for
which we refer the interested reader to Appendix~\ref{sec:stacktests}. From
these tests, we conclude that our measurements are reliable to within their
measured errors or upper bounds, and are not the result of various
possible sources of spurious signal.
\noindent\begin{figure}[hb!]
\centerline{\includegraphics[width=.5\txw]{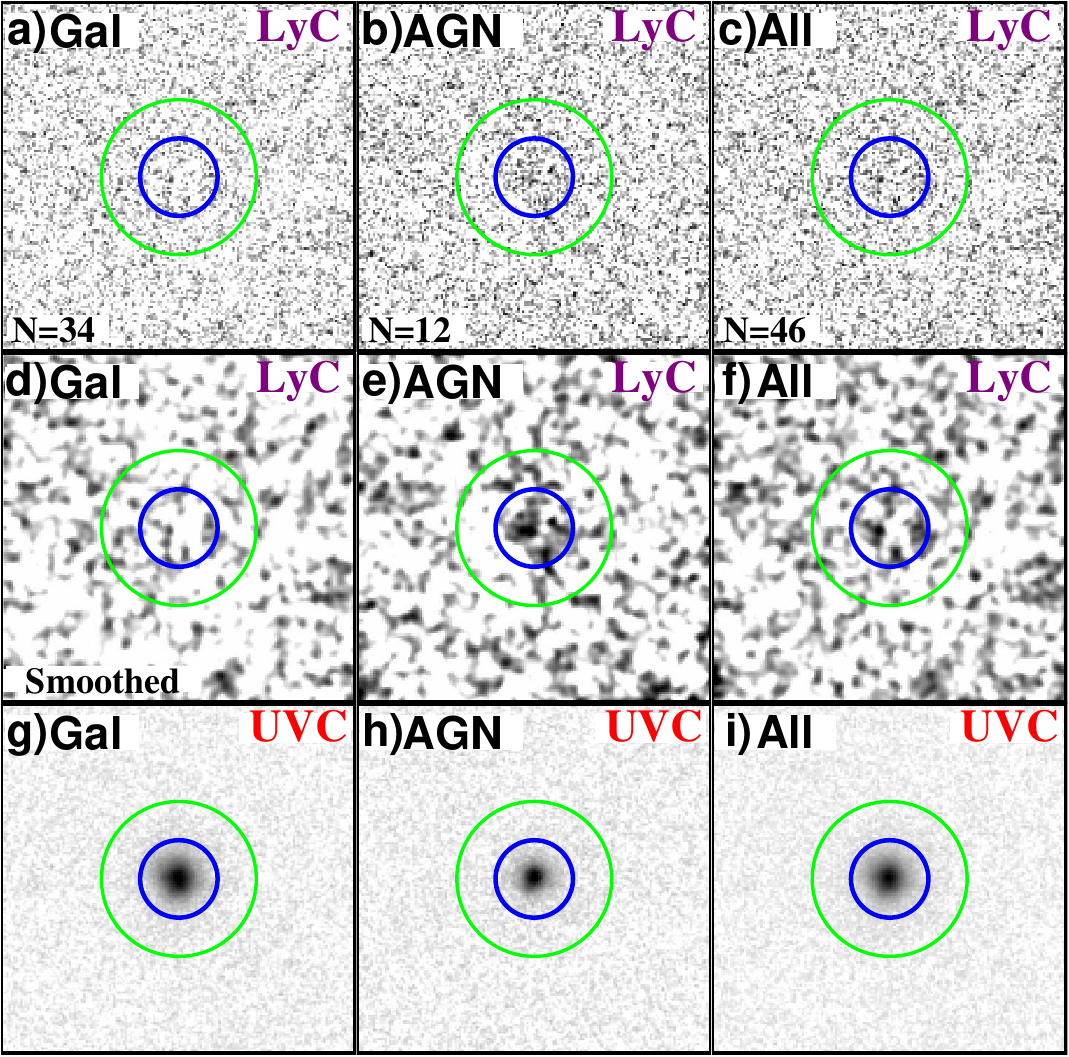} }
\caption{\noindent\small [\emph{Top Row}]: LyC stacks of all galaxies in
our sample with high quality spectra and reliable redshifts;  
[\emph{Middle Row}]: The same as the top row but convolved with a 1$\sigma$
Gaussian kernel. [\emph{Bottom Row}] The UVC counterparts of the top row;
\emph{a, d, and g} [Left column of panels:] Composite stacks of all galaxies
without AGN in our spectroscopic sample observed in the F225W, F275W, and
F336W filters; \emph{b, e, and h} [2nd column:] Composite stacks of all
galaxies hosting (weak) AGN; \emph{c, f, and i} [3rd column:] Composite stacks
of all 46 galaxies in our sample. These stacks represent the average
\emph{observed} LyC $F_{\nu}$ from all galaxies integrated from
2.3\,$\lesssim$\,$z$\,$\lesssim$\,4.1, scaled to a common zeropoint magnitude.
The blue and green circles have radii 0\farcs5 and 1\farcs0, respectively.
The measured SNR of the combined LyC emission in these stacks is $\sim$2.3, 0.7, and 3.9\,$\sigma$ for the stack
of all galaxies, all galaxies without AGN, and all galaxies with AGN,
respectively. The AGN stacks exhibit both a centrally concentrated and
extended component in their flux distributions, from contributions of a
central AGN point source and perhaps also from scattered photons
(Fig~\ref{fig:figure7}). These images suggest that LyC escape paths may be slightly
offset from a galaxy center, including point source emission from the AGN.
Given the random orientation of galaxies in each stack, this would explain the
faint, non-centrally concentrated, and extended morphology of the detected LyC
emission. \label{fig:figure10}} \end{figure}

\subsection{Super-Stacks of LyC Emission from Galaxies at 2.3\,$\le$\,$z$\,$\le$\,4.1} 
\label{sec:stackofstacks}

In order to determine at what SNR our observations can measure the LyC flux
from our total sample of galaxies that span the 2.3\,$\le$\,$z$\,$\le$\,4.1
redshift range, and probe the faintest LyC emission from our galaxies, we
construct a stack of the observed LyC flux in $f_{\nu}$ of \emph{all} the
galaxies in our sample as follows. We first extract the sub-images from the
F225W, F275W, and F336W and apply our sky-subtraction and
neighbor masking procedures, as outlined in \S\ref{sec:method}. We then
scale all sub-images in the stack to a common zeropoint and stack the
sub-images by a weighted average, as described in \S\ref{sec:method}. We
created stacks for the full sample of all galaxies and separate stacks for
the galaxies with and without AGN, as shown in Fig.~\ref{fig:figure10}. We
note that these ``Super-Stacks'' represent the average \emph{observed} LyC
flux from our sample integrated from $z$\,=\,2.3--4.1 through the various
sight-lines, and so the flux in these stacks will be dominated by the galaxies with
the brightest apparent LyC flux.

Due to the very low sky-background in the F225W and F275W filters
\citepalias{Windhorst2011}, the relative scaling of the count rates in the
sub-images slightly amplified the contribution of detector noise from these
filters in the stack. However, since the F336W filter is limited
by photon noise from the much brighter zodiacal background at these
wavelengths, these ``Super-Stacks'' therefore also have more significant
sky-background. We find that the stack of the total sample reaches a SNR
of $\sim$2.3$\sigma$, while the LyC stacks of galaxies with and without AGN achieve
SNR ratios of 3.9$\sigma$ and 0.7$\sigma$ in their UVC matched apertures, respectively.
We also find that the average apparent AGN LyC flux outshines that from galaxies without
AGN by a factor of ${ F_{\nu}^{\mathrm{AGN}} } / { F_{\nu}^{\mathrm{Gal}}
}$$\simeq$7.7 ($\Delta$\mAB$\simeq$2.2\,mag). 

Since these ``Super-Stacks'' were created solely for the purposes of visualization
and probing the SNR of the total observed LyC signal from our samples, we do
not perform any further analysis of the LyC flux measured in these stacks.
Because the absorption of ionizing photons by the IGM is non-linearly
dependent on redshift, modeling of the combined intrinsic LyC flux from
galaxies spanning 2.3\,$\lesssim$\,$z$\,$\lesssim$\,4.1 through various sight-lines
would become increasingly divergent, such that the resulting total
$\fesc$ value of these galaxies would be highly uncertain. 

This exceedingly faint LyC emission emerging from the stack of the 12
galaxies with AGN appears to have a flat spatial distribution that is
\emph{not} centrally concentrated. This may allude to the manner in which LyC
escapes from galaxies. In order to ionize the IGM, LyC photons must escape
through holes in the surrounding gaseous and/or dusty material between stars,
the central point source, and the line-of-sight IGM, which can be distributed
randomly within or around galaxies. With at most a few clear sight-lines per galaxy,
these stacked images suggest that some escape paths of LyC may be on average
somewhat offset from the galaxy center, \ie\ escaping more from the outskirts than
the centers of these galaxies. Given the random orientation of galaxies in each stack,
this would explain the faint, non-centrally concentrated, and extended morphology
of the detected LyC emission. This may indicate that LyC photons produced by
accretion disks in AGN escape from galaxies with weak AGN via scattering.
We discuss the radial profiles of galaxies further in \S\ref{sec:obsradprof}.

In our adopted Planck cosmology, the angular size scale decreases by
$\sim$16\% and the apparent fluxes dim by $\sim$75\% from $z$\,$\simeq$\,2.3
to $z$\,$\simeq$\,4.1. We deliberately did \emph{not} scale any of the pixel
values or resample the pixel scale to account for these these changes during
the stacking process over all redshifts, as we only created these super-stacks
to quantify the SNR of the average \emph{observed} LyC flux for our entire
sample. Using a varying pixel scale for each galaxy would have introduced
correlated inter-pixel resampling noise, which would also decrease the sky SB
limits and the effect of averaging over residual subtle systematics. Stacking
with the same plate scale for all redshifts also preserves the observed photon
statistics, which are needed for accurate sky-subtraction. Hence, resampling
all the images as a function of redshift would reduce the SNR of the resulting
stacked LyC signal. We note that the physical scale of the galaxies that we
stack changes by $\pm$4\% within each redshift bin and by $\pm$16\% for the
entire sample. This does not noticeably affect the LyC and UVC light profiles
in each of our three redshift bins, which are discussed in
\S\ref{sec:obsradprof}, but does ``blur'' the light distribution seen in the
super-stacks in Fig.~\ref{fig:figure10} by approximately these amounts in the
radial direction from the center. Fig.~\ref{fig:figure10} can thus only be
used to visualize the combined \emph{observed} LyC signal over the entire
redshift range $z$\,$\simeq$2.3--4.1, but cannot be used for further
quantitative analysis.

\section{LyC Escape Fractions, and Radial Surface Brightness Profiles}
\label{sec:fesc}

\subsection{Relative and Absolute LyC Escape Fractions for Stacks}
\label{sec:fescintro}

Estimating the escape fraction of LyC photons from galaxies, \fesc, is non-trivial,
as it requires modeling of their apparent \emph{intrinsic} LyC flux,
$F^{\rm int}_{\nu,\!\lyc}$, and the wavelength-dependent transmission of LyC
photons through the IGM, $\mathcal{T}^{\!\lyc}_{\igm}(z,\nu)$, for a galaxy at
redshift $z$. Quantitatively, the average \emph{observed} LyC flux  ($F^{\rm
obs}_{\nu,\!\lyc}$) from a stack of galaxies, measured by a photon counting
device such as a CCD, is given by:
\vspace{-6pt}
\begin{equation}  \langle F^{\rm obs}_{\nu,\!\lyc}\rangle =
\frac{1}{\mathrm{N_{gal}}}\sum_{\rm i=1}^{N_{\rm gal}} \frac{\int
T^{\!\lyc}_{\rm obs}(\nu)\mathcal{T}_{\igm}(z_{\rm i},\nu)f_{\rm esc,i}^{\rm
abs}F_{\nu\rm ,i}^{\rm int}(\nu)\frac{d\nu}{\nu}}{\int T^{\!\lyc}_{\rm
obs}(\nu)\frac{d\nu}{\nu}} \tag{1} \label{eq:fobs} \end{equation}

\noindent where $N_{\rm gal}$ denotes the number of galaxies in the stack,
$F_{\nu\rm ,i}^{\rm int}$ denotes the \emph{intrinsic} (i.e. produced)
stellar SED from galaxy `i', $f_{\rm esc,i}^{\rm abs}$ denotes the fraction of
the observed LyC flux that escaped from the galaxy into the IGM,
$\mathcal{T}_{\igm}(z_{\rm i},\nu)$ denotes the wavelength-dependent IGM
transmission curve for galaxy 'i' at redshift $z$, which we acquired from the
recent absorption models of \citet{Inoue2014}, and the $T^{\!\lyc}_{\rm
obs}(\nu)$ term denotes the combined transmission of the throughput from the
Optical Telescope Assembly (OTA), the filter throughput, and the Quantum
Efficiency (QE) of the detector used for the LyC observation (see
Fig.~\ref{fig:figure1} for the WFC3/UVIS total system throughput
curves).

We expect $f_{\rm esc}$ generally to vary between individual objects. However,
since we stack the observed LyC emission from all galaxies in our sample, we
simplify our analysis by assuming a constant $f_{\rm esc}^{\rm abs}$ value for
all galaxies within a given redshift bin. We denote this `sample averaged'
escape fraction with $\langle\fescabs\rangle$. We can then take $\langle
\fescabs\rangle$ outside of the sum in Eq~\ref{eq:fobs}, and write: 
\vspace*{-6pt}
\begin{equation}  \langle\fescabs\rangle=\frac{\langle F^{\rm
obs}_{\nu,\!\lyc}\rangle}{\frac{1}{\mathrm{N_{gal}}}\sum\limits_{\rm i=1}^{\rm
N_{gal}}\frac{\int T^{\!\lyc}_{\rm obs}(\nu)\mathcal{T}_{\igm}(z_{\rm
i},\nu)F_{\nu\rm ,i}^{\rm int}(\nu)\frac{d\nu}{\nu}}{\int T^{\!\lyc}_{\rm
obs}(\nu)\frac{d\nu}{\nu}}}, \tag{2a} \label{eq:fescabs}  \end{equation}

\noindent which can also be expressed as:
\vspace*{-4pt}
\begin{equation} \langle\fescabs\rangle=\frac{\langle F^{\rm
obs}_{\nu,\!\lyc}\rangle}{\frac{1}{\mathrm{N_{gal}}}\sum\limits_{\rm i=1}^{\rm
N_{gal}}F^{\rm int}_{\nu,\!\lyc,\rm i}}=\frac{\langle F^{\rm
obs}_{\nu,\!\lyc}\rangle}{\langle F^{\rm int}_{\nu,\!\lyc}\rangle}. \tag{2b}
\label{eq:fescabssimp} \end{equation}

A more thorough analysis should also take into account that the impact of the
IGM varies substantially around this mean for individual objects and sight-lines
\citep[see][]{Inoue2008,Nestor2011}. We determine $F_{\nu\rm ,i}^{\rm
int}(\nu)$ for each galaxy from the minimized $\chi^2$ fit
\citet[][BC03]{Bruzual2003} synthetic stellar population model SED, which was
fit to 4--6 non-ionizing continuum broadband WFC3/IR+ACS/WFC measurements
longwards of \Lya\ at the known \emph{fixed} redshift. These best fit SEDs
were allowed four degrees of freedom for the $\chi^2$ minimization at the fixed
spectroscopic redshift of each object, \ie\ the age, stellar mass, \AV, and
the exponentially decreasing SFR timescale ($\tau$). These were fit from a grid
of SEDs using a wide range in each of these parameters. Thus, the best fit SEDs
correspond to the observed $F_{\nu}$ of the galaxy \emph{with} \citet{Calzetti2000}
attenuation applied to the \emph{intrinsic SED}, thereby determining the best
fit \AV\ value. We use a \citet{Salpeter1955} IMF and adopt solar metallicities
for the SEDs.

In order to determine the absolute fraction of escaping LyC ($\fescabs$),
which compares the apparent flux of LyC photons produced by stars in the
galaxy to the observed LyC emission (\ie\ $F_{\nu,\!\lyc}^{\rm
obs}/F_{\nu,\!\lyc}^{\rm int}$), the effects from dust must be removed from
the SED to obtain its  \emph{intrinsic} LyC flux, $F_{\nu,\!\lyc,\rm i}^{\rm
int}(\nu)$. Since we began our SED fitting with the intrinsic stellar
population photospheric flux --- which was then reddened by a specified \AV\ value
using \citet{Calzetti2000} attenuation --- we simply use the initial
dust-free intrinsic stellar photospheric SED to calculate $F_{\nu\rm ,i}^{\rm
int}(\nu)$ for each galaxy. Thus, the \emph{absolute} escape fraction
quantifies the amount of LyC that is not absorbed by dust, the multiphase ISM,
or other sources of LyC absorption in the galaxy. 

The sample average escape fraction of LyC flux \textit{relative} to the
non-ionizing UVC flux ($F_{\nu,\uvc}$), denoted as \fescrel, is defined the as:
\vspace*{-6pt}
\begin{equation} \langle\fescrel\rangle\!=\!\frac{1}{\mathrm{N_{gal}}}
\sum\limits^{\rm N_{gal}}_{\rm i=1}\!\frac{F_{\nu,\uvc,\rm i}^{\rm
int}\big{/}\!\!F_{\nu,\!\lyc,\rm i}^{\rm int}}{F_{\nu,\uvc,\rm i}^{\rm
obs}\big{/}\!\!F^{\rm obs}_{\nu,\!\lyc,\rm
i}}\!\simeq\!\frac{1}{\mathrm{N_{gal}}}\frac{\langle F_{\nu,\!\lyc}^{\rm
obs}\rangle}{\langle F_{\nu,\uvc}^{\rm obs}\rangle}\!\sum\limits^{\rm
N_{gal}}_{\rm i=1}\!\frac{F_{\nu,\uvc,\rm i}^{\rm int}}{F_{\nu,\!\lyc,\rm
i}^{\rm int}} \tag{3a} \label{eq:fescrel} \end{equation}

\noindent Using Eq.~\ref{eq:fescabssimp}, we can further simplify this
expression as:
\vspace*{-10pt}
\begin{equation}
\langle\fescrel\rangle\!\simeq\!\langle\fescabs\rangle\frac{\sum\limits_{\rm
i=1}^{\rm N_{gal}}F^{\rm int}_{\nu,\uvc,\rm i}}{\sum\limits_{\rm i=1}^{\rm
N_{gal}}F^{\rm obs}_{\nu,\uvc,\rm
i}}=\langle\fescabs\rangle\Big{\langle}\!\frac{F^{\rm int}_{\nu,\uvc}}{F^{\rm
obs}_{\nu,\uvc}}\!\Big{\rangle}, \tag{3b} \label{eq:fescrelsimp}
\end{equation}\vspace{-6pt}

\noindent where $F_{\nu,\uvc,\rm i}^{\rm obs}$ is the observed UVC flux from
galaxy 'i' as measured in the ACS/WFC UVC filters (see \S\ref{sec:method})
and:
\begin{equation} F_{\nu,\uvc,\rm i}^{\rm int}=\frac{\int T^{\uvc}_{\rm
obs}(\nu)F_{\nu\rm ,i}^{\rm int}(\nu)\frac{d\nu}{\nu}}{\int T^{\uvc}_{\rm
obs}(\nu)\frac{d\nu}{\nu}} \tag{4} \label{eq:fuvci} \end{equation}

\noindent\begin{table*}[ht!]  \centering  \caption{Summary of \fesc\
Constraints \label{table:table3}}  \setlength{\tabcolsep}{4.2pt}
\begin{tabular}{rrrrrrrrr}      \toprule\\[-20pt]  \colhead{\zmean} & N$_{obj}$ &
\colhead{$\langle f_{\uvc}/f_{\!\lyc}\rangle_{\rm obs}$} & \colhead{$\langle
f_{\uvc}/f_{\!\lyc}\rangle_{\rm int}$} & \colhead{$\langle \mathrm{t_{\rm
age}}\rangle$}  & \colhead{\AV$_{med}$}  & \colhead{$\langle
\mathcal{T}_{\igm} \rangle$} & \colhead{$\langle\fescabs\rangle$} \\[-5pt]
\multicolumn{1}{c}{} & \multicolumn{1}{c}{} & \multicolumn{1}{c}{} & \multicolumn{1}{c}{} & \multicolumn{1}{c}{[yr]} & \multicolumn{1}{c}{[mag]} & \multicolumn{1}{c}{} & \multicolumn{1}{c}{[\%]} \\[-5pt]
\multicolumn{1}{c}{(1)} & \multicolumn{1}{c}{(2)} & \multicolumn{1}{c}{(3)} & \multicolumn{1}{c}{(4)} & \multicolumn{1}{c}{(5)} & \multicolumn{1}{c}{(6)} & \multicolumn{1}{c}{(7)} & \multicolumn{1}{c}{(8)} \\[-2pt] \midrule \\[-21pt]
\multicolumn{8}{l}{\sc Galaxies without AGN:} \\[-4pt]
2.350 & 17 & $27^{+61}_{-5}$ & $20.2\pm0.1$ & $10^{8.2^{+0.9}_{-0.3}}$ & $0.40^{+0.20}_{-0.40}$ & $0.326^{+0.062}_{-0.085}$ & $22^{+44}_{-22}$ \\[-5pt]
2.752 & 7  & $<98.0$  & $12.8\pm0.1$ & $10^{7.9^{+0.6}_{-0.1}}$ & $0.40^{+0.21}_{-0.02}$ & $0.218^{+0.102}_{-0.085}$ & $<53$ \\[-5pt]
3.603 & 10 & $<50.1$ & $15.6\pm0.2$ & $10^{8.5^{+0.6}_{-0.8}}$ & $0.0^{+0.4}_{-0.0}$ & $0.066^{+0.045}_{-0.033}$ & $<55$ \\[-2pt] 
\bottomrule\vspace*{-12pt}  
\end{tabular}  \begin{minipage}{\txw}{\small Table columns:
(1): Average redshift of each stack; (2): Number of objects in each redshift
bin, as in Table~\ref{table:table2}; (3): Average \emph{observed} flux ratio
$f_{\nu,\uvc}/f_{\nu,\!\lyc}$ and its $\pm$1$\sigma$ error range, as measured
from the LyC and UVC stacks in their respective apertures (see
\S\ref{sec:photometry} and Table~\ref{table:table2}); (4): Average
\emph{intrinsic} flux ratio $f_{\nu,\uvc}/f_{\nu,\!\lyc}$ and its
$\pm$1$\sigma$ error range, as derived from the BC03 best fit SED models
galaxies without AGN in each of our redshift bins (see \S\ref{sec:fescintro}
and Eq.~\ref{eq:intratio}); (5): Average age of the stellar populations from the
best fit BC03 models and their $\pm$1$\sigma$ standard deviations in years; (6):
Median dust extinction \AV\ and its $\pm$1$\sigma$ error range of the best fit
BC03 SED model (the median \AV\ is more representative, as the distributions of
each subsample is asymmetric; see \S\ref{sec:extinction}); 
(7): Average filter-weighted IGM transmission of all sight-lines and redshifts
in the stacks and their $\pm$1$\sigma$ standard deviations, calculated from
the \citet{Inoue2014} models; (8) ML and $\pm$1$\sigma$ or upper limit values of the Monte Carlo analysis of \fescabs\ in percent, \ie\ the escape fraction of LyC including effects from all components of the ISM and reddening by dust as described in \S\ref{sec:fescintro} (Eq.~\ref{eq:fescabssimp}) } \end{minipage}
\end{table*}

\noindent for UVC observations with a total system throughput of
$T^{\uvc}_{\rm obs}(\nu)$. Thus, the relative and absolute escape fractions
differ by a factor of $\Big{\langle}\!\frac{F^{\rm int}_{\nu,\uvc}}{F^{\rm
obs}_{\nu,\uvc}}\!\Big{\rangle}$\,=\,$\langle f_{\rm esc}^{\uvc}\rangle^{-1}$
for the total sample, which deviates from unity depending on the \AV\ and
$\chi^2$ values of the SED fits. The escape fraction of non-ionizing UVC
photons, $f^{\uvc}_{\rm esc,i}$, is related to the observed reddening in
galaxy 'i' as $f_{\rm esc}^{\uvc}$\,=\,$10^{-0.4\,A_{\uvc}}$. Note that
this term can be omitted from Eq.~\ref{eq:fescrel} when using the
\emph{intrinsic} (unreddened) model SEDs instead of the observed ones. We determine
the ratio of intrinsic fluxes of the LyC and UVC emission from all galaxies as:
\vspace{-6pt}
\begin{equation}
\Big{\langle}\frac{F_{\uvc}}{F_{\!\lyc}}\Big{\rangle}_{_{_{\!\!\!\rm
int}}}\!\!\!=\frac{\sum\limits^{\rm N_{gal}}_{\rm i=1}F_{\nu,\uvc,\rm i}^{\rm
int}}{\sum\limits^{\rm N_{gal}}_{\rm i=1}F_{\nu,\!\lyc,\rm i}^{\rm int}}
\tag{5} \label{eq:intratio} \end{equation}\vspace{-6pt}

\noindent without applying the $\mathcal{T}_{\igm}(z_{\rm i},\nu)$ term to
$F_{\nu,\!\lyc,\rm i}^{\rm int}$. The \textit{observed} LyC and UVC flux
ratios can be obtained by performing photometry on stacked images of the LyC
and UVC emission shown in figures \ref{fig:figure6}--\ref{fig:figure8}.

Compared to
the \emph{intrinsic} LyC flux, the \emph{intrinsic} stellar UVC flux of these
galaxies increases by a factor $2.98^{+0.08}_{-0.07}$
\citep[e.g.,][]{Siana2010}. We determine this factor as expressed in
Eq.~\ref{eq:intratio} from the intrinsic SEDs and filter curves used for LyC
and UVC observations, with their 1$\sigma$ dispersion, as listed in Col. 4 of
Table~\ref{table:table3}. Moreover, at
\zmean\,$\simeq$\,2.68, the \emph{filter weighted} average IGM transmission of
the redshifts in the stack is 0.247$^{+0.086}_{-0.085}$, which was determined
from the models of \citet{Inoue2014}, and is listed in Col. 7 of
Table~\ref{table:table3}. 

We also include an estimate of the SED error from
the 4--6 observed continuum data points that were used to fit each SED at its
fixed known redshift, \ie\ the filters that sample the continuum emission of
the galaxies longwards of \Lya\ up to WFC3/IR F125W, which are not affected by
the IGM, even at high redshift. The main uncertainty in the SED fitting is
therefore \emph{not} the $\chi^2$ values of fitted data, but the uncertainty
in the applied internal extinction values \AV\ to each SED, which is unknown.
We therefore do not include a dust correction \emph{error} in our calculations.
\citet{Calzetti2000} empirically derived the dust attenuation curves of nearby
starburst galaxies and found a total to selective extinction value of
$R_V\!=\!4.05\!\pm\!0.8$. Our SEDs were reddened with various dust screen \AV\
values by $F_{\nu}^{\rm obs}(\lambda)\!=\!F_{\nu}^{\rm
int}(\lambda)10^{-0.4A_{\lambda}}$, where:
$A_{\lambda}\!=\!{k(\lambda)\AV}/{R_V}$. The attenuation for wavelengths
shorter than 630\,\AA\ and longer than 2200\,\AA\ were extrapolated from the
interpolated slope of the endpoints of the attenuation curves. The applied
reddening does not include the uncertainty in the $R_V$ value. Estimating the
reddening error in the flux of our sample galaxies would require a more
extensive SED fitting analysis, which takes into account the equally probable
\AV\ values that fall within the measurement errors of the observed continuum
data points. We instead assume a single \AV\ value and vary the SED flux based
on their observed continuum errors. This error estimate is equivalent to
applying a convolution to the intrinsic SED LyC flux with a Gaussian kernel
(see \S\ref{sec:fescMC}), and so applying an additional dust correction error
would only increase the size of this kernel. Since the uncertainties in our
measurements are dominated by the variation of the IGM transmission of LyC at
the various redshifts and sight-lines, we did not attempt to quantify this
additional dust-correction error. Nevertheless, our calculated \fesc\ values
themselves (\S~\ref{sec:fescMC}) are not significantly affected by the \AV\ 
uncertainty, though their quoted $\pm$1$\sigma$ ranges would increase
somewhat. Further details on the adopted \AV\ distributions are given in
Appendix~\ref{sec:extinction}.

\subsection{Estimating the LyC Escape Fraction: Monte Carlo Analysis}
\label{sec:fescMC}

Since we cannot measure the amount of intrinsic LyC radiation
produced by stars within galaxies directly, we must use the best available
stellar population synthesis and IGM absorption models to estimate the
fraction of LyC that escapes from galaxies at high redshift in a statistical
way. We first find the best-fit BC03 model of each galaxy at their spectroscopically verified redshifts using 4--6 band photometric data points taken from the \citet{Skelton2014} photometric catalog. With the BC03 SED and the IGM transmission models of \citet{Inoue2014}, we
simulated the most likely \fesc\ values for the galaxies in our three redshift
bins with a Monte Carlo analysis, treating the various measured and modeled
fluxes and IGM transmission as discrete random variables (RVs), which
incorporates a range of possible values for these measurements. This method
also allows us to model the most likely apparent intrinsic flux from
individual galaxies in a stack, which can vary substantially around the mean
for galaxies at different redshifts observed through different IGM sight-lines
\citep{Inoue2008}. We model all observed stacked fluxes from galaxies using
Gaussian RVs for each pixel value in each sub-image, where the mean of the Gaussian
is the count-rate in the exposure, and the variance is equal to the sum of the
variance of the sky-background and the square of the corresponding pixel value from
the RMS map. These 151$\times$151 Gaussian RVs in each sub-image are combined by
weighted sums in the same way described in \S\ref{sec:method}, and flux is measured
within the UVC aperture for each realization of the stack.

For each galaxy, we take our best fit BC03 SED to obtain our dust-free
(intrinsic) model LyC flux measurement ($F^{\rm int}_{\nu,\!\lyc,\rm i}$), as
denoted in Eq.~\ref{eq:fescabs}, before correcting for IGM absorption. We then
approximate the error of this \emph{model} LyC flux by independently varying
the \emph{observed} continuum data points the SEDs were fit to within their
error bars, and refitting the SED by least squares, \ie\
$F_{\nu}^{\prime}(\nu)$\,=\,$\alpha F_{\nu}(\nu)$, where $\alpha$\,=\,$\mathbf
{F_{\nu}^r}\!\cdot\!\mathbf{F_{\nu}^m}/||\mathbf{F_{\nu}^m}||^2$ and
$\mathbf{F_{\nu,i}^r}$ and $\mathbf{F_{\nu,i}^m}$ are the randomized
\emph{observed} flux measurements and continuum band \emph{model} fluxes,
respectively. This process is equivalent to convolving the \emph{model} LyC
flux with a Gaussian kernel of width equal to the quadratic sum of the
\emph{observed} relative continuum errors. These continuum band \emph{model}
fluxes were calculated by convolving the apparent dust-attenuated SED with the
respective filter transmission curve, similar to Eq.~\ref{eq:fuvci}.

\noindent\begin{figure}[t!] \centerline{
\includegraphics[width=0.53\txw]{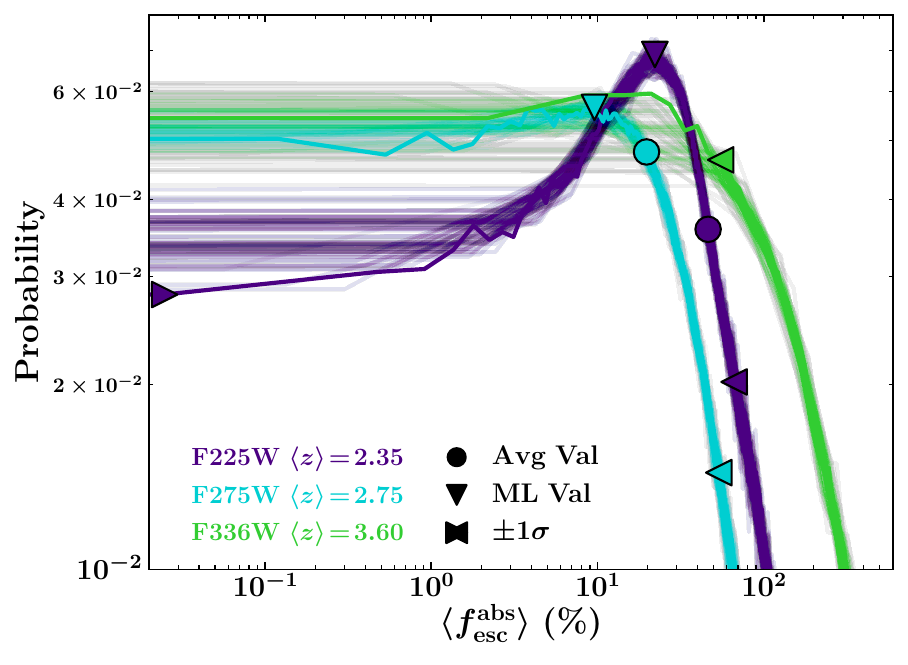}} \caption{\noindent\small
Probability mass functions (PMFs) of the \emph{absolute} \fesc\
values from the MC simulations described in \S\ref{sec:fesc} plotted against their relative probability. This analysis was performed 10$^3$ times using the measured
and modeled intrinsic stacked \emph{apparent} LyC flux and their
$\pm$1$\sigma$ ranges. We apply the IGM attenuation models of
\citealt{Inoue2014} to our modeled LyC fluxes. These \fescabs\ values were
optimally binned according to the Freedman-Diaconis rule (see
\S\ref{sec:method}). Downwards triangles and circles indicate the resulting ML
and average \fesc\ values in each probability distribution function,
respectively, while the left/right facing triangles indicate the $\pm$1$\sigma$
range around the mode.  \label{fig:figure11}} \end{figure}
Using the models of IGM transmission with updated absorber statistics from
\citet{Inoue2014}, which simulates the transmission of photons through the IGM
from 600\AA$<$$\lambda_{rest}$$<$1300\AA\ for $10^4$ lines-of-sight, we apply
the IGM attenuation to our \emph{model} LyC flux by convolving the SED with
the wavelength dependent IGM transmission coefficient curve at the redshift of
the galaxy for all $10^4$ simulated lines-of-sight. We then convolve the IGM
attenuated model flux with the model error that we calculated to obtain our
final model LyC flux (\ie\ $F^{\rm int}_{\nu,\!\lyc,\rm i}$). We then stacked
all of the IGM attenuated \emph{model} LyC fluxes of all galaxies in their
respective redshift bins to obtain our stacked \emph{model} LyC flux as
denoted in Eq.~\ref{eq:fescabs}. The stacked \emph{model} LyC flux was then
used to calculate \fescabs , and the dust attenuated \fescabs, as
shown in Eq.~\ref{eq:fescabssimp}. Since we performed a
non-correlated sum of the model LyC flux RVs to estimate this intrinsic
stacked LyC flux, we run this \fesc\ calculation for a total of $10^3$ trials,
which we combine in order to generate a statistically significant sample of
possible \fesc\ values. The probability mass function (PMF) of \fesc\ was then
calculated by optimally binning these \fesc\ samples according to the
Freedman-Diaconis rule (see \S\ref{sec:sky}), and normalizing by
$N_{samples}$ to give their relative probabilities. The full \fesc\ PMFs are shown in Fig.~\ref{fig:figure11} for
galaxies without AGN. The statistics of the sample array, \ie\ the ML values,
averages, and $\pm$1$\sigma$ error ranges were computed and are shown in
Fig.~\ref{fig:figure11} and Cols. (8) and (9) of Table~\ref{table:table3}.

Since each element in our sample represents a simulated possible
value of \fesc, we take the mode, or the maximum likelihood (ML) value of our
PMF as  the \fesc\ value representative of our galaxies at their average redshift,
as these values correspond to the escape fraction of the total flux from all
galaxies in each stack. The $\pm$1$\sigma$ error bars were computed from the inner 68\% of the PMF at equal probability, and upper limits were computed at (84\%) of the full dataset. The averages and
medians were computed from the full dataset as well. These \emph{model} LyC fluxes also represent
the lines-of-sight, where escaping LyC flux was transmitted through the IGM
before being absorbed by Lyman Limit Systems and Damped Lyman-$\alpha$ systems
within $\Delta$$z$\,$\simeq$\,0.5. The opaque lines-of-sight, where the IGM
transmission peaks near $\mathcal{T}^{\lyc}_{\igm}$\,$\simeq$\,0.01, represent
$\sim$30--40\% of our potential model LyC flux values. These lines-of-sight
result in higher \fesc, as the model LyC would have been attenuated by more
absorbers. However, $\sim$40--50\% of our lines-of-sight have average IGM
transmission values $\mathcal{T}^{\lyc}_{\igm}$\,$\gtrsim$\,0.4 (where the
transmission distribution is at a local minimum), and corresponds to the peak
of the \fesc\ PMFs, where the model LyC flux encountered fewer absorbers. These
lines-of-sight have a local maximum transmission near
$\mathcal{T}^{\lyc}_{\igm}$\,$\simeq$\,0.7, and about $\sim$0.3\% 
of these sight-lines can be as high as $\mathcal{T}^{\lyc}_{\igm}$\,$\simeq$\,0.85.

\subsection{Implications of the \fesc\ MC Results}
\label{sec:fescresults}

We list the results of our \fesc\ MC simulations in Table~\ref{table:table3}.
The average absolute escape fraction, $\langle\fescabs\rangle$, from galaxies
at various redshifts can be used to determine what fraction of LyC produced by
the stellar photospheres in those galaxies escapes, \ie\ is not absorbed by
interstellar neutral \ion{H}{1}, dust, etc., at their average redshift.
However, variations in IGM transmission can cause these values to become
highly uncertain when stacking LyC emission from galaxies over too broad of a 
redshift range. Thus, in order to ascertain any meaningful evolution in \fescabs,
we must stack galaxies at similar redshifts and compare their \fescabs\ values
from sample to sample. Then, any trends in the independent subsamples can be
used to constrain correlations of \fesc\ with galaxy properties or evolution
with redshift. Modeling these properties can also be used to determine their
impact on \fesc, and to see if trends in these properties with redshift can
affect the apparent evolution of \fesc\ with cosmic time. 

The galaxies selected in our \zmean\,$\simeq$\,2.35 and \zmean\,$\simeq$\,2.75
stacks have, on average, younger stellar populations and more dust than the
\zmean\,$\simeq$\,3.60 stack. The \fesc\ value for
galaxies \emph{selected} at \zmean\,$\simeq$\,3.60 are indicative of somewhat older
stellar populations (of $\sim$\,1 Gyr), but are not significantly affected by
the lower amount of dust observed in these galaxies. The
\zmean\,$\simeq$\,2.35 and \zmean\,$\simeq$\,2.75 stacks sample galaxies that
are undergoing a period of more active star-formation compared to the two
higher redshift samples, which may have led to the accumulation of more \ion{H}{1} gas and dust
in these galaxies, but also a brighter \textit{intrinsic} LyC flux. Thus,
these \fescabs\ values also imply that the ISM can absorb a larger fraction
of LyC flux from older stellar populations than from younger ones when comparing
\fescabs\ from older and younger stellar populations. 

Although young stellar populations can produce more intrinsic LyC than older ones, which then has a
higher probability of escaping the ISM, higher extinction from dust in the UVC may correlate to a reduced efficiency of LyC escape. Although dust is the dominant factor for attenuation for $\lambda$$>$912\AA, ionizing radiation is more strongly absorbed by neutral hydrogen due to the higher cross sectional area \citep{Richings2014}. 
LyC escape requires very low neutral Hydrogen column densities ($\mathrm{N_{H}}\!\!<\!\!10^{17}$). Since the amount of extinction from dust is strongly correlated to the column density of Hydrogen \citep{Bohlin1978,Fitzpatrick1999,Rachford2002}, higher extinction may then be indicative of low \fesc. This
apparent correlation of high dust extinction and low \fesc\ is consistent with the
results of several observational and analytical studies that investigate the
impact of various galactic parameters on \fesc\ \citep[e.g.,][]{Mathis1971,
Leitherer1995, Inoue2001, Bergvall2013}. 

\subsection{The Observed Radial Surface Brightness Profiles in UVC and LyC}
\label{sec:obsradprof}

\noindent\begin{figure}[b!]
\centerline{\includegraphics[width=0.48\txw]{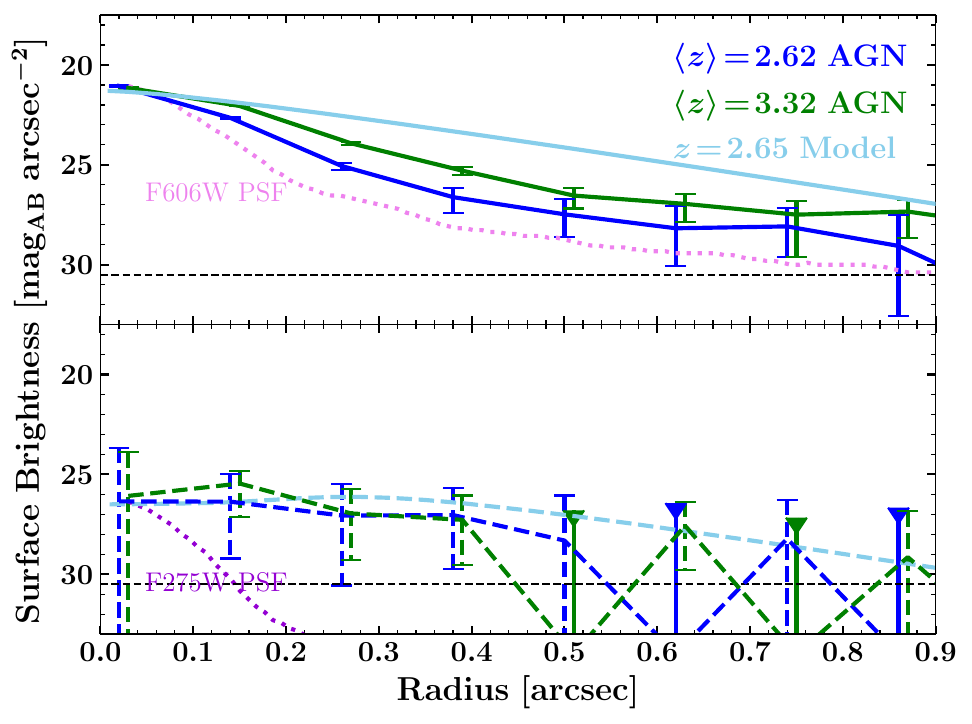}}
\caption{\noindent\small Radial surface brightness profiles of the non-ionizing
UVC signal (\emph{solid curves}) and the LyC signal (\emph{dashed
curves}) measured in the stacks (Fig.~\ref{fig:figure8}) for the galaxies
with AGN samples. The curves are color-coded according to their
mean redshift (filter): \zmean=2.68
(F606W and F275W; blue) and \zmean=3.49 (F775W and F336W; green). The \emph{observed}
PSF in F275W and F606W are indicated by dotted purple and pink curves, which
were normalized to the central SB of the corresponding LyC surface brightness
profiles. The horizontal black dashed line indicates the 1$\sigma$ sensitivity
limit for the LyC profile in F275W. Both UVC surface brightness profiles
are extended with respect to the corresponding PSF curves.  The observed LyC
stack SB profiles are also extended \emph{and} flatter than the UVC profiles,
which is also predicted from our LyC scattering model (\emph{light blue dashed
curve}), where scattering of the escaping LyC photons off electrons and/or dust
with a porous ISM spreads the LyC emission beyond the distribution of the stellar
UVC light (\emph{light blue solid curve}). The \emph{light blue solid curve} is
scaled from \emph{light blue dashed curve} by a single ratio of
$\frac{f_{\uvc}}{f_{\lyc}}$, which may depend on radius. See \S\ref{sec:modelradprof} for further
details of the model. \label{fig:figure12}} \end{figure}
The radial profiles of our LyC and UVC stacks from Fig.~\ref{fig:figure8} for F275W
and F336W are shown in Fig.~\ref{fig:figure12}. We construct all observed radial SB
profiles by summing successive annuli of 3 pixel radii beginning with the
central pixel, where each pixel is treated as a Gaussian RV with the mean set to
the pixel value in the stack and variance set to the sum of the variance from the
square of the pixel value in the corresponding RMS map, and the variance from the
sky-background. This allowed us to estimate uncertainties on a per pixel basis
for generating flux distributions of the sum of several pixels. The averages
and $\pm$1$\sigma$ errors or 1$\sigma$ upper bounds to these distributions are
indicated as vertical bars and downwards triangles, respectively.

The stacked UVC profiles are shown as solid curves, and those for LyC are
dashed. The \emph{observed} PSFs in the WFC3/UVIS F275W and ACS/WFC F606W
mosaics are indicated by dotted curves, normalized to the central surface
brightness of the corresponding LyC SB profile. The PSF in F336W is very
similar to the F275W PSF, so we do not plot it. These are available in Table~\ref{table:table1}
and Fig.~7\emph{b} of \citetalias{Windhorst2011}. Note that these PSFs measured
in the 0$\farcs$03 mosaics are undersampled. The 1\,$\sigma$ SB sensitivity
limit for the LyC profile in the F275W stack is indicated by a horizontal
dashed line at $\mu_{\rm AB}\simeq$30.5 \magarc. For the sensitivity limits of
the samples in the other filters, we refer the reader to Cols. 9 and 10 of
Table~\ref{table:tableA1} and to the discussion of systematics in the stacked
data in Appendices~\ref{sec:stacktests}-- \ref{sec:systematics}. These SB
sensitivity limits are consistent with the 1$\sigma$ sky-subtraction errors discussed in
\S\ref{sec:datareduction}.

Both UVC SB profiles are clearly extended with respect to their
corresponding filter PSFs, as expected for stacked galaxy radial light
profiles at $z$\,$\simeq$\,3--6 \citep[e.g.,][]{Hathi2008}. The much deeper
HUDF UVC stacks of \citet{Hathi2008} suggested a possible ``break'' (or slight
change in slope) near $r$\,$\gtrsim$0$\farcs$3--0$\farcs$4, from exponential
in the inner parts to a somewhat less steep profile in the outskirts. Our
stacked UVC light profiles do not clearly show a change in slope at
$r$\,$\gtrsim$ 0$\farcs$3--0$\farcs$4, since our (77--180 orbit) UVC stacks
are not nearly as deep as their $\sim$1680--4300 orbit stacks, and because of
our much more stringent method of masking neighbors.

Both LyC SB profiles are also clearly extended with respect to their
\emph{observed} PSFs, and remain extended to $r$\,$\simeq$\,0$\farcs$5,
beyond which errors in the sky-subtraction start to become substantial. The
very faint, flat, non-centrally concentrated appearance of the combined LyC
signal makes the extraction of its SB profile uncertain at larger radii. The
relatively flat LyC SB profiles may indicate a more complicated LyC escape
scenario, in which the light distribution of the LyC flux of a stack of
galaxies is largely dependent on the porosity of the ISM in those galaxies,
and/or the scattering processes that the LyC photons undergo before escape. 
We find that the UVC SB profiles are well fit to S\'ersic profiles of index of $n\!\simeq\!2.4\pm0.7$, while the LyC SB profiles could not converge to a S\'ersic fit but are better fit to straight lines with slope $\sim\!2.5\pm0.6$\,mag\,arcsec$^{-2}$ per arcsec. The difference in linear slope between UVC and LyC is $\sim$6\,mag\,arcsec$^{-2}$ per arcsec with a formal SNR of $\sim$2.8, so the LyC is therefore likely flatter. This may also be indicative of a decreasing LyC opacity with radius, as the LyC and UVC escape morphology differs radially.

After integrating these SB profiles as elliptical frustums between each
isophote, we find reasonable agreement with our photometric analysis (see
Table~\ref{table:table2}), although the flux represented by the radial SB
profiles is consistently fainter by $\sim$0.3$\pm$0.2 mag. This discrepancy is
expected, given that our SB profiles do not extend out to the larger
aperture sizes used in the photometry of the stacks, and therefore miss some
real LyC flux that might be present at larger radii and at very faint SB levels.

\subsection{Modeling the UVC and LyC Radial Surface Brightness Profiles}
\label{sec:modelradprof}

For the highest SNR measurements in the LyC stacks (\ie\ F275W/F336W), the
radial SB profile of escaping LyC flux appears to be flatter than the
corresponding \textit{non-ionizing} UVC profile (the dashed and solid colored
curves in Fig.~\ref{fig:figure12}, respectively).

A LyC SB profile that is measurably flatter than the corresponding UVC SB
profile could arise naturally in a porous ISM, in which the covering factor of
neutral gas decreases with increasing galacto-centric distance. To illustrate
this quantitatively, we consider the transfer of UVC and LyC photons
through simplified models of galaxies with a multiphase ISM.

To calculate this, we assume that the UVC sources are spatially extended and
characterized by a volume emissivity $\epsilon_{\uvc}(r)$. We assume an
exponential distribution with galactic radius:
$\epsilon_{\uvc}(r)$\,=\,$\epsilon_{\uvc,0}\,\exp(-r/r_0)$. The normalization
constant $\epsilon_{\uvc,0}$ and scale length of $r_0$ are obtained by
matching the observed SB profiles in Fig.~\ref{fig:figure12}. We further
assume that LyC emission traces the UVC emission. We attribute differences in
observed SB profiles to the fact that neutral clumps of gas are opaque to LyC
radiation, but not to UVC.

We also assume a (spherical) distribution of neutral gas clumps, which is
described completely by its covering factor, $f_{cov}(r)$\,$\equiv$\,$n_{\rm
c}(r)A_{\rm c}(r)$. Here, $n_{\rm c}$ and $A_{\rm c}(r)$ denote the number of
clumps and area of a clump at $r$, respectively. The covering factor $f_{\rm
cov}$ then denotes the probability that a sight-line intersects a clump per
unit length. For example, for clumps of fixed size that are outflowing at
\emph{an assumed constant} velocity $v$, we have a number density dependence
as $f_{cov}$\,$\propto$\,$n_{\rm c} \propto r^{-2}$ (we refer the reader to
\citet{Dijkstra2012} for a more detailed description of this covering factor).

The precise radial dependence of $f_{cov}$ is not known. However, when
$f_{cov}$ decreases with $r$ we generally expect increased LyC escape
fractions at larger galacto-centric distances. We consider two parameter
models for $f_{cov}=Ar^{-x}$, and fit for $A$ and $x$. Both $A$ and $v$ can
also depend on radius. Hence, $f_{cov}$ generally is some unknown power law of
$r$ (\ie\ r$^{-x}$), where $x$ typically ranges between 0 and 3. This
calculation shows that when sight-lines with low impact parameter see the
largest $f_{cov}$, we see a reduced escape fraction in these directions.

An interesting possibility is that the neutral gas clouds can theoretically
\emph{scatter} LyC photons: LyC photons penetrate the neutral clumps over an
average distance that corresponds to $\tau\simeq$1. Direct recombination to
the ground state produces LyC photons that can escape from the neutral cloud,
as the optical depth to the edge of cloud is $\tau$\,$\simeq$\,1. This
``scattering'' (absorption and re-emission of LyC photons occurs on the
recombination time scale inside the cloud) of LyC photons could further
flatten the predicted surface brightness profile. The possible effects of LyC
scattering can be expanded to include scattering off free electrons and dust
grains (which also differs between LyC and UVC).

While the dot-dashed curve in Fig.~\ref{fig:figure12} is only a single example
(matching our F275W LyC observations at \zmean$\simeq$\,2.62) of these model
LyC SB profiles, model predictions with similar parameter values fit the SB
profiles in the other redshift bin. With these models, we can integrate out
to larger impact parameters and get a constraint on the total escape fraction
that accounts for the difference in SB profiles. This procedure gives a
slightly larger value for \fesc\ than those given in Table~\ref{table:table3}
(This is, of course, almost identical to the constraints one would obtain
simply by extrapolating the surface brightness profiles to larger impact
parameters, and integrating over them.) 

\noindent\begin{figure}[b!] \centerline{
\includegraphics[width=0.48\txw]{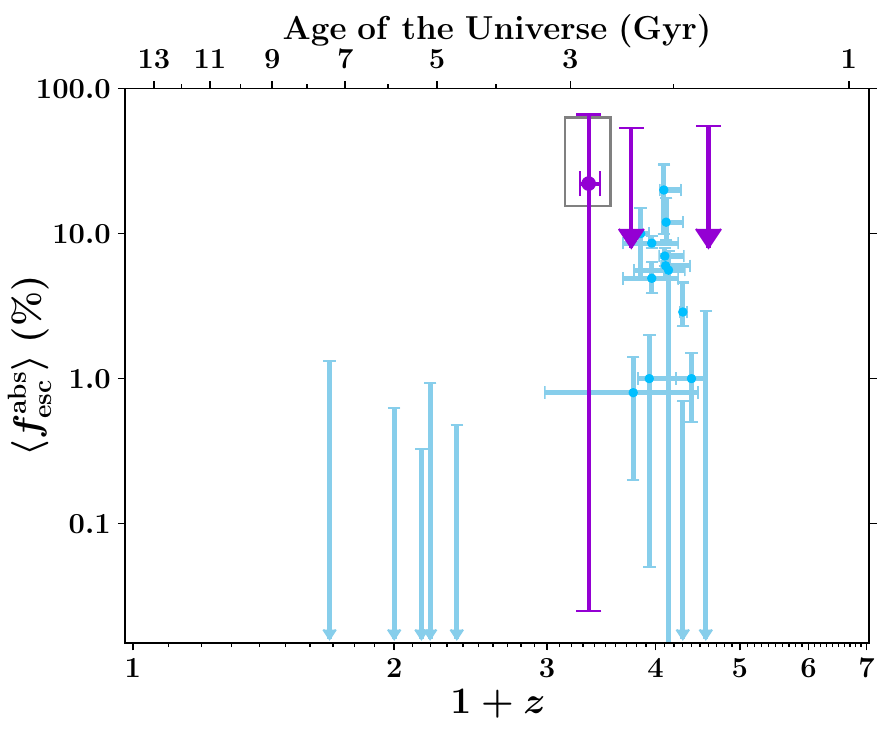} } \caption{\noindent\small
The \emph{absolute} LyC escape fraction for various galaxy samples as a
function of redshift. Plotted is our ML average \fescabs\ value with
their $\pm$1$\sigma$ range and 1$\sigma$ upper limits for our
galaxies \emph{without} AGN sample (purple filled circled and triangles)
taken from the probability mass functions of Fig.~\ref{fig:figure11}, generated
from our MC simulations described in \S\ref{sec:fesc}. The interquartile range of the \zmean=2.35 data is indicated by a box to highlight the high asymmetry of it's PMF. The blue points indicate
available published data as referenced in \S\ref{sec:intro}, some of which were
converted from quoted \fescrel\ values using extinction values from the literature
source (see \S\ref{sec:fescintro}, Eq.~\ref{eq:fescrelsimp}). All vertical error
bars are the $\pm$1$\sigma$ uncertainty on the \fesc\ values. Some errors were
converted from the quoted 2--3$\sigma$ uncertainties. Upper limits
are shown as blue downward triangles. Although the blue points represent galaxy
samples with different properties from our samples, and the quoted errors were
derived from uncertainties with different error assessment, the combined data
suggests a correlation of \fescabs\ with redshift, which may not be a simple power
law in (1+$z$). This compiled dataset does not rule out the possibility
that massive galaxies may have had high enough LyC \fesc\ values to complete
hydrogen reionization by $z$\,$\sim$\,6, if galaxies at $2\!\lesssim\! z\!\lesssim\!4$
and $z\!\gtrsim\!6$ are analogous. \label{fig:figure13}} \end{figure}     

\section{Discussion of Results}
\label{sec:discussion}
\subsection{Summary of Available Data on \fesc\ vs. Redshift}
\label{sec:fescvsz}

The constraints we place on \fesc\ are valid for the luminosity
range $M_{\rm AB}^{\uvc}$\,$\simeq$\,$-$21.1$^{+0.9}_{-0.5}$\,mag present in
the sample which was selected to have reliable spectroscopic redshifts (see
Fig.~\ref{fig:figure4}), with the two lowest redshift subsamples being dominated
 by relatively younger stellar populations with active star-formation
and significant dust extinction, the \zmean\,=\,3.60 subsample comprises mostly
lower extinction galaxies with somewhat older stellar populations. For faint
galaxies to have finished reionization by $z$\,$\simeq$\,6--7, their \fesc\
values need to be $\gtrsim$20\% \citep{Ouchi2009, Wilkins2011, Kuhlen2012}
beyond $z$\,$\simeq$\,6 \emph{and} their luminosities need to reach as faint as
$M_{\uvc}$\,$\simeq$\,--15 to --13\,mag \citep[e.g.,][]{Robertson2013}. Hence,
if faint (dwarf) galaxies contributed significantly to reionization at
$z$\,$\lesssim$\,6--7, one should consider how much their \fesc\ fraction
might have increased \emph{both} towards higher redshifts due to the expected
lower metallicities and lower dust extinction, \emph{and} at fainter
luminosities due to the larger impact that SN driven outflows have on lower 
mass dwarf galaxies \citep[e.g.,][]{Ricotti2000, Razoumov2007, Wise2009,
Fernandez2011}. Given that our spectroscopic selection samples luminous
galaxies in all three redshift bins, our LyC detections can only constrain
the first possibility, which we discuss here.

In Fig.~\ref{fig:figure13}, we plot our ML and 1$\sigma$
upper bound \fescabs\ values generated from the MC simulation listed in
Table~\ref{table:table3} for galaxies without AGN activity (purple
filled circles and triangles, respectively). We show the interquartile range of the \zmean=2.35 \fesc\ data to emphasize it's highly asymmetric PMF, which has more data below the ML point. We also plot similarly derived,
\fescabs\ data available in published work summarized in \S\ref{sec:intro} as light
blue points, with upper limits indicated as blue triangles. The light blue \fesc\ points
indicate galaxies with restframe 1500\AA\ luminosities close to those sampled
in Fig. 5a--5c (\ie\ $\langle M_{AB}\rangle$ $\simeq$--21.1$^{+0.9}_{-0.5}$ mag). The
dependence of the \fesc\ values of galaxies on luminosity is not well determined,
but no clear dependence on luminosity is obvious in Fig.~\ref{fig:figure13}. We
will therefore discuss the redshift dependence of \fesc\ here for the
luminosities sampled in Fig.~\ref{fig:figure4}.

We first converted the published \fescrel\ values to \fescabs\ when necessary
using the quoted extinction values from the literature source. We note that
these \fesc\ values were derived from different observational analyses,
including both space and ground based spectra and imaging, with different
object selection, reduction techniques, error assessment, and application of
IGM attenuation models. We plot only the quoted \fesc\ values from the literature source most analogous to this study, i.e. those derived from their full, stacked sample. Some of the published errors may not account for the
same uncertainties that we address in \S\ref{sec:fescMC}. When necessary, we
converted the quoted published uncertainties to 1$\sigma$ error bars, so they
are comparable to our results in Table~\ref{table:table3}.

Although the \fesc\ values plotted in Fig.~\ref{fig:figure13} were derived
with different methods, the current data appears to suggest a correlation of
\fesc\ with redshift. However, any such relation may not be a simple power
law in (1+$z$). Several authors \citep{Inoue2006, Razoumov2010, Finlator2012,
Kuhlen2012, Becker2013, Dijkstra2014} have suggested that redshift averaged
\fesc\ values for galaxies may increase significantly with redshift, possibly
as steeply as $\propto$(1+$z$)$^3$\,--\,(1+$z$)$^5$. This only holds only for
$z$\,$\lesssim$\,7, beyond which the implied escape fraction would approach 100\%
for the upper bound, but decrease monotonically at lower redshift
\citep[e.g.,][]{Razoumov2010}. If the (1+$z$)$^{\kappa}$ exponent values were
as steep as $\kappa\simeq$2.0, this prediction would provide \fesc\ values
at $z$\,$\gtrsim$\,6 in excess of $\sim$30\%, as required for hydrogen
reionization to have completed by $z$\,$\sim$\,6 \citep{Robertson2013}.
However, none of the simple (1+$z$)$^{\kappa}$ power laws for \fesc\ seem to be consistent with the data points in Fig.~\ref{fig:figure13}
to within their stated 1$\sigma$ errors.

\subsection{A Redshift Dependence Faster than (1+$z$)$^{\kappa}$?}
\label{sec:scurve}

Since the plotted 21 independent data points in Fig.~\ref{fig:figure13}
deviate from published power laws, \emph{no} single (1+$z$)$^{\kappa}$
curve seems to fit all the \fescabs\ data for galaxies without AGN. We
therefore suggest the possibility that \emph{a more sudden decrease} of
\fescabs\ with redshift may instead have to be considered. The
combined data in Fig.~\ref{fig:figure13} suggests, however, that \fescabs\ may
have declined by a factor of nearly $\sim$10 from $\gtrsim$20\% at
$z$\,$\gtrsim$\,2 to $\sim$1\% at $z$\,$\lesssim$\,2. These low \fesc\ values
at $z$\,$\lesssim$\,2 are predicted by some cosmological radiative transfer
models as well, which  also require a ``steep rise'' in \fesc\ at
$z$\,$\gtrsim$\,2 for massive galaxies to reionize the Universe
\citep[e.g.,][]{Khaire2015}, and have also been suggested in studies of the
\Lya\ escape fraction over redshift \citep[e.g.,][]{Blanc2011}.

Fig.~\ref{fig:figure13} indicates that the sudden decrease
in \fescabs\ may have occurred within the epoch of $z$$\sim$2, or
within about $\pm$1\,Gyr of the observed peak in the cosmic star-formation
history (SFH) \citep{Madau1996, FaucherGiguere2008, Cucciati2012,
Burgarella2013}. This period may indicate the epoch where the universe
transitions from infall/merger driven star-forming galaxies at
2$\lesssim$\,$z$\,$\lesssim$6 to a more passively evolving universe 
by giant galaxies at $z$\,$\lesssim$\,1--2 \citep{Driver1998}. This transition
may have resulted in dust and gas rapidly accumulating in the disks and
central bulges of forming galaxies, with a SN rate that has progressively less
impact on clearing gas and dust from the galaxies that are steadily growing in
mass with cosmic time. It is possible that this process may have
caused \fescabs\ to rapidly drop over a relatively narrow interval of cosmic
time in luminous galaxies, as massive LyC producing stars formed during the
period of high SFR become either SNe II or AGB stars, which then enrich the
ISM with dust within $\sim$1\,Gyr \citep{Mathis1990,Bekki2015}. The infall of hydrogen in
these galaxies could have then caused \fescabs\ to decrease substantially \citep{Rauch2011,vandeVoort2012}. 
The subsequent increase of dust can then prevent the collapse of cold gas by photoelectric heating from
stars or AGN in the galaxy \citep{Krumholz2012,Forbes2016}. This would then lead to a
decrease in the galaxy's SFR, as feedback from heating inhibits the formation
of new massive stars \citep[e.g.,][]{Inoue2001,Inoue2001a}. The decline in SFR
would also lead to a decreasing SN rate \citep{Botticella2012}, further
preventing the escape of LyC, as there would have been fewer clear channels
produced by SN for the LyC to escape. LyC produced by AGN can be absorbed by
gas and dust in the disk of the galaxy itself, depending on viewing angle.
When galaxies produce stronger AGN outflows, more of their LyC radiation may
escape approximately \emph{perpendicular to} the galactic disk
\citep[e.g.][]{Windhorst1998,Reunanen2003}, which contributes to maintaining
the ionized state of the IGM, as AGN begin to dominate the ionizing background
at $z$\,$\lesssim$\,3.

\subsection{The Role of Galaxies with Weak AGN in Reionization}
\label{sec:AGN}

Fig~\ref{fig:figure8} shows the stacked LyC and UVC images of the known
galaxies with AGN in our sample. The \zmean=2.374 stack only includes two
AGN with a LyC flux of \mAB\,$>$\,27.91 mag (UVC aperture). The
\zmean\,$\simeq$\,2.61 and 3.32 samples contain 7 and 3 stacked
AGN with measured LyC fluxes of \mAB\,$\simeq$\,28.3 and 27.42 mag with
SNR$\sim$2.7 and 2.5, respectively. These fluxes are typically more
luminous in LyC and have higher SNR than their non-AGN counterparts, despite
having fewer stacked galaxies. This is most likely due to LyC originating from
the central accretion disk, made visible by stronger AGN outflows when viewed
under the right angle. AGN outflows can also increase the porosity of the ISM
in its host galaxy \citep[e.g.,][]{Silk2005}, thereby increasing \fesc\ of the
LyC produced by stars, which further contributes to the total measured LyC flux
from that galaxy.

The stacks in Fig~\ref{fig:figure6} suggest some variety of LyC morphologies,
though the UVC images exhibit more compact light profiles compared to the non-AGN
stacks in all three cases (see the discussion in
\S\ref{sec:modelradprof}--\ref{sec:obsradprof}). The \zmean\,$\simeq$\,2.62
stack is the most extended of the AGN both in LyC and UVC, which is most
likely due to the increased sensitivity to fainter flux at low redshift, with
a central bright point source from radiation escaping along the observed
line-of-sight. The radial dependence of the LyC SB profile for this stack may be
due to the viewing angle of the AGN relative to the direction of the escaping
LyC radiation, or possibly due to the fact that the LyC undergoes a more
complex escape process, where photons can be reflected off of relativistic
electrons in the AGN corona and accretion disk, or by hot dust in the torus
via Thomson and/or inverse Compton scattering \citep[e.g.,][]{Haardt1993}. The
\zmean\,$\simeq$\,3.32 AGN LyC stack appears to be more point-like, indicating
that these observed LyC photons may be escaping predominantly along the
line-of-sight, which is supported by the presence of broad emission lines in their
spectra, although the more extended LyC emission may not be visible
due to the average SB of these AGN at higher redshift being dimmed by an
additional $\sim$61\% from \zmean\,$\simeq$\,2.62 to \zmean\,$\simeq$\,3.32.

Fig.~\ref{fig:figure4} shows that the \emph{average} UVC luminosities of
``Galaxies with weak AGN'' in our sample is about the same, or somewhat
fainter than that of galaxies without AGN. Their average luminosity in
Fig.~\ref{fig:figure4} \MAB\,$\simeq$\,--20.4$\pm$0.9\,mag at
$z$\,$\simeq$\,2.3--4.1 does not indicate clearly QSO dominated luminosities or
SEDs. Table~\ref{table:table2} shows that the LyC flux measured from the
stacks of (weak) AGN at all redshifts is typically \emph{brighter} than
galaxies without AGN. Thus, precise modeling of the intrinsic LyC emission
must include the contribution of flux emitted by, or reprocessed from, the
AGN accretion disk. The SED of the AGN accretion disk may be more complicated
than a simple blackbody curve, as the SED must account for the broad and
narrow emission line regions, as well as energy lost to relativistic jets
and photons scattered/absorbed by the corona and central torus and non-AGN
dust, which is also viewing angle dependent. Because we cannot fit both
stellar+AGN SED models to the 4--6 band continuum data currently available
for galaxies with weak AGN, we do not calculate escape fractions for these
galaxies. Since the SEDs of these galaxies are likely dominated by stellar light at
the non-ionizing wavelengths, their \fesc($z$) correlation may be similar
to the trend seen from the escape fractions of galaxies without AGN.
From the compact appearance in some of our stacked images --- and from the
fact that they are on average brighter than galaxies without AGN --- the
LyC flux in galaxies with weak AGN may be dominated by light originating from their
accretion disks.

Further data and modeling is needed to better constrain \fescabs($z$) for both
galaxies and weak AGN to confirm these observed trends. The current data for
AGN may be consistent with a more modest drop in \fescabs($z$) than for
galaxies that may have occurred close to the peak in the epoch of AGN activity
around $z$\,$\simeq$\,2.5 (\eg\ \citealt{Fontanot2007, Croom2009, Ikeda2011},
\citeyear{Ikeda2012}). Since AGN activity can affect the SFRs, it is possible
that when AGN outflows started to ramp up after the peak in the cosmic
star-formation history at $z$\,$\simeq$\,2 \citep{Springel2005, Hopkins2006}, their
outflows cleared enough paths in the host galaxy ISM to increase \fescabs\ of
a possibly AGN induced top-heavy stellar population IMF.

Because galaxies far outnumber AGN, and despite being fainter in LyC on average,
their \fescabs\ values suggest that galaxies may have produced sufficient LyC
radiation to maintain reionization at $z$\,$\gtrsim$\,3, while AGN likely
dominated in the production of ionizing LyC flux at $z$\,$\lesssim$\,2--3. Even though our spectroscopically selected sample of galaxies outnumber the \emph{weak} AGN by a factor of $\sim$3 (see col. 4 of Table~\ref{table:table2}), the total ionizing flux from AGN is brighter than that from galaxies \emph{without} AGN by $\sim$7.7$\times$12/34$\sim$2.7.

The current samples are still very small, and clearly need further
confirmation through much larger samples, both through deep UV/optical imaging
of wider HST fields and through spectroscopy on fields with high quality
existing HST data. Further theoretical work is needed to outline exactly how
quickly \fesc\ may have increased towards higher redshifts \emph{and} at
fainter luminosities, as well as at lower metallicities and lower extinction
at higher redshifts, while producing enough escaping LyC photons from faint
galaxies to finish and maintain reionization at $z$\,$\lesssim$\,6--7.

There is already a significant issue in accounting for reionization with the
faint galaxy population observed via cluster lensing at $z$\,$\simeq$\,9. At
redshifts larger than 8, the Hubble Frontier Fields reveal a strong drop in
rest-frame UV luminosity density \citep[e.g.,][]{Ishigaki2015}. Hence, it is also
possible that one may need to consider an additional source of reionizing
photons beyond $z$\,$\simeq$\,6--7. This source might include feedback on both
the IGM ionization and clumpiness via hard ionizing photons from high mass
X-ray binaries \citep[e.g.,][]{Mirabel2011}. Other astrophysical sources such
as \ion{Population}{3} stars or mini AGN seem strongly constrained via
chemical evolution \citep{Kulkarni2014} and the X-ray background
\citep{Dijkstra2004}. It is possible that \fesc\ may evolve with redshift
and/or with galaxy properties (\eg\ mass, \AV, SFR, and/or age). 

\section{Conclusions} \label{sec:conclusion}

We studied LyC emission that may be escaping from galaxies using improved
\HST\ WFC3 of the ERS fields in three filters, where LyC may be
observed from galaxies at $z$\,$\simeq$\,2.3--4.1. The data that we used in
our analysis was drizzled with the much more accurate 2013 WFC3 geometric
distortion correction tables, which resulted in the correction of
significant astrometric offsets that remained in earlier ERS UVIS mosaics.
The WFC3 ERS UV images were taken in 2009 September, when the CTE was still at
a level where faint flux could still be measured without significant losses.
We verified that any loss in CTE is not the primary limitation to our
measurements (see Appendix~\ref{sec:CTE}).

We extracted sub-images centered on galaxies with high quality
spectroscopically measured redshifts from the ERS mosaics, and averaged the
LyC flux of those galaxies. We payed careful attention to the removal of
potentially nearby contaminating objects and low level variations in the UV
sky-background during this stacking process. We ensured that no significant
amount of contaminating flux longwards of the Lyman-break
($\lambda$\,$>$\,912\AA) was included in our stacks. We performed a series of
critical tests to ensure our stacking procedure was not affected by various
systematics in the mosaic images. All of these tests are described in
Appendices~\ref{sec:stacktests}-- \ref{sec:systematics}. The following are our
main findings:

\smallskip\noindent (1) Our measurements of the average LyC flux in the stacks for galaxies at $z$\,$\simeq$\,2.3--4.1 is summarized in
Table~\ref{table:table2}. We find that the LyC flux of faint galaxies at
\zmean\,$\simeq$\,2.35, 2.69, and 3.54 is generally constrained at the
$<$1--3$\sigma$ level, in typical image stacks of 13--19 objects in the
WFC3/UVIS F225W, F275W, and F336W filters, respectively. These upper limits
corresponds to total LyC fluxes of \mAB\,$\gtrsim$\,28.1--29.0 mag. The LyC flux
of weak AGN is detected to be brighter on average at $z$\,$\simeq$\,2.3--3.5,
but over $\sim$2--10$\times$ fewer objects per stack.

\smallskip\noindent (2) The combined LyC emission averaged over the three
filters suggests an overall LyC flux distribution that is non-centrally
concentrated, which may be explained by a radial dependence in the ISM
porosity and/or scattering of the LyC photons. We find that the LyC flux from
AGN is flatter than its UVC counterpart. This may
suggest a complex escape process that may be determined by the distribution
and extent of neutral (dusty) gas clouds within a porous multiphase ISM.

\smallskip\noindent (3) From our best fit BC03 SED models fit to \HST\
continuum observations longwards of \Lya, the observed LyC flux corresponds to
an \emph{average} absolute LyC escape fraction constrained to
\fescabs\,$\sim$22$^{+44}_{-27}$\% at \zmean$\simeq$2.4 and $\lesssim$55\% at
\zmean$\simeq$2.8--3.6. While the error bars on the implied \fesc\ values in
each of the three redshift bins remain large, within the error bars,
the data suggest an increasing trend of \fesc\ with redshift at $z$\,$\gtrsim$2.

\smallskip\noindent (4) The available published \fesc\ data for galaxies
may suggest \emph{a more sudden increase} in \fescabs\ with redshift
that occurred around $z$\,$\sim$2. For galaxies,
the steepest drop in \fesc\ occurs at $z$\,$\lesssim$\,2, near the peak of
the cosmic star-formation history within an interval of $\pm$1\,Gyr from
this peak in cosmic time. 

\smallskip\noindent (5) If galaxies \emph{without} AGN at $z$$\sim$2--4 are analogous to those at $z$$\gtrsim$6, the upper limits to their \fescabs\
values suggest that they may have had a sufficient LyC escape fraction to reionize the IGM by
$z$\,$\gtrsim$\,6. The SEDs of galaxies with weak AGN is likely dominated by 
\emph{stellar light} in the non-ionizing continuum. Galaxies with weak AGN outshine galaxies \emph{without} AGN in our sample by a
factor of $\sim$7.7, or \mAB$\sim$2.3\,mag. Hence, while galaxies without AGN likely began and
maintained cosmic reionization at $z$\,$\gtrsim$\,3, galaxies with (weak) AGN
likely dominated the contribution to the cosmic ionizing background and
maintain reionization at $z$\,$\lesssim$2--3, although the role of massive
galaxies without AGN may not have been negligible at $z$\,$\lesssim$2.

The transition from galaxy dominated reionization to weak AGN reionization appears
to have occurred at $z$\,$\sim$\,2--3, \ie\ right around the peak in
the cosmic SFR \citep{Madau1996}, which may indicate the epoch where the
universe transitions from infall/merger driven SFGs at
2\,$\lesssim$\,$z$\,$\lesssim$\,6 to a more passively evolving universe
 by giant galaxies at $z$\,$\lesssim$\,1--2. This may result in gas
and dust rapidly accumulating in the disks and nuclei of forming galaxies,
combined with a SN rate that has progressively less impact on clearing
gas/dust in galaxies that are steadily growing in mass with cosmic time. The
accumulating \ion{H}{1} gas and decreasing SFR may have caused \fescabs\ to
\emph{rapidly drop} over a relatively narrow interval of cosmic time
($\sim$1.5\,Gyr), as the LyC flux heats the dust and inhibits the formation of
new massive stars. When AGN outflows began to increase after the peak in the
cosmic star-formation history at $z$$\sim$2, their outflows may have cleared
enough paths in the ISM of host galaxies to enhance the fraction of escaping
LyC radiation produced by massive stars and from the accretion disk, resulting
in AGN beginning to dominate the ionizing background at $z$\,$\lesssim$\,2.

\smallskip\noindent (6) Further data on LyC \fesc\ are essential for both
galaxies and weak AGN to confirm both their trends in \fescabs($z$). The
current samples are still very small, and clearly need further confirmation
through much larger samples, both through deep imaging of wider HST fields in
the UV and through deeper spectroscopy on fields with high quality existing
HST data, \eg\ with the \JWST\ FGS/NIRISS grisms and with NIRSpec
\citep{Gardner2006}. Further theoretical work is needed to outline exactly how
quickly \fesc\ may have increased towards higher redshifts \emph{and} at
fainter luminosities, as well as at lower metallicities and lower dust
extinction at higher redshifts, while producing enough escaping LyC photons
from faint galaxies to complete and maintain reionization.

\section*{Acknowledgments}

This paper is based on Early Release Science observations made by the WFC3
Scientific Oversight Committee. We are grateful to the Director of the Space
Telescope Science Institute, Dr. Matt Mountain, for generously awarding
Director's Discretionary time for this program. Finally, we are deeply
indebted to the crew of STS-125 for refurbishing and repairing \HST. We thank
Drs. George Becker, Renyue Cen, Nimish Hathi, Anne Jaskot, Karen Olson,
Michael Rutkowski, Mr. Jacob Vehonsky and the referee for their useful comments and
suggestions for this work. Support for \HST\ programs GO-11359, AR-13877, and AR-14591
was provided by NASA through grants from the STScI, which is operated by the
Association of Universities for Research Inc., under NASA contract NAS
5-26555.


\bibliography{ms}


\include{app}

\end{document}

%% file: app.tex

\newcommand{\AVmed}   {\ensuremath{A^{med}_{V}}}
\newcommand{\AFUV}   {\ensuremath{A_{FUV}}}
\newcommand\EBminV   {\ensuremath{E_{B-V}}}
\newcommand{\ripf}   {\ensuremath{r_{\rm ip-f}}}
\newcommand{\rph}    {\ensuremath{r_{\rm ph}}}
\newcommand{\Vst}    {\ensuremath{m^{\rm Star}_{\rm V}}}
\newcommand{\IphV}   {\ensuremath{I^{\rm ph}_{\rm V}}}
\newcommand{\IphNUV}  {\ensuremath{I^{\rm ph}_{\rm NUV}}}
\newcommand{\SBstV}   {\ensuremath{SB^{\rm Star}_{\rm V}}}
\newcommand{\SBstNUV}  {\ensuremath{SB^{\rm Star}_{\rm NUV}}}
\newcommand{\SBzodiV}  {\ensuremath{SB^{\rm Zodi}_{\rm V}}}
\newcommand{\SBzodiNUV} {\ensuremath{SB^{\rm Zodi}_{\rm NUV}}}
\newcommand{\SBskyNUV} {\ensuremath{SB^{\rm Sky}_{\rm NUV}}}
\newcommand{\ZPv}    {\ensuremath{ZP_{\rm V}}}
\newcommand{\ZPnuv}   {\ensuremath{ZP_{\rm NUV}}}
\newcommand{\IstV}   {\ensuremath{I^{\rm Star}_{\rm V}}}
\newcommand{\IzodiV}  {\ensuremath{I^{\rm Zodi}_{\rm V}}}
\newcommand{\IstNUV}  {\ensuremath{I^{\rm Star}_{\rm NUV}}}
\newcommand{\IzodiNUV} {\ensuremath{I^{\rm Zodi}_{\rm NUV}}}

\appendix
\twocolumngrid

\section{Critical Testing of our Stacking Procedure and LyC Measurements}
\label{sec:stacktests}

We performed various tests on our data to assess the robustness of our LyC
stacking method and detections (\S\ref{sec:stacking}). We test the impact of
several sources of possible systematic effects or spurious signal and evaluate
the overall level of confidence in our quoted uncertainties on the resulting 
measured LyC signal.
\subsection{Detection Tests of the LyC Stacking and Measurements }
\label{sec:dettests}

\begin{figure*}[t]
\centerline{\includegraphics[width=0.68\txw]{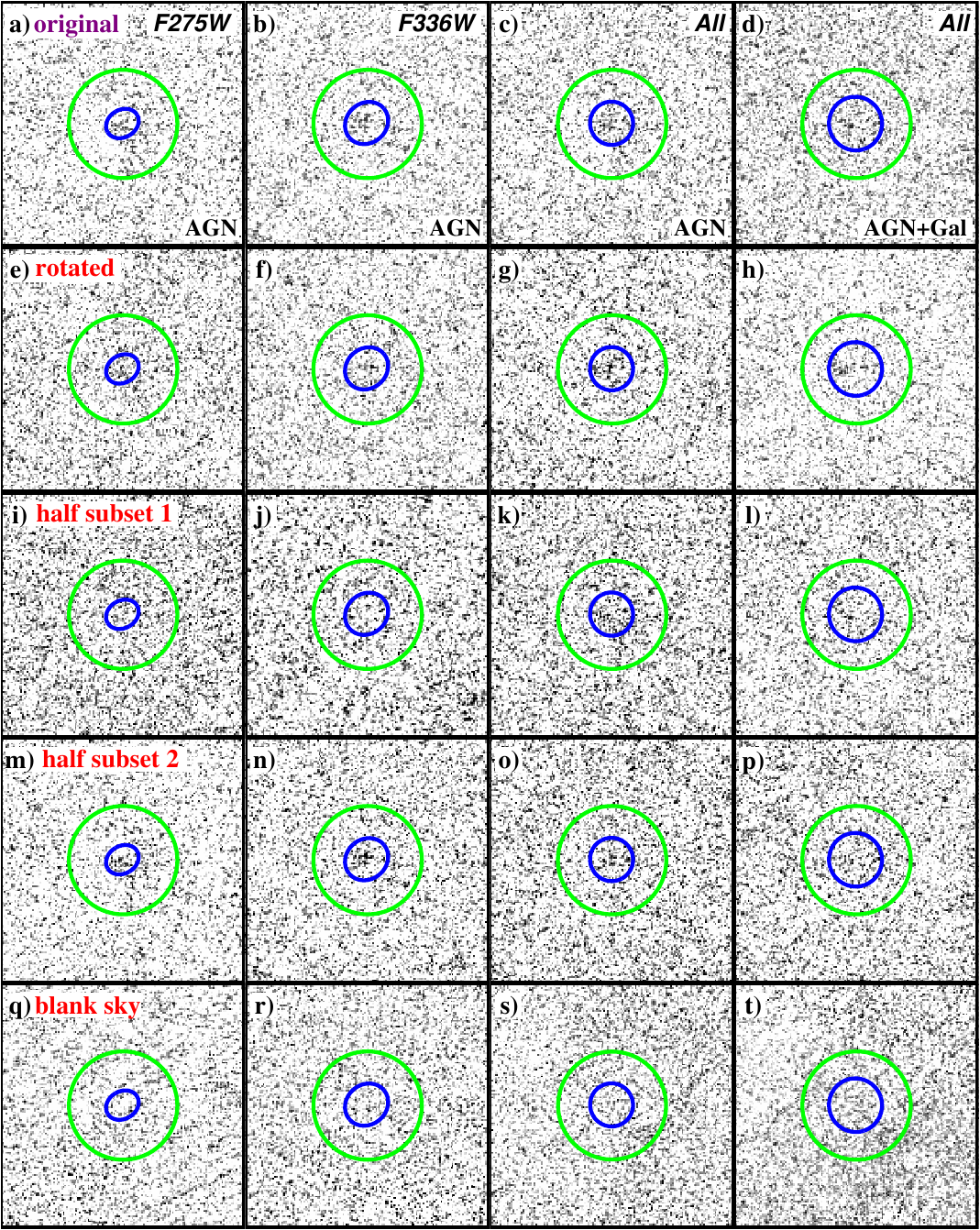}}
\caption{\noindent
Tests for systematics in our stacking procedure for our stacks that had
detections $>$2$\sigma$. For reference, we show the original stacks that
were unaltered in the stacking process in panels [\emph{a}] through [\emph{d}].
In panels [\emph{e}] through [\emph{h}] we present the result of our test with
all images rotated over random multiples of 90$^{\circ}$ before stacking; in
panels [\emph{i}] through [\emph{l}] we show the result of randomly dividing
our sample in two and stacking the first independent subset of images; in panels
[\emph{m}] through [\emph{p}] we show the complementary stack for the remainder
of the images; and in panels [\emph{q}] through [\emph{t}] we show that no excess
LyC signal is detected in stacks of equal numbers of \emph{random} blank sky
areas. {The first column display stacks for tests on the F275W AGN sample, the
second for the F336W AGN sample, the third for all AGN in our sample, and the
fourth for our entire sample of galaxies with and without AGN, where \emph{all}
indicates the F225W, F275W, and F336W filters.} The meaning of blue and green
ellipses is as in Fig.~\ref{fig:figure6}--\ref{fig:figure8}.
\label{fig:figureA1}}
\end{figure*}

To verify that residual astrometric offsets and/or trapped electron trails do
not significantly affect our measurements, we first \emph{randomly} rotate
each of the individual sub-images by integer multiples of 90$^\circ$, then
repeat the stacking described in \S\ref{sec:method}, as these systematics would
be directionally dependent. In Fig.~\ref{fig:figureA1}, we compare our
spectroscopic samples to the original LyC stacks (panels \emph{a}--\emph{d})
with the results of our rotation tests (panels \emph{e}--\emph{h}). The detected
LyC signal in the rotated stacks does indeed remain consistent with our
original stacks within 0.13$\pm$0.15 mag (see Table~\ref{table:tableA1}).
In the case of F275W, rotation of the images actually \emph{improves} the
SNR of the measurement (by $\sim$5\%), while the measured photometry remains
within $\sim$0.1\,mag from that of the original (unrotated) stack. Such an
improvement in SNR without significantly affecting the detected flux is most
likely due to small scale residual gradients in the background of the individual
sub-images, which after the random rotations would average out in a stack. For
the F336W filter, this is not the case, and the measured 1$\sigma$
upper limit actually decreased (by $\sim$0.4\,mag) upon random rotation,
suggesting that randomizing the surrounding sky may slightly dilute the flux,
as small scale fluctuations in the brighter background of this filter may
combine with the central flux in the image (see Table~\ref{table:tableA1}).
The F225W ERS images were all taken at the end of each orbit --- closest to
the Earth's limb --- since the F225W sky-background was expected to be the
faintest. Hence, slight amounts of Earth-shine may have contributed to somewhat
larger sky-gradients in the F225W mosaics than in the other filters. 

We also extract from the WFC3/UVIS ERS mosaics random patches of
\emph{blank sky}, equal in number to the number of galaxy sub-images used for
the stacks for each filter. These are combined using the same stacking method,
and have the \emph{same} sensitivity to LyC emission as the galaxy stacks. We
present the resulting blank sky stacks in the bottom row (panels
\emph{q}--\emph{t}) of Fig.~\ref{fig:figureA1} and their blank-sky
``photometry'' is tabulated in Table~\ref{table:tableA1}. No significant
signal is detected at $\geq$28.6\,mag (\ie\ the central aperture ``flux''
is present at the $\sigma$$<$1 level) in any of the blank sky stacks,
which implies that our measured LyC signal in the other cases
discussed above is most likely real, and associated with objects that
were selected at longer (UVC) wavelengths. 
\noindent
\subsection{\SExtractor\ versus ``Tic-tac-toe'' Photometry}
\label{sec:tttphot}

It may be of some concern that our use of \SExtractor\ generated UVC
apertures for photometric measurements might cause us to miss some amount of
extended and faint LyC flux at larger galactocentric radii, which may
be smaller than desired.

\begin{figure*}[ht!]
\centerline{
\includegraphics[width=\txw]{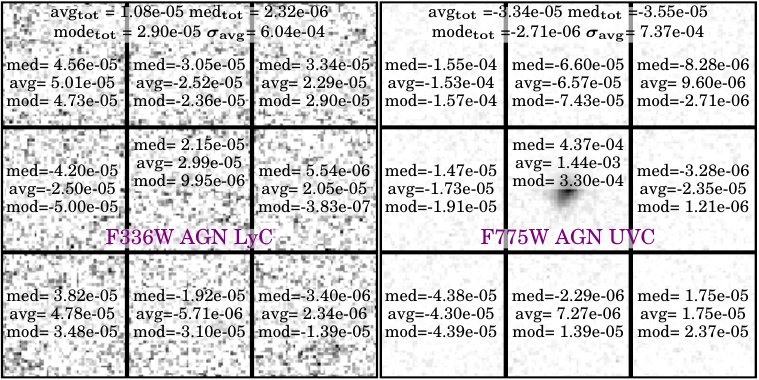}
}
\caption{\noindent\small
Example of a ``tic-tac-toe'' sky-background analysis of the 151$\times$151 pixel
F336W AGN LyC and F775W UVC stacks. Units displayed are in e$^{-}$/s for each
respective box, and full stack statistics are displayed at the top of the image.
[\emph{left 4 panels}] and Stacks of all galaxies and AGN for each filter
and a ``Super-stack'' of the entire sample [\emph{right 4 panels}]. 
Any large scale gradients in the residual sky-background left in the drizzled
images are $\gtrsim$5.2--40$\times$ fainter than the faint remaining sky
background residuals determined in Fig.~\ref{fig:figure3}. Such gradients in
the sky will be fainter than $\sim$32.1, 32.0, and 32.4 \magarc\ across
the 1$\farcs$53 ``tic-tac-toe'' aperture in the WFC3/UVIS F225W, F275W, and
F336W stacks. Residual UV sky-gradients appear, therefore, fainter than the LyC
signal where this can be measured, as discussed in the text. Not all
``tic-tac-toes'' used in Table~\ref{table:tableA1} are shown here.
\label{fig:figureA2}} \end{figure*}

We therefore analyzed the flux and SNR within the central 51$\times$51 pixels
(1$\farcs$53$\times$1$\farcs$53) of a ``tic-tac-toe'' 9-segment grid with
respect to the background level, corrected for residual gradients, determined
from the eight surrounding segments. An example is shown in 
Fig.~\ref{fig:figureA2}. The final columns of Table~\ref{table:tableA1} compare
the SNR measured using our ``tic-tac-toe'' photometry and those measured within
the \SExtractor\ LyC apertures. We measure similar fluxes --- at slightly lower
SNR --- using the ``tic-tac-toe'' aperture compared to the smaller \SExtractor\
fitted apertures and measurements, validating their robustness against modest
variations in choice of aperture size, and the specific details of the
sky-background subtraction. 

The details of our ``tic-tac-toe'' photometry results are shown in
Table~\ref{table:tableA1}, which serves to verify the quantities listed in
Table~\ref{table:table2} for the \SExtractor\ UVC apertures used for LyC
detections in \S\ref{sec:photometry}. Col.~1--3 of Table~\ref{table:tableA1}
are the same as Col.~1, 2 \& 4 of Table~\ref{table:table2}; Col.~4 lists the
measured ERS sky-background \emph{before} the sky-level itself was removed
from each exposure in the AstroDrizzle reduction (for details, see
\citetalias{Windhorst2011} Table 2); Col.~5--6 lists the sky-subtracted LyC
and UVC fluxes and their formal 1$\sigma$ errors in the central square
aperture from the ``tic-tac-toe'' photometry; Col.~7--8 lists the resulting
total LyC and UVC AB magnitudes measured over the full 
1$\farcs$53$\times$1$\farcs$53 ``tic-tac-toe'' apertures, which are to be
compared with the same quantities derived from the \SExtractor\ apertures in
Col.~6 and 9 of Table~\ref{table:table2}; Col.~9 of Table~\ref{table:tableA1}
lists the corresponding \emph{average} LyC and UVC SB values inside the
1$\farcs$53$\times$1$\farcs$53 apertures; Col.~10 lists the 1$\sigma$ SB error
on this value \emph{implied} by the observed total SNR of the LyC or UVC flux,
assuming a fully flat SB distribution inside the central ``tic-tac-toe'' aperture
(see \S\ref{sec:obsradprof}); Col.~11 shows the SNR predicted by the WFC3/UVIS and ACS/WFC 
CCD equations as a check of the observed SNR derived from the ``tic-tac-toe''
analysis, which is listed in Col.~12. 
\noindent

When averaged over the 24 cases listed in Table~\ref{table:tableA1}, the SNR
predicted by the CCD detector properties is about 0.83$\pm$0.62 (rms)$\times$
the observed SNR that was derived from the ``tic-tac-toe'' photometry. This is
likely because the SNR predicted from the CCD equation has an uncertain
component related to how detector read noise and dark current noise affect the
drizzled images in a \emph{correlated} fashion. These noise components are only
measured in the original CCD images (see \citetalias{Windhorst2011} and
\citet{Dressel2015} for a discussion of all WFC3 CCD parameters). For the CCD
equation prediction of the drizzled mosaics, we assumed that this term would
scale with the original/drizzled pixel ratio. Depending on how large a fraction
of the read noise and dark current is correlated on scales of a few pixels,
this may not be exactly true. In any case, with the above assumption, the 
predicted SNR for the 1\farcs53$\times$1\farcs53 ``tic-tac-toe'' apertures
is close to the SNR derived from the observed parameters in the ``tic-tac-toe''
photometry. More specifically, the observed ``tic-tac-toe'' SNR values in Col.~12
of Table~\ref{table:tableA1} are generally close to the SNR derived from the
photometry using the smaller \SExtractor\ UVC apertures (see
\S\ref{sec:photometry} and Table~\ref{table:table2}). The exception is the
F336W photometry, where the ``tic-tac-toe'' SNR appears to be lower. This
is likely due to the larger amount of higher sky-background and its sky-background
gradients included in the ``tic-tac-toe'' aperture in the WFC3/UVIS F336W
filter.

Table~\ref{table:table1} lists the measured 1$\sigma$ SB sensitivity over a
2$\farcs$0 diameter aperture in our drizzled WFC3 mosaics. For example,
the F275W filter has a single mosaic 1$\sigma$ SB sensitivity of $\sim$29.82 
\magarc\ when measured over a 2$\farcs$0 diameter aperture. \emph{If} the
sky-background was completely flat without any gradients (but see the
discussion in Appendix~\ref{sec:gradients}) then in a 11 object galaxy
stack with 8 surrounding ``tic-tac-toe'' apertures, the quality of the
sky-subtraction, or the 1$\sigma$ SB sensitivity, could in principle be as good
as 29.82 mag + 2.5 log($\sqrt{(1.53/2)^2\!\times\!11\!\times\!8}$)=32.0\,\magarc.
Here, we scaled the 1$\sigma$ SB sensitivity with the aperture size, and the
rms deviation of the surrounding sky with the number of sub-images and number
of available surrounding sky boxes (each of which have their own well measured
residual sky values and sky rms; see Fig.~\ref{fig:figureA2}). All of these
values are determined \emph{after} image stacking, and are used in predicting
the ``tic-tac-toe'' SNR in Table~\ref{table:tableA1}. 

This formal limit is better than the measured 1$\sigma$ SB sensitivity
value derived from the ``tic-tac-toe'' photometry in Table~\ref{table:tableA1}
(30.3\,\magarc), which could be due to some correlated or systematic noise
present in the actual data. Since subtle residual gradients
remain in the sky-background, as discussed in Appendix~\ref{sec:gradients}, the
real 1$\sigma$ SB sensitivity limit that can be achieved is \emph{not as faint}
as 32.0\,\magarc. We suggest in Appendix~\ref{sec:gradients} that
uncertainties from subtle residual sky-background gradients in the LyC stacks
limit our photometry to $\sim$30.3\,\magarc, so that any LyC SB values
measured to be fainter than this should be considered with this limit in mind.
For that reason, we do not plot any of the SB values in the LyC light profiles
of Fig.~\ref{fig:figure10} to levels fainter than 30.3\,\magarc. As quantified
in Appendix~\ref{sec:tttqual}, the ``tic-tac-toe''
photometry does confirm the \SExtractor\ photometry from the smaller LyC and UVC
apertures discussed in \S\ref{sec:photometry}, to within their respective
errors. 
\noindent\begin{table*}
\centering
\footnotesize
\caption{Photometric ``Tic-Tac-Toe'' Analysis of the LyC and UVC from Galaxies
and AGN, and Other Photometry Tests\label{table:tableA1}}
\setlength{\tabcolsep}{5pt}
\begin{tabular}{lccccclcclll}
	\toprule \\[-19pt]
Filter                   & $z$-range               & N$_{objects}$              &
$\mu_{\rm sky}$          & $\!\!S_{\rm LyC/UVC}$   & $\mathrm{m.e.}_{S_{\rm LyC/UVC}}$ &
$S_{\rm LyC/UVC}$        & $\sigma_{S}$            & $\mu_{\rm LyC/UVC}$        & 
1-$\sigma_{\mu}$         & $SNR$                   & $SNR$                      \\ [-5pt]
                         &                         &                            & 
[mag/"$^{2}$]            & [\emin/pix/s]           & [\emin/pix/s]              &
[\mAB]                   &  [\mAB]                 & [mag/"$^{2}$]              & 
[mag/"$^{2}$]            & (Pred.)                 & (Obs.)                     \\[-5pt]
(1)                      & (2)                     & (3)                        &
(4)                      & (5)                     & (6)                        &
\multicolumn{1}{c}{(7)}  & (8)                     & (9)                        &
(10)                     & \multicolumn{1}{c}{(11)} & \multicolumn{1}{c}{(12)} \\[-2pt]                 
\midrule \\[-18pt]
\multicolumn{12}{l}{\sc LyC - Galaxies with AGN:}\\[-4pt]
F275W & 2.559--3.076 &  7 & 25.64 & 1.48$\times$10$^{-5}$ & 4.52$\times$10$^{-5}$ & 27.71 & 0.33 & 28.59 & 29.88 & ~~~~3.25 & ~~~~3.29 \\[-5pt]
F336W & 3.132--3.917 &  3 & 24.82 & 1.91$\times$10$^{-5}$ & 1.21$\times$10$^{-4}$ & 27.97 & 0.69 & 28.85 & 29.35 & ~~~~0.91 & ~~~~1.58 \\[-4pt]
\multicolumn{12}{l}{\sc UVC - Galaxies with AGN:}\\[-4pt]
F606W & 2.559--3.076 &  7 & 22.86 & 1.61$\times$10$^{-3}$ & 1.41$\times$10$^{-5}$ & 24.97 & 0.01 & 25.85 & 31.0 & ~~~~149. & ~~~~114. \\[-5pt]
F775W & 3.132--3.917 &  3 & 22.64 & 1.48$\times$10$^{-3}$ & 1.47$\times$10$^{-5}$ & 24.24 & 0.01 & 25.12 & 30.1 & ~~~~77.5 & ~~~~100. \\[-4pt]
\multicolumn{12}{l}{\sc LyC - All Galaxies:}\\[-4pt]
F225W & 2.302--2.450 & 19 & 25.46 & 4.89$\times$10$^{-6}$ & 2.54$\times$10$^{-6}$ & 28.82 & 0.56 & 29.70 & 30.41 & ~~~~2.90 & ~~~~1.92 \\[-5pt]
F275W & 2.559--3.076 & 14 & 25.64 & 6.65$\times$10$^{-6}$ & 3.17$\times$10$^{-6}$ & 28.58 & 0.52 & 29.46 & 30.27 & ~~~~2.91 & ~~~~2.10 \\[-5pt]
F336W & 3.132--3.917 & 13 & 24.82 & 5.86$\times$10$^{-6}$ & 6.04$\times$10$^{-6}$ & $>$29.29 & --- & $>$30.13 & 30.10 & ~~~~(1$\sigma$) & ~~~~--- \\[-5pt]
All   & 3.132--3.917 & 46 & 25.46 & 4.02$\times$10$^{-6}$ & 2.14$\times$10$^{-6}$ & 29.04 & 0.58 & 29.92 & 30.60 & ~~~~1.76 & ~~~~1.87 \\[-4pt]
\multicolumn{12}{l}{\sc UVC - All Galaxies:}\\[-4pt]
F606W & 2.302--2.450 & 19 & 22.86 & 2.90$\times$10$^{-3}$ & 8.99$\times$10$^{-6}$ & 24.34 & .003 & 25.22 & 31.49 & ~~~~181. & ~~~~322. \\[-5pt]
F606W & 2.559--3.076 & 14 & 22.86 & 2.19$\times$10$^{-3}$ & 9.95$\times$10$^{-6}$ & 24.64 & .005 & 25.52 & 31.38 & ~~~~118. & ~~~~220. \\[-5pt]
F775W & 3.132--3.917 & 13 & 22.64 & 1.13$\times$10$^{-3}$ & 7.52$\times$10$^{-6}$ & 24.53 & .007 & 25.41 & 30.85 & ~~~~76.2 & ~~~~150. \\[-5pt]
All   & 3.132--3.917 & 46 & 22.86 & 2.51$\times$10$^{-3}$ & 7.49$\times$10$^{-6}$ & 24.49 & .003 & 25.37 & 31.69 & ~~~~292. & ~~~~335. \\[-4pt]
\multicolumn{12}{l}{\sc LyC - All Galaxies --- Rotated by random n$\times$90$^\circ$:}\\[-4pt]
F225W & 2.262--2.450 & 19 & 25.46 & 5.05$\times$10$^{-6}$ & 2.55$\times$10$^{-6}$ & 28.79 & 0.55 & 29.67 & 30.41 & ~~~~2.98 & ~~~~1.98 \\[-5pt]
F275W & 2.481--3.076 & 14 & 25.64 & 7.01$\times$10$^{-6}$ & 3.18$\times$10$^{-6}$ & 28.52 & 0.49 & 29.40 & 30.26 & ~~~~3.07 & ~~~~2.20 \\[-5pt]
F336W & 3.110--4.149 & 13 & 24.82 & 4.92$\times$10$^{-7}$ & 6.06$\times$10$^{-6}$ & $>$29.67 & --- & $>$30.32 & 30.32 & ~~~~(1$\sigma$) & ~~~~--- \\[-5pt]
All   & 3.110--4.149 & 46 & 25.46 & 4.10$\times$10$^{-6}$ & 2.14$\times$10$^{-6}$ & 29.01 & 0.57 & 29.89 & 30.60 & ~~~~1.79 & ~~~~1.92 \\[-4pt]
\multicolumn{12}{l}{\sc Random empty sky:}\\[-4pt]
F225W & 2.262--2.450 & 19 & 25.46 & 2.52$\times$10$^{-6}$ & 2.74$\times$10$^{-6}$ &$>$29.63 & ---  &$>$30.42 & 30.33 &~~~~(1$\sigma$)&~~~~--- \\[-5pt]
F275W & 2.481--3.076 & 14 & 25.64 & 6.39$\times$10$^{-6}$ & 3.71$\times$10$^{-6}$ &$>$28.62 & ---  &$>$29.50 & 30.09 &~~~~(1$\sigma$)&~~~~--- \\[-5pt]
F336W & 3.110--4.149 & 13 & 24.82 & 1.18$\times$10$^{-6}$ & 6.03$\times$10$^{-6}$ &$>$29.22 & ---  &$>$31.87 & 30.10 &~~~~(1$\sigma$)&~~~~--- \\[-5pt]
All   & 3.110--4.149 & 46 & 25.46 & 3.94$\times$10$^{-6}$ & 2.38$\times$10$^{-6}$ &$>$29.05 & ---  &$>$29.94 & 30.49 &~~~~(1$\sigma$)&~~~~--- \\[-2pt]
\bottomrule
\vspace*{-8pt}
\end{tabular}
\begin{minipage}{\txw}{\small (1) WFC3 filter; (2) redshift range (as in
Table~\ref{table:table2}); 3) Number of galaxies with reliable spectroscopic
redshifts used in each stack; (4) sky surface brightness
in AB \magarc\ using EXPTIME and MDRZSKY from the FITS header, corrected for
the number of tiles in the mosaic; (5)--(6) Average sky-subtracted flux and mean
error thereon in \emin\,pix$^{-1}$\,s$^{-1}$ over $N$ sub-images in the central
51$\times$51 pixel ``tic-tac-toe'' aperture (see Appendix~\ref{sec:tttphot});
(7)--(8) Total LyC and UVC flux and error thereon, expressed as AB mag;
(9)--(10) Average LyC and UVC surface brightness and 1$\sigma$ error thereon,
both in AB \magarc; (11) SNR of LyC or UVC flux in Col.~(5), predicted from the
WFC3/UVIS or ACS/WFC CCD equation; (12) SNR of LyC or UVC detection in
Col.~(5), observed from the total sky-subtracted LyC (row 1--4) or UVC (row
5--8) flux and corresponding sky subtraction error in Col.~(6). (These are to
be compared with the SNR of LyC or UVC detections within the \SExtractor\ UVC
apertures listed in Table~\ref{table:table2}).}
\end{minipage}
\end{table*}

\subsection{Quality of \SExtractor\ versus ``Tic-tac-toe'' Photometry}
\label{sec:tttqual}

When compared to our main photometry using \SExtractor\ UVC apertures
in Table~\ref{table:table2}, the ``tic-tac-toe'' photometry in
Table~\ref{table:tableA1} shows the following: for galaxies, the measured
difference in flux within the LyC detected apertures and the
1\farcs53$\times$1\farcs3 ``tic-tac-toe'' aperture is $\Delta$(LyC-TTT)=0.55,
when averaged over the F275W and F336W LyC filters for AGN.
{This difference is not significant, which is} possibly another sign of the
very flat LyC SB profiles discussed in \S\ref{sec:obsradprof}. 

In the ``tic-tac-toe'' exercise, all contaminating neighbors were removed from
the central 1\farcs53$\times$1\farcs53 ``tic-tac-toe'' aperture.
Nonetheless, it is also possible that some contaminating very low SB flux
(invisible in the individual sub-images, even in our deepest WFC3 IR images)
from nearby neighbors could have leaked into the larger ``tic-tac-toe''
apertures. In any case, the above numbers show that the total contaminating
flux is likely $\lesssim$0.3 mag, since any contaminating flux should be far
smaller in the much smaller LyC apertures (see also \S\ref{sec:photometry}). 

The ``tic-tac-toe'' photometry in Table~\ref{table:tableA1} also allows us to
further quantify the three other critical tests that were qualitatively
discussed in Appendix~\ref{sec:dettests}. For the stack of all galaxies, the
``tic-tac-toe'' photometry of the sub-images that were
randomly rotated by 90$^\circ$ shows a difference with the ``unrotated'' image
stacks of $\Delta$(Rotated--Unrotated)\,=\,+0.13$\pm$0.15\,mag (m.e.) when
averaged over the four LyC filters. In other words, the ``tic-tac-toe''
photometry of the rotated stack is consistent with that of the
corresponding unrotated stack, showing that --- to within the errors --- our
stacking method yields reproducible and consistent LyC fluxes.

No signal was detected (at the AB$\gtrsim$28.6 mag level) in any of the random 
blank-sky central ``tic-tac-toe'' apertures, when compared to the average
surrounding sky boxes. These ``blank-sky'' stacks thus serve as a check on the
quality of the UV sky-background subtraction, and the effects of any subtle
remaining sky-gradients. They also show that in random sky areas, no
significant amount of flux from cosmic rays residuals that might have not been
completely removed during the drizzling process were added in at the at the
AB$\simeq$28.6 mag level (see Fig.~\ref{fig:figureA1}).
Hence, these blank-sky stacks are our best check that even for N=3--6 one third
to half-orbit exposures per filter, the bulk of our stacked, very faint LyC
signal at \mAB$\sim$27.4--28.6 mag (Tables~\ref{table:table2} and
\ref{table:tableA1}) is not due to residual unfiltered cosmic rays or noise
peaks. If this were true, then these random blank-sky stacks would have shown
as significant false signal at \emph{similar} \mAB\ levels as the real LyC
detections, which was not the case. 

\section{Possible Sources of Contaminating Non-ionizing Flux}
\label{sec:contaminants}

\subsection{In-filter Red-leak of Non-ionizing Flux}
\label{sec:red-leak}

The WFC3/UVIS filters were designed to minimize the transmission of
photons with wavelengths higher or lower than their specified cutoffs (see 
Fig.~\ref{fig:figure1}(a)). However, as seen in Fig.~\ref{fig:figure1}(b), a
small amount of flux red-ward of the Lyman Limit from galaxies observed in these
filters with redshifts in the ranges of Table~\ref{table:table1} can still leak
into the filter and contaminate LyC observations with non-ionizing UVC flux.
The lower redshift bounds in Table~\ref{table:table1} were carefully chosen such
that \emph{no} light $>912$\AA\ is sampled below the filter's red edge. The
filter red edge is defined as $<0.5\%$ of the filter's peak transmission. For
galaxies at the higher redshifts in the ranges of Table~\ref{table:table1}, and
especially those at higher redshifts than the designated upper bound, the
contribution from UVC ``red-leak'' can become the dominant source of photons
measured in the filter, as the portion of the spectrum intended for LyC 
observation becomes exceedingly faint at shorter wavelengths and the
non-ionizing continuum remains roughly constant at longer wavelengths. Thus, in
order to accurately measure LyC photometry and escape fractions from these
samples of galaxies, we must verify that the flux measured from our sample is
dominated by LyC photons. 

Since we cannot directly measure the fraction of non-ionizing flux leaking into
the filter from the observation, we estimate this value by modeling the
contribution of LyC and UVC to the observed flux from the total sample. Using
SEDs fit from continuum observations of our galaxy sample and { \it average }
line-of-sight IGM transmission models (see \S\ref{sec:fescintro}), we calculate
the average UVC ``red-leak'' of our observation in the WFC3/UVIS F225W, F275W, and
F336W filters by comparing the total flux integrated in the
entire filter, and the total flux integrated below the Lyman Limit of each
galaxy. We calculate this value as:\vspace{-7pt}

\begin{equation*}
\resizebox{0.99\hsize}{!}{%
  $\frac{F_{\nu}^{\uvc}}{F_{\nu}^{\!\lyc}} = 1 - \displaystyle\sum\limits^{\rm N_{gal}}_{\rm i=1} \frac{\int\limits_{\lambda_1}^{\rm \lambda_{912}}\mathcal{T}_{\rm obs}^{\!\lyc}(\nu)\mathcal{T}_{\igm}(z_{\rm i},\nu)F_{\nu,i}(\nu) \frac{d\nu}{\nu}}{\int\limits_{\lambda_1}^{\rm\lambda_{912}}\mathcal{T}_{obs}^{\!\lyc}(\nu)\frac{d\nu}{\nu}}\Bigg/\frac{\int\limits_{\lambda_1}^{\lambda_2}\mathcal{T}_{\rm obs}^{\!\lyc}(\nu)\mathcal{T}_{\igm}(z_{\rm i},\nu)F_{\nu,i}(\nu)\frac{d\nu}{\nu}} {\int\limits_{\lambda_1}^{\lambda_2}\mathcal{T}_{obs}^{\!\lyc}(\nu)\frac{d\nu}{\nu}}$%
  }
\end{equation*}

\noindent where $\lambda_1$ and $\lambda_2$ are the minimum and maximum
wavelengths of the full filter transmission curve, $\lambda_{912}$ is the
observed wavelength of the Lyman Limit of the galaxy, $F_{\nu}$ is the SED flux
of the galaxy in erg\,s$^{-1}$\,cm$^{-2}$\,Hz$^{-1}$, $\mathcal{T}_{\igm}$ is
the average line-of-sight IGM transmission at the redshift of the galaxy, and
$\mathcal{T}_{\rm obs}^{\!\lyc}(\nu)$ is the combined throughput of the filter,
detector quantum efficiency (QE), and optical telescope assembly (OTA). This
value quantifies the fraction of flux we measure from these galaxies in the
filter intended for LyC observations that is non-ionizing. For the F225W,
F275W, and F336W filters, the \emph{percentage of total} ``red-leak''
photons that contribute to the measured LyC flux of our sample are
$\sim$\,0.65\%, 0.64\%, and 0.19\% respectively. That is, less than
1\% of the anticipated and measured LyC flux itself could be red-leak flux from
longwards of 912\AA. From these values, it is clear that LyC observations of
galaxies at the redshift ranges indicated in Table~\ref{table:table1} with
their respective filters are dominated by LyC photons. We also note that our MC
analysis of the observed LyC flux from these galaxies accounts for these
``red-leak'' photons, in order to make appropriate corrections for low level
non-ionizing contamination of the order of $\sim$\,0.28\%. 

\subsection{UVIS filter Pinholes }
\label{sec:pinholes}

Pinholes are very small voids in the coating on the surface of a filter. These
voids appear usually due to poor adhesion of the coating in these regions where
particulate matter on the surface of the filter is coated over when the
substrate is cast, or from mechanical abrasion or chemical interactions when
the filter is in use. Several of the WFC3/UVIS filters have pinholes, so we
must make sure that none of the LyC flux that we measure is due to out-of-band
flux leaking in through the filter in an area where such a pinhole exists. Most
of the obvious pinholes were known before WFC3's launch, and the filters with
the fewest pinholes were chosen for flight \citep{Dressel2015}. To the best of
our knowledge, the number of pinholes did not increase during the 7 years that
the WFC3 filters were on the ground. Visible pinholes on the selected filters
were painted over when possible \citep{Baggett2006}. Any remaining pinholes not
painted over before launch are likely $\lesssim$0.2 mm in diameter, or they
would have been treated before final instrument assembly. Unfortunately, no
record was kept of any less obvious pinholes in the flight filters that were
not painted over before launch. Remaining pinholes could cause subtle
field dependent red-leaks and very low level sky gradients, which we quantify
here.

We need to first estimate how large the footprint and the amplitude of any
pinhole red-leak on the WFC3 CCDs could be. The HST f/24 beam gets re-imaged
inside WFC3 to f/31 \citep{Dressel2015}, so that the plate scale on the WFC3
UVIS detector changes from 3\farcs 58/mm to 2\farcs 77/mm (\ie\
206265/(2400$\times$31) ''/mm). The WFC3 UVIS Marconi CCDs have 15\micron\ pixels,
so the two 2k$\times$4k CCD arrays are about 61 mm in physical size. The
WFC3/UVIS F225W filter is in Selectable Optical Filter Assembly (SOFA) filter
wheel 3, F336W in filter wheel 4, and F275W in filter wheel 6 out of 12, where
wheel 12 is closest to the CCD's. The average location of these three filter 
wheels is about 190$\pm$25 mm away from the focal plane (Fig. 2.1 of
\citealt{Dressel2015}), which we hereafter refer to as the ``center of the
SOFA''. The $\pm$25 mm indicates the approximate range over which these three
UVIS filter wheels are mounted inside the SOFA. Each SOFA filter is 57.3 mm
square and $\lesssim$5.151 mm thick \citep{Baggett2006}, as fabricated by the
filter vendor to the specifications defined by the WFC3 Scientific Oversight
Committee and Instrument Product Team. 

The SOFA is about 1/3 of the way between the focal plane and the pupil, which
is the anamorphic asphere mirror inside WFC3 that corrects for the spherical
aberration in HST's primary mirror. Fortunately therefore, all pinholes in the
WFC3 UVIS filters will be severely out of focus, since the filters are so far
from the focal plane. We first need to calculate how large the pupil of each
image is in the filter plane. The anamorphic asphere mirror has a diameter of
about 25 mm and is about 630 mm away from the CCD. It is about 440 mm from the
center of the SOFA, so that the \emph{radius} of the image pupil at the filter
distance is about \ripf$\simeq (190/630)\cdot 25/2\simeq$3.77 mm. Hence, the
image pupil at the filter is about 7.54/57.3$\simeq$13\% of the filter size. 

Next, we need to estimate how large the footprint and the amplitude of any
pinhole flux on the WFC3 CCDs could be. If WFC3's f/31 beam goes through a
pinhole with an r$\gtrsim$0.08 mm radius (\ie\ $\gtrsim$0.5$\times$5.151/31)
in a $\lesssim$5.151 mm thick UV filter about 190 mm in front of the CCD, this
pinhole will affect a beam with an opening angle $\theta$=90$^\circ$--atan(31)
$\simeq$ 1.85$^\circ$ projected onto the CCD. As viewed from the CCD, the
remainder of the pixels outside this beam will not view the sky through the
pinhole. At the CCD, the circular beam that \emph{is} affected by this pinhole
will have a 190/31 $\simeq$ 6.1 mm $\simeq$ 17\farcs 0 $\simeq$ 430 pixel
radius on the CCD, and so its diameter will cover about 20\% of the WFC3
CCD FOV. To avoid internal reflections in the camera, the UVIS CCDs are tilted
by $\sim$21$^\circ$ with respect to the axis of the beam, so the projected
footprint of each pinhole is actually about 430/cos(21)$\simeq$460 pixels in
radius. In other words, the footprint projected by the pinhole on the CCD is
very large, and will significantly dilute the extra SB signal projected
through the pinhole. In the limit, a much smaller pinhole (r$<<$0.08) mm would
not see the entire f/31 beam, and will thus act like a pinhole camera that
illuminates the entire CCD, diluting the extra SB that goes through this
smaller pinhole even more. 

If the interference or AR coating were not present in a pinhole for one of our
WFC UVIS filters, Fig.~\ref{fig:figure1}\emph{b} shows that it could have a
significant increase in local throughput, or less in case that the local
defect was only partial in transmission. In a worst case, the pinhole would
act like a F606W or F775W WFC3 filter at that location, if the
OTA$\times$Filter-Throughput$\times$CCD-QE at those wavelengths reached the
WFC3 maximum in the F606W filter of $\sim$28\% (see Fig. 3.2 and 5.2 of
\citealt{Dressel2015}). In such cases, the Zodiacal sky in the beam illuminated
by the pinhole would be much higher than seen in our UVIS filters, possibly as
high as that in our broadband optical filters, or slightly higher if the
pinhole were fully transparent. 

The brightest object in the WFC3 ERS has $V$\,$\simeq$\,17\,\mAB\ and the most
commonly seen objects are faint galaxies with $V$\,$\simeq$\,26--27\,\mAB\
(Fig. 12 of \citetalias{Windhorst2011}). Hence, their collective SB is well
below the average Zodiacal background, which is 23.7--22.6 \magarc\ (see Col.~4
of Table~\ref{table:tableA1}). Hence, for all practical purposes, the pinhole
contribution will just be the full white light Zodiacal background if its
throughput reaches the maximum total throughput of $\sim$28\%. Following
Table~\ref{table:tableA1}, we assume that the white light Zodi would have on
average a SB$\simeq$22.9 \magarc\ through such a pinhole, which is roughly the
observed value in F606W (see \citetalias{Windhorst2011}). We will also consider
the case of a single V$\simeq$17 \mAB\ star shining behind a pinhole, as well
as the integrated sky SB derived from the faint galaxy counts in Fig. 12 of
\citetalias{Windhorst2011} to \mAB$\lesssim$26\,mag. In F606W, the latter
reaches $10^5$ galaxies/0.5\,mag/deg$^2$ to \mAB$\lesssim$26\,mag, and in
F275W, they reach $3\times 10^4$ galaxies/0.5\,mag/deg$^2$ to
\mAB$\lesssim$25.5 mag. On a per square arcsecond basis, the integrated sky SB
from faint galaxies is therefore $\sim$29.3 \magarc\ in F606W and $\sim$30.1 
\magarc\ in F275W, respectively, \ie\ fully negligible compared to the
Zodiacal sky SB values from Table~\ref{table:tableA1} of \SBzodiV$\simeq$22.9
and \SBzodiNUV$\simeq$25.5 \magarc\ in these filters, respectively. Hence, 
only the Zodiacal light and the effective SB of the occasional bright star
behind the filter would be the main sources of pinhole contamination. All
calculations below are done in terms of surface brightness (SB in \magarc) or
Intensity ($I$ in relative counts/sec). 

In a slow f/31 beam, when the pinhole is larger than the minimum size to
transmit through the filter, the total white light transmitted would increase
proportionally to the pinhole area compared to the total area of the image
pupil at the filter (\rph/\ripf)$^2$, both measured in mm. For an untreated
pinhole with an assumed \rph$\simeq$0.1 mm, we can now estimate the increase in
sky SB contribution from this pinhole over a r=460 pixel radius on the CCD. In
relative units, this is increase is:
\begin{equation} 
\IphV = F \cdot [\rph^2/\ripf^2] \cdot [\IstV+\IzodiV]
\label{eq:pinholeSBV} 
\end{equation}
with \IstV=$10^{-0.4(\SBstV-\ZPv)}$, \IzodiV=$10^{-0.4(\SBzodiV-\ZPv)}$, and:
\begin{equation} 
\IphNUV = [(\ripf^2-\rph^2)/\ripf^2] \cdot [\IstNUV+\IzodiNUV]
\label{eq:pinholeSBNUV}
\end{equation}
\IstNUV=$10^{-0.4(\SBstNUV-\ZPnuv)}$, \IzodiNUV=$10^{-0.4(\SBzodiNUV-\ZPnuv)}$.
Eq.~\ref{eq:pinholeSBV} describes the relative counts of optical white
light through the pinhole, and Eq.~\ref{eq:pinholeSBNUV} the relative counts
for the uncorrupted NUV sky SB. Note that the optical white light
\SBzodiV$\simeq$22.9 \magarc\ is compared here to the WFC3 F606W zeropoint of
\ZPv=26.08 mag (\ie\ \emph{not} the ACS ZP in this case), and the original UV
\SBzodiNUV$\simeq$25.5 \magarc\ (Col 4 of Table~\ref{table:tableA1}) is
compared to the combined F225W and F275W zeropoints of $\sim$24.1\,mag. The
factor F$\simeq$(10000-4000)/2000)$\simeq$3 reflects that the pinhole could
transmit three broadband filters worth of white light from 4000-10,000\AA\ (see
Fig.~\ref{fig:figure1}\emph{b}). We find approximately the same values if we 
instead use the WFC3 white light filter F200LP and its zeropoint ZP=27.36 mag,
and set the factor F=1. The limit of \rph$\simeq$\ripf\ would describe a
hypothetical pinhole so large that it transmits full white light over the
entire image pupil at the filter, which now acts like a wide $V$ band filter. In
that case, both equations still give the correct results:
Eq.~\ref{eq:pinholeSBV} describes its wide $V$ band SB, and
Eq.~\ref{eq:pinholeSBNUV} describes its now vanishingly small NUV SB. 

At first, we ignore the terms with \IstV\ and \IstNUV\ due to a bright star
near a pinhole. The SB from the pinhole in $V$ and NUV then are
\IphV$\simeq$0.039 and \IphNUV$\simeq$0.28 in the same relative units, 
respectively. The ratio \IphV/\IphNUV is 0.14, so that about 14\% of
white light background through the pinhole gets added to the UV sky:
\begin{equation} 
\SBskyNUV = \SBzodiNUV -2.5log(\IphV/\IphNUV) 
\label{eq:pinholeUVsky} 
\end{equation}
The error on this is at least 25/190$\sim$0.13 mag, depending how far the
SOFA filter is from the CCD. That is, a full white light pinhole footprint
could add $\sim$14$\pm$2\% to the sky SB over an annulus with
a diameter about 20\% of the 61 mm CCD area. This is the worst case --- a
smaller pinhole that doesn't project the entire f/31 beam through the filter
could instead add a much fainter SB over the whole chip. 

If we also add the effect from a \Vst$\simeq$17\,mag, $m^{\rm Star}_{\rm
NUV}$\,$\simeq$\,18 mag star whose image pupil in the filter plane
illuminates the pinhole, we must first correct for the fact that the light from
this point source is now spread out by factor of (\ripf/\rph)$^2$ at the
filter, so we must add 2.5log(3.77/0.1)$^2$ to the star's point source flux of
\mAB$\sim$17--18\,mag in Eq.~\ref{eq:pinholeSBV}--\ref{eq:pinholeSBNUV} to get
the equivalent SB from the star that actually affects the pinhole, expressed in
the appropriate relative units. Note that the Zodiacal background and
integrated galaxy counts do not have this problem, since they are already
expressed as proper SB in Eq.~\ref{eq:pinholeSBV}--\ref{eq:pinholeSBNUV}. These
numbers are \SBstV$\simeq$24.9 and \SBstNUV$\simeq$25.8\magarc.
Eq.~\ref{eq:pinholeSBV}--\ref{eq:pinholeSBNUV} converts these SB numbers to
relative fluxes, then adds them linearly. With a $V$$\sim$17\,mag star, the
white light SB from the pinhole now grows from \IphV$\simeq$0.039 to
\IphV$\simeq$0.046. This is only 17\% larger than just the light from Zodiacal
light alone, since the star is so much more spread out behind the filter. The
NUV comparison term remains at \IphNUV$\simeq$0.28. Hence, an out of focus
image of a $V$$\sim$17\,mag star behind the pinhole adds $\sim$17$\pm$2\% of
white light background to the NUV sky.

To first order, both the proper NUV light and any white light pinhole flux 
would get flat-fielded away, although the pinhole regions
would have a different color of the sky-background than the regular UV sky. The
WFC3 UVIS Marconi CCDs have QE curve that is fairly flat as a function of
wavelength, with most QE values between 2000-8000\AA\ peak around QE$\sim$70\%
with a range of $\pm$10\%. This peak actually lies within the F275W filter. We
therefore did the experiment to flat-field a WFC3 F606W image with a high SNR
flat-field taken in the F275W filter, and --- owing to WFC3's flat QE curve ---
this did not result in a residual gradient larger than about 10\% of sky
corner-to-corner across the whole CCD frame. Below we will assume a worst case
of $\sim$10\%. 

We can use this result to compute the additional sky gradient that an unpainted
pinhole with r$\simeq$0.1 mm could have caused in our UV images. When a pinhole
adds about 14--17\% extra flux to the regular UV Zodiacal sky over a 920 pixel
diameter circle on the CCD, the residual sky gradient that the improper
flat-fielding induces is at most 0.6--0.8\% of the \emph{total} Zodiacal sky
over 920 native pixels, or about 404 drizzled pixels. Across our sub-image size
of 71$\times$71 drizzled pixels, this residual sky gradient --- if present in
our object sub-images --- is thus less than $\sim$0.22-0.26\% of the local
Zodiacal sky. That is, the error in the UV sky of Eq.~\ref{eq:pinholeUVsky}
from sky gradients induced by the partial improper flat-fielding of any pinhole
white light is:
\begin{equation}
\sigma_{Sky} \simeq \SBskyNUV+2.5log(\IphV/\IphNUV \cdot 0.1 \cdot 71/404) 
\label{eq:pinholeUVskyerror}
\end{equation}
For our average UV sky of 25.36 \magarc\ (which is now brightened by --0.14
mag due to the extra pinhole flux), the residual sky-gradients left by pinholes
in 71$\times$71 drizzled pixels are thus fainter than 25.36--2.5 log(0.22\%)
$\simeq$31.9 \magarc\ and fainter than $\simeq$31.7 \magarc\ if a $V$$\simeq$17
mag star is also nearby the pinhole. These pinhole induced systematics are at
worst slightly brighter than those possibly caused by subtle residual
gradients left at the 32.3 \magarc\ level across the CCD's due to remaining
errors in the bias, dark frames, or flat-fields, as discussed in \S\ref{sec:sky}
and Appendix~\ref{sec:gradients}). Of course, the latter may affect our stacks
everywhere in the CCD mosaics, while the former occur only in sporadic
(although unknown) locations.

We investigated if the effects of any pinhole red-leaks were in fact seen in the
ERS data, since obvious pinholes would appear as large donut shaped objects in
the drizzled images. No obvious defects with $\gtrsim$14--17\% increased
transmission were seen in the raw ERS data on scales of $\sim$920 native
pixels, although this is hard to see due to cosmic rays and the shallow depth
of individual images. Partial transmission defects might exist at lower
levels. As discussed in \S\ref{sec:method}, we inspected all LyC sub-images
individually, and removed the ones with suspected increased noise due to
residual cosmic rays, structure in the weight maps due to drizzled image
borders, and other image defects. The objects removed from the LyC stacking all
appear to have higher image rms due to the proximity of structure in the
weight maps due to drizzled image borders. No obvious enhancements in the LyC
signal were seen due to the proximity of bright stars. 

We find that both red-leak flux from UV filter pinholes and other subtle
calibration errors may result in residual sky gradients of order
$\sim$32--32.3 \magarc\ across our stacking sky boxes. Unless these effects
can be removed through further refinement of the WFC3 calibration techniques,
residual systematic subtraction errors of order $\simeq$32 \magarc\ may well
pose a fundamental limit to the LyC stacking method of WFC3 data. This is why
the light profiles in Fig.~\ref{fig:figure12} cannot be extended to SB levels
much fainter than AB$\simeq$32 \magarc\ (see also the discussion in
Appendix~\ref{sec:tttphot} and Table~\ref{table:tableA1}).

Finally, if our LyC detections were in fact caused by subtle sky gradients on
r$\simeq$920 pixel scales, they should have been \emph{preferentially} located
near the same physical regions of the CCD, as would have been the case with CTE
effects if these had been already significant in Fig.~\ref{fig:figureC1}. We
saw no evidence of our strongest LyC candidates being located in the same
physical CCD region of each exposure. On the contrary, our strongest LyC
emitters are known ``Galaxies with weak AGN'' (Table~\ref{table:table2}), and
show no correlation in their observed position on CCDs. On average, their LyC
flux is brighter than that of our ``Galaxies without AGN''. The AGN LyC flux is
sometimes compact ($\lesssim$a few pixels; Fig.~\ref{fig:figure8}) --- as opposed
to that of the galaxies (see Fig.~\ref{fig:figure7}) --- and therefore occurs on
too small of an optical scale to be caused by the severely out of focus red-leak
through pinholes. Thus, since the physical location of the AGN on the CCDs is
spatially uncorrelated, their enhanced LyC flux is unlikely associated with
residual sky-background gradients from pinholes. Had that been the case, a
number of our much more numerous ``Galaxies without AGN'' would likely have
also been exposed near pinhole induced sky gradients and shown the same amount
of red-leaked flux, which is not the case. Hence, local residual sky-background
gradients due to pinhole induced red-leak enhancements are not likely to have
affected our LyC measurements, at least not to the level of AB$\simeq$32\,\magarc. 

\subsection{Estimating LyC Contamination from Objects Below the $\chi^2$ Image Detection Limit}
\label{faintcontam}
{Here we estimate the potential non-ionizing contamination to our LyC stacks
from interloping objects below the $\chi^2$ image detection limit. As seen in
\S\ref{sec:method}, the deepest $\chi^2$ images allow us to locate possible
low-redshift contaminants in sub-images to AB$\sim$27.5\,mag, such that they can
be removed from our stacks so they cannot contribute any flux to our photometry.
However, the possibility that fainter objects which may remain undetected in $\chi^2$
detection image must be addressed, since those objects could potentially contribute
some flux within the \SExtractor\ aperture photometry. 

As an example, the F336W stack LyC photometry is taken within an aperture
area of $\sim$0.5\,arcsec$^2$ (see Table \ref{table:table2}). To assess this
fainter contaminating flux, we need to estimate to the total F336W stack sky-surface
brightness from objects undetected at AB$\gtrsim$27.5\,mag. For this, we will use
the galaxy galaxy counts of \citet{Driver2016} from 20 filters ranging from
$\lambda$$\simeq$0.15--500\,\micron. At nearly all wavelengths, their normalized
differential counts (Fig. 8 or A1 in the on-line version only) converge with a
well determined slope of $\frac{\Delta\rho_{\mathrm{L}}}{\Delta\mathrm{m}}$$\simeq$--0.177.
The total sky-surface brightness contributed by each magnitude bin in the F336W
counts peaks at AB$\simeq$24\,mag. The faintest galaxy counts that contribute in
F336W are the UVUDF counts in the HUDF \citep{Teplitz2013}, which reach AB$\sim$28\,mag.
\citet{Driver2016} performed MC tests to determine the uncertainty in the extrapolated
total sky-signal, which is $\lesssim$20\% in F336W.

We will extrapolate this converging signal with the same slope as measured
between AB$\simeq$24 and AB$\simeq$28\,mag to arbitrarily fainter
fluxes, e.g. from 27.5\,mag to 38\,mag. This is the F336W flux that a very dim
(\MAB$\simeq$--10 mag) galaxy would have at z$\simeq$3, where the distance
modulus in Planck 2016 cosmology is DM=47.47\,mag. The actual F336W sky-brightness
in \citet{Driver2016} drops from $\sim$10$^{-28.3}$\,W\,Hz$^{-1}$\,m$^{-2}$\,deg$^{-2}$\,(0.5 mag$^{-1}$)
at AB=27.5 mag to $\sim$10$^{-30.2}$ W\,Hz$^{-1}$\,m$^{-2}$\,deg$^{-2}$\,(0.5 mag$^{-1}$)
at AB=38.0\,mag. Over the 21 contributing 0.5 mag-bins from AB=28 to AB=38\,mag, this
sky-integral is $\sim$10$^{-28}$\,W\,Hz$^{-1}$\,m$^{-2}$\,deg$^{-2}$ or
1.85$\times$10$^{-9}$\,Jy\,arcsec$^{-2}$, or 30.73\,mag\,arcsec$^2$. 

Within the 0.5\,arcsec$^2$ \SExtractor\ aperture, the contribution of contamination
from {\it unresolved, unseen} galaxies between AB=27.5--38\,mag amounts to a total
integrated flux of AB=31.5\,mag. This is well below the level of our AGN LyC detections
in Table~\ref{table:table2} and also fainter than our 1$\sigma$ sky-subtraction errors
in in Table~\ref{table:tableA1} (Column 10). Any such contaminating flux from
unresolved objects at AB$\gtrsim$27.5\,mag will also be present in the eight
neighboring sky-apertures in Fig.~\ref{fig:figureA2}, and so would be statistically
subtracted to first order. Thus, after subtracting all detectable contaminating
neighbors at AB$\lesssim$27.5\,mag using the $\chi^2$ images, statistically the LyC signal
is not significantly affected by contaminating objects below the HST $\chi^2$ image
detection limit of AB$\sim$27.5\,mag.

\section{Sources of Systematic Uncertainties}
\label{sec:systematics}

\subsection{Impact of Gradients in the Residual Sky-Background}
\label{sec:gradients}

Subtle gradients still exist in the new ERS UV mosaics, but at a much reduced
level from the v0.7 ERS mosaics of \citetalias{Windhorst2011}. These are
$\lesssim$3--5\% of the Zodiacal sky values from \emph{corner to corner} across
each of the eight individual 4096$\times$4096 pixel CCD images that were
drizzled onto the UVIS mosaics. This subtle gradient pattern was not very
discernible, but appears to be similar in each of the 8 full WFC3/UVIS CCD
frames in the ERS to a good approximation, and roughly linear across each CCD.
The cause of these remaining gradients could be subtle residual errors in the
on-orbit master bias frames, in the delta-flat corrections used in the recent
WFC3 pipeline reduction, and/or from variation in exposure time or background
noise across the drizzled mosaic \citep{Baggett2012, Mack2013}. These
remaining gradients are too faint to accurately map and remove from individual
WFC3/UVIS UV exposures prior to drizzling, and removal of inaccurately measured
gradients would introduce additional unintended errors in the mosaics. We
therefore will assess the effects that these 3--5\% global gradients have on
the $\lesssim$151 pixel scales at which local sky-subtraction is performed in
the LyC image stacks.

Dividing each LyC stack into our 9 segment ``tic-tac-toe'' grid, we determine
the sky-background level and uncertainty in each of the 8 segments around the
central box that contains the LyC candidate itself, which we exclude from the
sky-background calculation. We compute these by optimally binning the count
rates of the 8 outer segments, then we fit a normal distribution to the inner
quartile of this data, taking the average of the fitted distribution as the 
sky value and the +1$\sigma$ value to be the 84\% of the pixel histogram. We
estimate the gradients in the stacks from the rms value of the fitted average 
count rate in each segment. As shown in the examples of our ``tic-tac-toe''
photometric analysis in Fig.~\ref{fig:figureA2} and tabulated in
Table~\ref{table:tableA1}, we find that any residual sky-background
\emph{gradient} left in the image stacks is $\sim$5.2--40$\times$ ($\sim$1.8--4
mag) fainter than the residual sky-background numbers derived from
Fig.~\ref{fig:figure3}. This then implies that \emph{any} gradients in the
local sky-background in the LyC image stacks (containing 13--19 objects each)
are fainter than $\sim$32.3, 32.1, and 32.5 \magarc\ in WFC3/UVIS F225W,
F275, and F336W across the 
4$\farcs$53\,$\times$\,4$\farcs$53 extent of each sub-stack, respectively. 

These numbers are consistent with the aforementioned $\sim$3--5\% linear
gradient across each of the full WFC3/UVIS mosaic images \emph{before
drizzling}, and corresponds to a $\lesssim$0.2\% error in the sky-subtraction
across typical 151$\times$151 pixel sub-image stacks. For a UV sky brightness of
$\mu_{\rm sky}$$\sim$\,25.5 \magarc\ (see Table~\ref{table:tableA1}), this
amounts to a sky-subtraction error of $\sim$32.3 \magarc\ across a 151$\times$151
pixel stack. 
One possible source of such gradients are residual dark current subtraction
errors. \citet{Rafelski2015} show that the 2009 WFC3/UVIS dark current may vary
between 0.00045\emin/s and 0.00035\emin/s across the CCDs (black curve in their
Fig. 15). From experience, the quality of the calibration files is such that
these gradients are typically subtracted at the level of (conservatively) 
$\sim$20\% of the gradient itself. That is, this dark current subtraction error
across the 151 pixel ``tic-tac-toe'' sub-images (out of 4096
pixels across the two CCDs) will amount to a residual sky subtraction error in the
subimages of approximately: --2.5 log
[0.20$\times$((0.00045/0.00035)-1)*151/4096]$\simeq$6.7 mag below sky. This could
then leave a residual dark current gradient on top of the UV zodiacal sky (25.5
\magarc) of 25.5+6.7 = 32.2 \magarc, consistent with the limits given above.
This level of uncertainty in the local sky-background level may
pose a fundamental limit on the sensitivity and accuracy of any LyC (surface)
photometry, which is slightly fainter than that potentially imposed by
pinholes. The systematic nature of these residual gradients also explains the
slight improvement in SNR noted in Appendix~\ref{sec:dettests} when the
individual F225W sub-images are rotated by random multiples of 90$^{\circ}$.
These residual gradients are also much fainter than the measured LyC signal
(see Table~\ref{table:tableA1}). 

\subsection{Assessment of Possible WFC3/UVIS CTE degradation $\lesssim$4
Months After Launch}
\label{sec:CTE}

Charge Transfer Efficiency (CTE) degrades over time due to high energy cosmic
ray collisions with the detector, and from encounters with relativistic protons
and electrons during \HST's frequent passages through the South Atlantic
Anomaly. Particle damage to the silicon of the CCD can cause areas where
electrons become trapped in the detector's crystal lattice during readout of
the array. The WFC3/UVIS detectors suffer from a CTE loss of $\sim$0.1 mag per
year. After several years in orbit, faint objects ($\lesssim$300\emin) can
lose up to 50\% of their flux during readout \citep{Noeske2012, Bourque2013}. 
CTE degradation can also cause charge trails to be visible in the images,
caused by the delayed release of trapped electrons during readout. Partial
recovery and correction of CTE losses in post-processing of the images is only
possible for brighter sources \citep{Anderson2010,Massey2014}. Flux from very
faint objects cannot be corrected in this manner, as their low electron counts
are lost in the background noise of the detector. Because the WFC3 UV data
were taken less than four months after Shuttle Servicing Mission SM4 that
installed WFC3 onto \HST, the WFC3/UVIS ERS data may not yet suffer from
significant CTE losses. 
%
\noindent\begin{figure}[htp!]
\centerline{
 \includegraphics[width=0.5\txw]{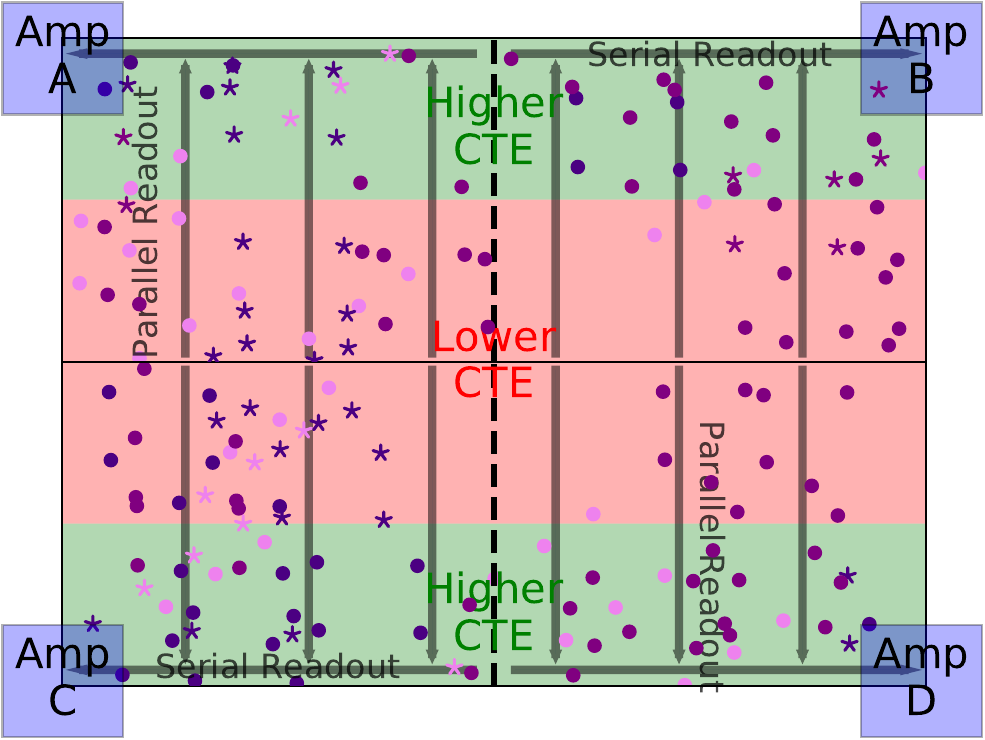}
}
\caption{\noindent\small
Detector layout and observed location of galaxy sub-samples. WFC3/UVIS F225W,
F275W, and F336W images. The two 2k$\times$4k CCDs are shown
as the rectangular panels that stretch from X=1 to 4096 and Y=1 to 2051 pixels.
Within each WFC3 CCD, the green regions indicate the half of each CCD quadrant
that is closest to the corresponding read-out amplifier along the parallel
read-out direction, while the red regions indicate the half that is furthest
away and will undergo more transfers along the array. The colored circles mark
the positions in physical WFC3/UVIS CCD coordinates of each of the
$z$\,$\simeq$2.3--4.8 galaxies with HQ spectra in our sample. Filled circles
mark objects where visible flux is seen in the individual LyC sub-images. The
sub-samples of objects in the green and red regions are referred to as our
``high-CTE'' and ``lower-CTE'' sub-samples, respectively. Averaged over all
objects, the difference between the stacked LyC
signal in the ``high-CTE'' are compared to the ``lower-CTE'' area is
$\Delta$(High-CTE--Lower-CTE) $\simeq$+0.22 mag. This suggests that
the CTE induced systematics are not yet larger than the rms error and other
systematics in the photometry. Since the circles (galaxies without AGN)
and stars (galaxies with AGN) are fairly uniformly
distributed across the CCD, CTE degradation must have not yet been a major
limitation in detecting the faint LyC signal four months after the WFC3
installation. 
\label{fig:figureC1}}
\end{figure}
\noindent\begin{table*}[htp!]
\centering
\caption{Assessment of the Impact of CTE Effects for Galaxies with HQ+IQ spectra\label{table:tableC3}}
\setlength{\tabcolsep}{2pt}
\begin{tabular}{lccrccccc}
	\toprule \\[-19pt]
\colhead{Filter} & \colhead{$z$ range} & \colhead{\zmean} &
\colhead{$N_{\rm obj}$} & \colhead{m$_{\rm LyC}$}   & \colhead{SNR$_{\rm LyC}$} & 
\colhead{A$_{\rm LyC}$} & \colhead{m$_{\rm UVC}$}   & \colhead{SNR$_{\rm UVC}$} \\[-5pt]
\colhead{(1)}           & \colhead{(2)}             & \colhead{(3)}           &
\colhead{(4)}           & \colhead{(5)}             & \colhead{(6)}           & 
\colhead{(7)}           & \colhead{(8)}             & \colhead{(9)} \\[-2pt]
\midrule \\[-21pt]
\multicolumn{9}{l}{\sc All Galaxies, Higher CTE Area:}\\[-4pt]
All & 2.298--4.149 & 2.785 & 30 & 28.78 & 1.41 & 0.76 & 24.60 & 445 \\[-4pt]
\multicolumn{9}{l}{\sc All Galaxies, Lower CTE Area:}\\[-4pt]
All & 2.276--4.120 & 2.764 & 28 & 28.56 & 1.67 & 0.75 & 24.63 & 409 \\[-2pt]
\bottomrule
\vspace*{3pt}
\end{tabular}
\begin{minipage}{\txw}{\small Table columns are as for Table~\ref{table:table2}.}
\end{minipage}
\end{table*}

The CTE may be lower in some regions further away from the amplifiers at
the beginning of readout due to any existing electron traps first being filled nearest
to the amps during readout. However, this effect is very non-linear, as the traps have
an unknown probability of capturing and releasing electrons per each charge
transfer, so electrons further away from from the amps might encounter more
traps if the captured electrons are released before the last columns are read
out. Thus, the \emph{cumulative} CTE will be higher on average closer to the
amps when combining all CTE effects along the \emph{parallel} readout direction.

Fig.~\ref{fig:figureC1} provides an overview of the WFC3/UVIS readout
configuration \citep[see Fig. 6.14 of][]{Dressel2015}. The two 2k$\times$4k WFC3/UVIS
CCDs are shown as rectangular
panels in Fig.~\ref{fig:figureC1}, displaying the total pixel arrays from
X=1--4096 and Y=1--2051 pixels. Within each WFC3/UVIS chip in
Fig.~\ref{fig:figureC1}, the green band indicates the half of each CCD 
\emph{closest} in the parallel read-out direction to the four corresponding
amplifiers (labeled as Amp A, B, C, D, respectively), where the cumulative CTE
degradation effects from CR and relativistic particle hits accumulated on-orbit
are expected to be \emph{lowest}. We refer to this higher CTE region as the
``higher CTE'' area. The red bands in each chip indicate the half of each WFC3
CCD \emph{furthest} from the amplifier in the $Y$ read-out direction, where the
effects from CTE \emph{degradation} would be more substantial. The cumulative
CTE value itself from these areas will be lower compared to the ``higher CTE''
regions, since its signal will have on average been read-out through more
charge transfer rows. We refer to this region as the ``lower CTE'' area. 

The colored circles and stars in Fig.~\ref{fig:figureC1} show, \emph{in physical
WFC3/UVIS CCD (X,Y) coordinates}, where all our galaxies without and with AGN,
respectively, at $z\simeq$2.3--4.1
were observed in all individual exposures within the CCDs for 
WFC3/UVIS. We note that the galaxy
sample was specifically selected to have high reliability in spectroscopic
redshifts (see \S\ref{sec:selection} and Tables~\ref{table:table2}---\ref{table:table3}),
therefore any apparent
spatial correlation of the objects in the CCD coordinates is not due to
CTE degradation, but from the spectroscopic object selection.

There are four drizzled exposures in both WFC3/UVIS F225W and F275W, and three
exposures in WFC3/UVIS F336W. Every object with a spectroscopic
redshift to AB$\lesssim$25 mag is therefore plotted up to 11$\times$ in the
WFC3/UVIS in Fig.~\ref{fig:figureC1} to monitor their actual locations in the
individual WFC3/UVIS F225W, F275W, and F336W exposures. The somewhat apparent
clustering of ERS galaxies with the most
reliable redshifts in Fig.~\ref{fig:figureC1} is caused by this repetitive 
plotting of the individual locations of the galaxies in each exposure,
overemphasizing the appearance of any real clustering in the mosaic. It is
possible that the way the VLT and other spectroscopic masks were configured to
observe galaxies for redshift measurements --- while maximally avoiding
spectral overlap --- further introduced some apparent clustering in a
particular region of Fig.~\ref{fig:figureC1}. Since the VLT spectroscopic masks
are comparable in size to the WFC3 FOV, it is possible that the objects who
received the most integration time during spectroscopy were more preferentially
selected to be on one side of the FOV to minimize the spectral
overlap. 

With these two separate regions subdivided by their aggregate CTE, we can now
determine if and how much the LyC measurements vary based on their location
within each chip, and if CTE was already significantly degraded in the September
2009 WFC3/UVIS mosaics less than four months after WFC'3s launch. Similar to
our randomly subdivided test stacks in Appendix~\ref{sec:dettests}, we combined
these two sub-samples in each filter to compare their photometry, background,
and noise levels. Some objects appear both in the ``higher CTE'' and
``lower CTE'' areas of Fig.~\ref{fig:figureC1}, which occurred due to
the dithering of the individual WFC3/UVIS ERS exposures. These objects were
added to the corresponding sub-stacks for assessing the effects from
CTE degradation. Fig.~\ref{fig:figureC1} show that about as many
objects with spectroscopic redshifts are located in the green (``higher CTE'')
areas as in the red (``lower CTE'') areas, allowing us to make a quantitative
comparison between the two subsets to assess the possible effects of
differential CTE across the detector.

The results for the two CTE sub-samples are shown in the bottom two sets
of rows in Table~\ref{table:tableC3}. From these measurements, as a function of
relative position on the CCDs, it is clear that each of these two CTE
sub-samples still yields LyC detections, with correspondingly
larger errors due to the smaller number of objects in each sub-sample. In
particular, the LyC flux in the ``lower CTE'' sub-sample remains at least as
significant as it is for the ``higher CTE'' sub-sample: on average, the LyC flux
in the ``lower CTE'' sub-sample is 0.22 mag brighter than in
the ``higher CTE'' sub-sample (Table~\ref{table:table2}). This differences in
photometry between the ``higher CTE'' and ``lower CTE'' sub-samples --- as
averaged over the four filters --- is not significant, but its sign is
such that the ``lower CTE'' sub-sample is actually somewhat brighter than the
``higher CTE'' sub-sample. \emph{Therefore this difference does not point to
significant WFC3 CTE effects as of Sept. 2009.} The difference also varies
fairly randomly between the four filters, showing no general trend with
$Y$ position on the CCD's. Small differences in the observed LyC flux from the
two different regions on each CCD may also be caused by some intrinsic
variation of LyC \fesc\ between the two sub-samples, and/or by the small number
statistics available in both sub-samples in general. Thus, any CTE induced
systematics four months after WFC3's launch are not yet larger than the random
errors in the faint LyC signal. This is consistent with the decline in CTE of
$\sim0.1$ mag/year measured by \citep{Noeske2012, Bourque2013}. 

The AGN stacks are typically brighter than the stacks without AGN. This 
information allows us to do a simple
statistical analysis to see if CTE degradation has already affected the
fraction of objects with, on average, brighter LyC flux between the ``higher CTE''
and ``lower CTE'' areas of the WFC3 CCDs. For this, we used the total of 12 AGN
and 36 galaxies with no AGN (Table \ref{table:table2}) covered
by WFC3 UVIS images in the filters F225W, F275W, and F336W,
respectively. Since these were taken in 4, 4, and 3 dither points each, this
resulted in 4$\times$2+4$\times$7+3$\times$3=45 stars in
Fig.~\ref{fig:figureC1} for AGN, and
4$\times$17+4$\times$7+3$\times$10=126 circles for galaxies, or a total of
171 points, 82 of which are in the
``higher CTE'' area, and 89 in the ``lower CTE'' area. 

Of these points indicated in Fig.~\ref{fig:figureC1}, 21/45 or
$\sim$47\% occurred in the ``higher CTE'' area and 24/45 or 53\% occurred in the
``lower CTE'' area. Since the circles are fairly uniformly distributed
across the both individual WFC3 CCDs, and AGN exist in
regions furthest away from the readout amplifiers, CTE must have been high
enough for these faint detections to survive the multiple transfers during
readout without becoming trapped in the detector. Hence, the WFC3 UVIS data has not
suffered from CTE effects that affected the
possible fraction of LyC detections within 4--5 months after launch for WFC3. 

Both of these tests suggest that, less than four months after WFC3's launch,
CTE degradation in the UVIS chip is not yet at a level that significantly
affects the readout and subsequent photometry of faint flux to within the
errors of our measurements, nor has it biased the distribution of objects with
marginal individual LyC detections across the CCD. If this had been the case,
we would have seen significantly larger differences in LyC flux across the two
CCD detectors in the parallel readout direction, and would in fact have seen
faint objects fully disappear into the noise in the ``lower CTE'' regions in
Fig.~\ref{fig:figureC1}, in addition to visible charge trails for all brighter
objects, none of which appear in our data.

\section{Modeling and Uncertainties}
\label{sec:model-errors}

\subsection{SED Fitting Uncertainties due to Extinction }
\label{sec:extinction}

\noindent\begin{figure}[ht!]
\centerline{\includegraphics[width=0.50\txw]{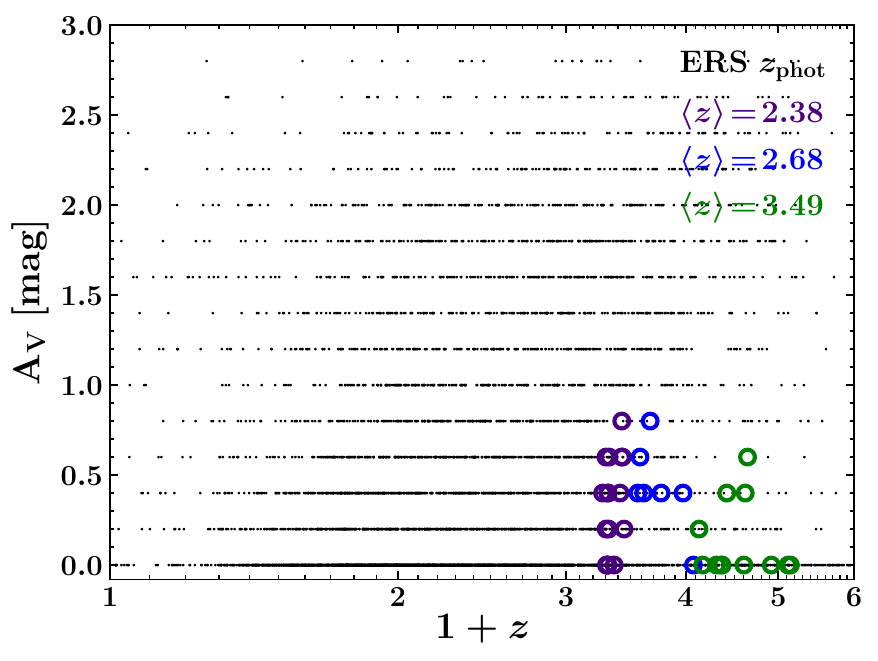}}
\caption{\noindent Distribution of dust extinction \AV\ values from best fit 
\citet{Calzetti2000} dust attenuated BC03 SEDs for all galaxies in the 10 band
ERS data (black dots), compared to the spectroscopic samples used for the stacks
in the three indicated redshift bins (colored open circles for galaxies).
\label{fig:figureD1}}
\end{figure}
In order to obtain accurate estimates for LyC and UVC dust extinction and
subsequent absolute LyC escape fractions, we must adopt the best available dust
attenuation models for galaxies with accurate redshifts and no contaminating
AGN signatures when performing the minimized $\chi^2$ stellar SED fits that
results in the most likely \AV\ values from the observed panchromatic ERS data.
From SED fitting of \citet{Calzetti2000} attenuated BC03 models using the
10 band WFC3+ACS photometry of all $\sim$6900 galaxies at
2\,$\lesssim$\,$z$\,$\lesssim$\,6 within the GOODS-S ERS field
(\citealt*{Windhorst2010}; \citetalias{Windhorst2011}, see also
\S\ref{sec:fescintro}), we find that most ERS galaxies have
0.0\,$\lesssim$\,\AV\,$\lesssim$\,1.0\,mag.

Fig.~\ref{fig:figureD1} shows the distribution of dust extinction
\AV\ values from best fit SEDs for all galaxies in the 10 band ERS data (small
black dots), compared to our spectroscopic sample in the four indicated
redshift bins. The SED fitting sampled the \AV\ parameter space in 0.2 mag
intervals. Table~\ref{table:table3} lists
the \AVmed\ values and their $\pm$1$\sigma$ ranges a function of redshift.

Fig.~\ref{fig:figureD1} shows that the \AV\ values of our galaxy
samples with spectroscopic redshifts are consistent with those found for the
entire ERS sample of 6900 galaxies with 10 band fitted photometric redshifts to
\mAB$\lesssim$27\,mag. This implies that our galaxies \emph{are} sampling the
available parameter space of \AV\ values \emph{at their approximate redshifts}. 
Since the \AV\ uncertainty in the SED fits is unknown, the MC simulated \fescabs\
values in \S\ref{sec:fescintro} and \S\ref{sec:fescMC} do not include an \AV\
uncertainty, even though they utilize the extinction corrected (intrinsic)
SEDs. This implicit \AV\ error is one of the dominant errors in the
\fescabs\ calculation, but is less important than the IGM transmission variations
in the MC derived \fesc\ values (see \S~\ref{sec:fescMC} and
\S\ref{sec:fescresults}). We also note that there may exist a degeneracy
between the \AV\ values and the ages of the best fit SEDs, which would also add
to the uncertainty of the \AV\ values. The \AV\ induced error can be as large as
the combined uncertainty of the photometric observations that we fit, and can
further increase the overall uncertainty of the SED fit. Furthermore, the error
in $R_V$ derived by \citet{Calzetti2000} ($R_V$=4.05$\pm$0.80) is not
propagated into the SED fit, which would also increase the \AV\ error. However,
the uncertainty in IGM transmission, which is primarily due to variations in
sight-lines and redshifts in the stacks \citep[see][]{Inoue2008}, dominates the
error in the MC \fesc\ values, and so an additional \AV\ error would only
slightly increase their $\pm$1$\sigma$ values, as discussed in
\S\ref{sec:fescMC} and Table~\ref{table:table3}.

Table~\ref{table:table3} also shows that \AVmed, for the samples with 
spectroscopic redshifts, increases from $\sim$0 mag at
z$\simeq$3.1--4.1 to $\sim$0.4 mag at z$\simeq$2.3--3, consistent with the
behavior seen in the entire ERS sample as a gradual increase in reddening
towards the lower redshifts, when the stellar populations have aged more and
produced more dust over cosmic time. Hence, the median \AV\ values and their
rms for galaxies appears to increase at the lower redshifts.

\subsection{LyC and UVC Surface Brightness Model Details}
\label{sec:sbapp}

For the UVC and LyC SB models in \S\ref{sec:modelradprof} and
Fig.~\ref{fig:figure12}, we generate the UVC surface brightness as following: 
\begin{equation}
SB_{\uvc}(b) =\int_{-\infty}^{\infty} ds \hspace{1mm} \epsilon_{\uvc}(b,s).
\end{equation} 
where $b$ denotes the impact parameter, $s$ denotes the line-of-sight
coordinate, and $\epsilon_{\uvc}(b,s)$ denotes the emission rate of UVC photons
per unit volume at $(b,s)$. The distance from the galaxy $r$ is defined as
$r=\sqrt{b^2+s^2}$ (Note that $2rdr = 2sds$ and
$ds=\frac{r}{s}dr=\frac{r}{\sqrt{r^2-b^2}}dr$, since $r^2 = s^2 +b^2$). We then obtain
the LyC surface brightness from:
\begin{equation}
SB_{\!\lyc}(b) =\int_{-\infty}^{\infty} ds \hspace{1mm} \epsilon_{\rm
UVC}(b,s)\frac{f_{\!\lyc}}{f_{\uvc}}\fesc(b,s).
\end{equation} 
where $\fesc(b,s)$ denotes the fraction of LyC photons that can escape from
$(b,s)$, and the factor $\frac{f_{\rm LyC}}{f_{\rm UVC}}$ simply rescales the
flux at UVC frequencies to that at LyC.

In our model of a clumpy ISM $f_{\rm c}(r)$ denotes \emph{the number of
self-shielding clumps per unit length at $r$} \citep[see][]{Dijkstra2012}. We
therefore find that $f_{\rm c}(r)dr$ denotes the number of self-shielding
clumps along a differential length $dr$. We assume that each clump is optically
thick to ionizing photons. In this case, the escape fraction from $(b,s)$ is
simply the probability of finding \emph{at least} one clump on a sight-line to
$(b,s)$, $P_{\rm clump}(b,s)$, which is given by:
\begin{equation}
P_{\rm clump}(b,s)=1-P_{\rm no clump}(b,s)=1-\exp[-N_{\rm clump}(b,s)],
\end{equation}
where in the last step we assumed that the number of clumps along a given
line-of-sight follows a Poisson distribution with mean $N_{\rm clump}(b,s)$.
This mean is given by: 
\begin{equation}
N_{\rm clump}(b,s)=\int_{-\infty}^{s} ds' \hspace{1mm} f_{\rm c}(b,s').
\end{equation}

\begin{table*}[t!]
\centering
\caption{List of Individual Galaxies reliable spectra}
\small
\setlength{\tabcolsep}{2pt}
\begin{tabular}{lccccrr}
\toprule \\[-19pt]
\colhead{RA \ (J2000.0)} & \colhead{Dec \ (J2000.0)}  & \colhead{$z$} & \colhead{$m_{J,{\rm AB}}$} & \colhead{$M_{\!1500\mathrm{\mathring{A}}}^{\mathrm{AB}}$} & \colhead{($V\!-\!I$)} & \colhead{AGN?}\\[-5pt]
\colhead{[deg]}  & \colhead{[deg]} & \colhead{\null} & 
 \colhead{[mag]} & \colhead{[mag]} & \colhead{[mag]} & \colhead{\null} \\[2pt]
\midrule \\[-19pt]
53.006583 & $-$27.724170 & 2.7212 &  22.737  & $-$21.20 & 0.131    & yes \\[-5pt]
53.008846 & $-$27.724348 & 2.7260 &  23.919  & $-$19.72 & 0.811    & yes \\[-5pt]
53.012648 & $-$27.747244 & 2.5730 &  23.698  & $-$19.33 & 0.683    & yes \\[-5pt]
53.013515 & $-$27.755235 & 3.2171 &  24.381  & $-$21.30 & $-$0.023 & yes \\[-5pt]
53.013889 & $-$27.756827 & 2.3170 &  23.705  & $-$21.32 & $-$0.098 & no  \\[-5pt]
53.014539 & $-$27.727922 & 3.1320 &  24.114  & $-$21.12 & 0.283    & no  \\[-5pt]
53.020573 & $-$27.742150 & 3.4739 &  23.417  & $-$22.23 & 0.169    & yes \\[-5pt]
53.020927 & $-$27.770185 & 3.9170 &  24.484  & $-$21.60 & 0.598    & no  \\[-5pt]
53.033326 & $-$27.782577 & 2.6123 &  24.043  & $-$20.11 & 0.241    & yes \\[-5pt]
53.034441 & $-$27.698210 & 2.4700 &  24.552  & $-$19.77 & 0.094    & yes \\[-5pt]
53.035231 & $-$27.744125 & 4.1486 &  25.388  & $-$20.91 & 0.559    & no  \\[-5pt]
53.040821 & $-$27.719068 & 2.3021 &  22.509  & $-$21.80 & 0.133    & no  \\[-5pt]
53.042456 & $-$27.737862 & 2.3036 &  22.956  & $-$21.25 & 0.097    & no  \\[-5pt]
53.062429 & $-$27.735634 & 2.6730 &  25.336  & $-$19.01 & 0.220    & no  \\[-5pt]
53.065221 & $-$27.742901 & 2.6160 &  23.539  & $-$21.40 & 0.061    & no  \\[-5pt]
53.065771 & $-$27.695980 & 3.6433 &  24.415  & $-$21.15 & 0.381    & no  \\[-5pt]
53.072558 & $-$27.744441 & 2.6503 &  24.402  & $-$20.86 & $-$0.108 & yes \\[-5pt]
53.078023 & $-$27.731020 & 2.4160 &  23.373  & $-$20.82 & 0.165    & no  \\[-5pt]
53.078800 & $-$27.693745 & 2.3060 &  24.363  & $-$20.29 & $-$0.033 & no  \\[-5pt]
53.079284 & $-$27.691368 & 2.4352 &  23.072  & $-$21.01 & 0.143    & no  \\[-5pt]
53.095384 & $-$27.687524 & 3.3565 &  24.771  & $-$20.82 & 0.230    & no  \\[-5pt]
53.100815 & $-$27.715987 & 2.2980 &  23.162  & $-$20.15 & 0.142    & yes \\[-5pt]
53.102783 & $-$27.759367 & 2.3115 &  23.792  & $-$21.15 & $-$0.046 & no  \\[-5pt]
53.113001 & $-$27.745551 & 2.3230 &  24.272  & $-$19.73 & 0.081    & no  \\[-5pt]
53.117831 & $-$27.734305 & 3.2560 &  22.767  & $-$21.90 & 0.408    & yes \\[-5pt]
53.120610 & $-$27.736585 & 3.3680 &  24.630  & $-$21.25 & 0.186    & no  \\[-5pt]
53.121611 & $-$27.672921 & 3.3083 &  24.994  & $-$20.76 & 0.095    & no  \\[-5pt]
53.131718 & $-$27.669018 & 3.0762 &  23.951$^\dagger$  & $-$21.88 & 0.133    & no  \\[-5pt]
53.134558 & $-$27.690656 & 2.3200 &  22.968  & $-$21.72 & 0.057    & no  \\[-5pt]
53.134819 & $-$27.713359 & 2.4300 &  23.143  & $-$20.87 & 0.184    & no  \\[-5pt]
53.138759 & $-$27.700469 & 2.4500 &  23.285  & $-$20.51 & 0.250    & yes \\[-5pt]
53.144489 & $-$27.728071 & 2.2760 &  24.519  & $-$19.73 & 0.115    & no  \\[-5pt]
53.145431 & $-$27.698008 & 2.3129 &  23.781  & $-$21.00 & $-$0.015 & no  \\[-5pt]
53.145621 & $-$27.685249 & 2.7708 &  24.084  & $-$21.09 & 0.121    & no  \\[-5pt]
53.149223 & $-$27.748588 & 2.5658 &  23.770  & $-$21.27 & $-$0.036 & no  \\[-5pt]
53.149815 & $-$27.697213 & 3.6180 &  23.903  & $-$21.83 & 0.315    & no  \\[-5pt]
53.151291 & $-$27.742911 & 3.4173 &  23.943  & $-$21.44 & 0.426    & no  \\[-5pt]
53.157430 & $-$27.709016 & 2.9752 &  23.636  & $-$21.57 & 0.219    & no  \\[-5pt]
53.158912 & $-$27.742675 & 2.3277 &  22.417  & $-$21.19 & 0.230    & no  \\[-5pt]
53.161953 & $-$27.722657 & 2.4490 &  23.563  & $-$21.10 & 0.087    & no  \\[-5pt]	
53.167996 & $-$27.711349 & 2.5845 &  23.776  & $-$21.14 & 0.064    & no  \\[-5pt]
53.168265 & $-$27.741939 & 4.1200 &  25.013  & $-$21.11 & 0.490    & no  \\[-5pt]
53.174442 & $-$27.733297 & 2.5760 &  24.139  & $-$19.99 & 0.235    & yes \\[-5pt]
53.181805 & $-$27.729920 & 2.3168 &  23.211  & $-$20.77 & 0.136    & no  \\[-5pt]
53.182798 & $-$27.705269 & 2.3682 &  24.187  & $-$21.02 & $-$0.092 & no  \\[-5pt]
53.182838 & $-$27.734909 & 2.4284 &  22.744  & $-$20.89 & 0.316    & no  \\	
\bottomrule
$\dagger$ from \citet{Hsieh2012}
\vspace*{3pt}
\end{tabular}
\end{table*}